\documentclass[aps,prd,groupedaddress,preprintnumbers]{revtex4}
\newcommand{\lbl}[1]{\label{eq:#1}}
\newcommand{\rf}[1]{(\ref{eq:#1})}
\newcommand{\vs}[1]{\rule[- #1 mm]{0mm}{#1 mm}}
\newcommand{\be}{\vs{2}\begin{equation}}
\newcommand{\eq}{\vs{2}\begin{equation}}
\newcommand{\en}{\\[2mm]\end{equation}}
\newcommand{\ee}{\\[2mm]\end{equation}}
\newcommand{\bea}{\begin{eqnarray}}
\newcommand{\ena}{\end{eqnarray}}
\newcommand{\eea}{\end{eqnarray}}

\newlength{\dhatheight}

\usepackage{epsf,epsfig,psfrag,graphics,color}

\begin{document}

\renewcommand{\theequation}{\Roman{section}.\arabic{equation}}


\preprint{LPT-ORSAY/15-08}

\preprint{CPT-P001-2015}

\title{On some aspects of isospin breaking in the decay $K^\pm \to \pi^0 \pi^0 e^\pm \stackrel{_{(-)}}{\nu_e}$}

\author{V. Bernard}
\email{bernard@ipno.in2p3.fr}
\affiliation{Groupe de Physique Th\'eorique, Institut de Physique Nucl\'eaire\\
B\^at. 100, CNRS/IN2P3/Univ. Paris-Sud 11 (UMR 8608), 91405 Orsay Cedex, France}

\author{S. Descotes-Genon}
\email{sebastien.descotes-genon@th.u-psud.fr}
\affiliation{Laboratoire de Physique Th\'eorique, CNRS/Univ. Paris-Sud 11 
(UMR 8627), 91405 Orsay Cedex, France}

\author{M. Knecht}
\email{knecht@cpt.univ-mrs.fr} 
\affiliation{Centre de Physique Th\'eorique,
CNRS/Aix-Marseille Univ./Univ. du Sud Toulon-Var (UMR 7332)\\
CNRS-Luminy Case 907, 13288 Marseille Cedex 9, France}

\date{\today}

\begin{abstract}
\noindent
Two aspects of isospin breaking in the decay $K^\pm \to \pi^0 \pi^0 e^\pm \stackrel{_{(-)}}{\nu_e}$
are studied and discussed. The first addresses the possible influence of the phenomenological
description of the unitarity cusp on the extraction of the normalization of the form factor from
data. Using the scalar form factor of the pion as a theoretical laboratory, we find that this determination
is robust under variations of the phenomenological parameterizations of the form factor.
The second aspect concerns the issue of radiative corrections. We compute the radiative corrections
to the total decay rate for $K^\pm \to \pi^0 \pi^0 e^\pm \stackrel{_{(-)}}{\nu_e}$ in a setting
that allows comparison with the way radiative corrections were handled in the 
channel $K^\pm \to \pi^+ \pi^- e^\pm \stackrel{_{(-)}}{\nu_e}$. We find that once radiative
corrections are included, the normalizations of the form factor as determined experimentally
from data in the two decay channels come to a better agreement. The remaining
discrepancy can easily be accounted for by other isospin-breaking corrections, mainly
those due to the difference between the masses of the {\it up} and {\it down} quarks.

\end{abstract}

\maketitle

\section{INTRODUCTION} \label{intro}
\setcounter{equation}{0}

The program of analysing  $K_{\ell 4}$ decays of the charged kaon conducted
by the NA48/2 collaboration at the CERN SPS has so far been very successful.
In the $\pi^+\pi^-$ channel of the electron mode, $\ell = e$
[the decay $K^\pm \to \pi^+ \pi^- e^\pm \stackrel{_{(-)}}{\nu_e}$ 
will henceforth be refered to as $K^{+-}_{e4}$], it has led, besides 
a more precise determination of the corresponding branching ratio and hadronic form factors 
\cite{Batley:2012rf}, to a very accurate determination of the $\pi\pi$ $S$-wave scattering lengths $a_0^0$ and
$a_0^2$ \cite{Batley:2007zz,Batley:2010zza}, that constitutes a stringent test of the QCD prediction obtained within
the framework of chiral perturbation theory \cite{Weinberg:1966kf,Gasser:1983kx,Colangelo:2000jc,Colangelo:2001df}.

More recently, the results concerning an analysis of the data obtained
in the $\pi^0\pi^0$ channel of the electron mode 
[the decay $K^\pm \to \pi^0 \pi^0 e^\pm \stackrel{_{(-)}}{\nu_e}$ 
will henceforth be referred to as $K^{00}_{e4}$] have also become
available \cite{Batley:2014xwa}.
Although the number of events is lower [$\sim 6.5 \cdot 10^4$ events in the
$K^{00}_{e4}$ mode vs. $\sim 10^6$ events for $K^{+-}_{e4}$],
this allows for some cross checks at the level of the structure of one of the form factors,
that is identical for the two channels in the isospin limit. The normalisation of this common 
form factor, as measured in the two channels, reads \cite{Batley:2012rf,Batley:2014xwa}
\bea
\vert V_{us} \vert f_s [K^{+-}_{e4}] &=& 1.285 \pm 0.001_{\rm stat} \pm 0.004_{\rm syst} \pm 0.005_{\rm ext}
,
\nonumber\\
( 1+ \delta_{EM} )\vert V_{us} \vert f_s [K^{00}_{e4} ] &=& 1.369 \pm 0.003_{\rm stat} \pm 0.006_{\rm syst} \pm 0.009_{\rm ext}
.
\lbl{exp_comp}   
\eea
Ignoring for the time being the correction factor $\delta_{EM}$ (we will discuss radiative
corrections below), the difference of the two values, as compared to the value measured in the 
$K^{+-}_{e4}$ channel, amounts to 6.5\% in relative terms. 
This might be considered as a small difference, but given the uncertainties, it is, in statistical 
terms, quite significant. Adding all errors in quadrature \footnote{Obtaining the values
in Eq. \rf{exp_comp} from the measurements \cite{Batley:2012rf,Batley:2014xwa} of the corresponding 
branching ratios involves the lifetime $\tau_{K^\pm}$ of the charged kaon, whose uncertainty contributes 
to the``external" error bars.
The ratio in Eq. \rf{discrepancy}, however, does not depend on $\tau_{K^\pm}$ anymore, which lowers
the contribution of the ``external" uncertainties to Eq. \rf{discrepancy}. At the level of 
precision shown, this does not impinge on the uncertainty in Eq. \rf{discrepancy}.
We are indebted to B. Bloch-Devaux for drawing our attention
to this point.} this gives 
\be
( 1+ \delta_{EM} ) \frac{f_s [K^{00}_{e4}]}{ f_s [K^{+-}_{e4}]}
= 1.065 \pm 0.010
.
\lbl{discrepancy}
\ee

It seems difficult to ascribe a variation of 6.5\% to the radiative correction factor $\delta_{EM}$ alone.
While in some regions of phase space radiative corrections can reach the $\pm 10$\%
level, they usually sum up to $\pm 1$\% in the decay rate.
The radiative corrections to the $K^{+-}_{e4}$ decay mode have been
discussed in several places \cite{Vesna04,Cuplov:2003bj,Stoffer:2013sfa}
at one-loop precision in the low-energy expansion. But no comparable
study has been done for the $K^{00}_{e4}$ decay mode. There exists 
an older, less systematic, analysis \cite{NuovoCim} that covers the corrections due to
virtual photon exchanges and real photon emission, which could provide
the relevant contributions at a first stage, but its practical use
is somewhat limited, since the expressions given there are not always
very explicit, and moreover need to be checked.
Furthermore, not all radiative corrections occurring in the charged $K_{e4}^{+-}$ channel
\cite{Vesna04,Cuplov:2003bj,Stoffer:2013sfa} have been taken into account
in the analysis of the experimental data. These
additional radiative corrections could affect $f_s [K^{00}_{e4}]$
and $f_s [K^{+-}_{e4}]$ in different ways, and make up for another
part of the discrepancy.

If $\delta_{EM}$ alone does not explain the discrepancy \rf{discrepancy},
one has to look for other sources of isospin-breaking effects. These can
be due to the difference between the {\it up} and {\it down} quark masses $m_u$ and $m_d$,
conveniently described by the parameter $R$, with $1/R = (m_d - m_u)/(m_s - m_{ud})$,
where $m_s$ is the mass of the {\it strange} quark, whereas $m_{ud}$ denotes
the average mass of the {\it up} and {\it down} quarks, $m_{ud} = (m_u + m_d)/2$.
For instance, at lowest order in the chiral expansion, one has \cite{Vesna04,Nehme:2003bz}
\begin{equation}
\frac{f_s [K^{00}_{e4}] }{ f_s [K^{+-}_{e4}]} = \left( 1 + \frac{3}{2 R} \right)
.
\lbl{R-corr}
\end{equation}
Barring contributions of higher-order corrections, values of $R$ as small as \cite{Aoki:2013ldr} $R = 35.8(1.9)(1.8)$ can  
account for about two thirds of the effect in Eq. \rf{discrepancy}.

Finally, there are also isospin breaking effects induced by
the mass difference between charged and neutral pions. 
Most notable from this point of view is the presence of a unitarity 
cusp \cite{Batley:2014xwa} in the form
factor describing the amplitude of the $K^{00}_{e4}$ mode. The interpretation
of this cusp is by now well understood, and as in the case of the $K^\pm\to\pi^0\pi^0\pi^\pm$
decay \cite{{BudiniFonda61,Cabibbo:2004gq,Cabibbo:2005ez,Gamiz:2006km,Colangelo:2006va}}, 
it arises from the contribution of a $\pi^+\pi^-$ intermediate state in the unitarity sum [for a 
general discussion of the properties of the $K_{e4}$ form factors from the point of view
of analyticity and unitarity, see Ref. \cite{Bernard:2013faa} and references therein].

This cusp contains information on the combination $a_0^0 - a_0^2$ that describes
the amplitude for the process $\pi^+\pi^- \to \pi^0 \pi^0$ at threshold. Although this
information probably cannot be extracted from the $K^{00}_{e4}$ data in a way as statistically significant as 
the determination from the cusp in $K^\pm\to\pi^0\pi^0\pi^\pm$ \cite{NA48-Kpi3}, 
it is nevertheless important to include a correct description of this cusp
in the parameterisation of the form factor used to analyse the data. This necessity
has been demonstrated in full details in the case of the $K^\pm\to\pi^0\pi^0\pi^\pm$ decay,
and it is to be expected that the same attention to these matters 
should be paid also in the analysis of the $K^{00}_{e4}$ data.
Failure to do so may introduce a systematic bias which would make the comparison with the 
information on the form factor extracted from the $K^{+-}_{e4}$ data spurious to some extent.

It is the purpose of the present note to address some of these issues.
In a first step, we investigate the possible influence that
various parameterisations of the form factors could have
on the outcome of the analysis. In order to control
inputs and ouputs fully, we choose to work with a simplified model, where
the exact form factors are known from a theoretical point of view, and 
where one can assess the effects of various choices of parameterisations for the form factors used in 
order to analyse the numerically generated data [that we will henceforth refer to as pseudo-data]. This framework
is provided by the scalar form factors of the pions, defined as
\begin{eqnarray}
\langle \pi^0(p_1) \pi^0(p_2) \vert {\widehat m}({\overline u}u + {\overline d}d)(0)\vert\Omega\rangle
 &=&
 + F_S^{\pi^0}(s)
\quad { }
[s\equiv (p_1 + p_2)^2]
,
\nonumber
\\ 
\langle \pi^+(p_{\mbox{\tiny{$ +$}}}) \pi^-(p_{\mbox{\tiny{$ -$}}}) 
\vert {\widehat m}({\overline u}u + {\overline d}d) (0)\vert\Omega\rangle
 &=&
- F_S^{\pi}(s) 
\quad { }
[s \equiv (p_{\mbox{\tiny{$ +$}}} + p_{\mbox{\tiny{$ -$}}} )^2 ]
.
\lbl{eq:F_S}
\end{eqnarray}
Expressions of these form factors, with isospin-breaking contributions due to the difference
of masses between charged and neutral pions included, have been recently obtained in \cite{DescotesGenon:2012gv}
up to and including two loops in the low-energy expansion.
We will use these expressions 
in order to generate pseudo-data, which we can then submit to analysis, using various parameterisations
for the form factors, inspired by those in use for the analyses of the $K^{+-}_{e4}$
and $K^{00}_{e4}$ experimental data. The reason for working with the scalar form factors is at least
twofold. First, the form factors, with isospin-breaking effects included, are known at two loops
in both channels, whereas in the $K_{e4}$ case, only the form factors in the channel with two
charged pions have been studied at the same level  of accuracy as far as isospin-breaking corrections 
are concerned \cite{Bernard:2013faa} [see Ref. \cite{Colangelo:2008sm} for a systematic study at one loop].
Second, the $K_{e4}$ form factors depend on two more kinematical variables, besides the di-pion invariant mass.
The scalar form factors depend only on the latter, and offer therefore a simple kinematical
environment, so that the issues we wish to focus on can be addressed without unnecessary
additional complications.

In a second step, we address the issue of radiative corrections to the
total decay rate of the decay $K^\mp \to \pi^0 \pi^0 e^\mp \stackrel{_{(-)}}{\nu_e}$.
Our intent here is not to develop a full one-loop calculation,
at the same level of precision as those that exist for the decay channel
into two charged pions \cite{Vesna04,Cuplov:2003bj,Stoffer:2013sfa}.
We rather aim at  providing a simple estimate for the radiative corrections to the
total decay rate, much in the spirit of Refs. \cite{NuovoCim} and \cite{Bystritskiy:2009iv}
or, on a more general level, of Ref. \cite{Isidori:2007zt}. This will allow
us to assess how much of the discrepancy \rf{discrepancy} has to be ascribed
to other isospin-breaking effects in the form factors, such as discussed above.

The remainder of this study is then organised in the following way.
First, we give (Section \ref{sec:theory}) a theoretical discussion of the structure of the scalar
form factors of the pions using the explicit expressions obtained
in Ref. \cite{DescotesGenon:2012gv}. We will thus adapt the discussion
of Ref. \cite{Batley:2014xwa} to the case at hand. Working on this analogy will
allow us to give an assessment of some additional assumptions regarding
the structure of the form factors implicitly made in Ref. \cite{Batley:2014xwa}.
Next, we generate pseudo-data (Section \ref{sec:analyses}) using the known two-loop expressions of the form
factors, that we then analyze using various phenomenological parameterisations,
that do not necessarily comply with the outcome of Section \ref{sec:theory}.
The purpose here is to discuss in a quantitative way the possible systematic biases 
that can be induced by these different choices. The last part of Section \ref{sec:analyses}
addresses the determination of the combination $a_0^0 - a_0^2$ of
$S$-wave scattering lengths. Radiative corrections, aiming at an
estimate of the correction factor $\delta_{EM}$ in Eq. \rf{exp_comp},
are discussed in Section \ref{sec:rad_corr}. We first compute
radiative corrections to the $K_{e4}^{00}$ decay rate in a similar set-up
to the one used for the treatment of the data in the $K_{e4}^{+-}$ channel,
in order to obtain a meaningful comparison between the two channels.
Then we compute the effects of additional photonic corrections, not
included in this treatment. Finally, we end our study with a summary and conclusions.
Two Appendices contain technical details relevant for the discussions
in Section \ref{sec:theory} and in Section \ref{sec:rad_corr}, respectively.

\section{Describing the cusp: theory} \label{sec:theory}
\setcounter{equation}{0}

According to the general analysis of Ref. \cite{Cabibbo:2005ez},  
the occurence of both $\pi^0\pi^0$ and $\pi^+\pi^-$ intermediate
states at different thresholds leads to the following structure for the scalar 
form factor of the neutral pion ${F}_S^{\pi^0}(s)$ [$M_\pi$ stands for the charged-pion mass,
whereas $M_{\pi^0}$ is the mass of the neutral pion]:
\begin{eqnarray}
e^{-i \delta (s)} {F}_S^{\pi^0}(s) \ =\  
\left\{
\begin{array}{l}
{\cal F}_0^{\pi^0} (s) - i {\cal F}_1^{\pi^0} (s)  \quad [s \ge 4 M_\pi^2 ]\\
\\ 
{\cal F}_0^{\pi^0} (s) + {\cal F}_1^{\pi^0} (s)  \quad\,\, [4 M_{\pi^0}^2 \le s \le 4 M_\pi^2]
\end{array}
\right.
,
\lbl{F_decomp}
\end{eqnarray}
where  ${\cal F}_0^{\pi^0} (s)$ is a function 
of $s$ that is smooth as long as no other threshold, corresponding to higher
intermediate states, is reached. Here $\delta (s)$ represents a phase. It can 
be chosen arbitrarily, as long as it is also a smooth function of $s$. 
The cusp at $s = 4 M_\pi^2$ observed in the differential decay rates 
corresponding to this simplified [as compared to $K_{e4}$] situation then results 
from this decomposition, since
\begin{eqnarray}
\left\vert {F}_S^{\pi^0}(s) \right\vert ^2 \ =\  
\left\{
\begin{array}{l}
\left\vert {\cal F}_0^{\pi^0} (s) \right\vert ^2 + \left\vert {\cal F}_1^{\pi^0} (s) \right\vert ^2
- 2 {\rm Im}\,\left[ {\cal F}_0^{\pi^0} (s) {\cal F}_1^{\pi^0 *}(s) \right]
\quad [s\ge 4 M_{\pi}^2] \\
\\
\left\vert {\cal F}_0^{\pi^0} (s) \right\vert ^2 + \left\vert {\cal F}_1^{\pi^0} (s) \right\vert ^2
+ 2 {\rm Re}\,\left[ {\cal F}_0^{\pi^0} (s) {\cal F}_1^{\pi^0 *}(s) \right]
\quad [4 M_{\pi^0}^2 \le s \le 4 M_\pi^2]
\end{array}
\right.
.
\lbl{cusp_general}
\end{eqnarray} 
Apart from the dependence with respect to the second
kinematical variable $s_e$, the empirical parameterisation used for 
the fit to the $K_{e4}^{00}$ data, Eq. (9.1) in Ref. \cite{Batley:2014xwa}, 
complies with this general representation provided [the variable $S_\pi$
used in this reference corresponds to the variable $s$ used here]:

1)~${\cal F}_1^{\pi^0} (s)$ is parameterised as a {\it real} constant times ${\hat\sigma} (s)$,
with
\begin{equation}
{\hat\sigma} (s) = \sqrt{\left\vert 1 - \frac{4 M_\pi^2}{s} \right\vert }
= \sqrt{\left\vert \frac{q^2}{1 + q^2} \right\vert }
, 
\end{equation}
for $4 M_{\pi^0}^2 \le s \le 4 M_\pi^2$, or $q^2 \le 0$, with $s = 4 M_\pi^2 ( 1 + q^2)$.

2)~${\cal F}_1^{\pi^0} (s)$ is set to zero (its value for $s = 4 M_\pi^2$) for $s\ge 4 M_\pi^2$ ($q^2 \ge 0$)

3)~For $4 M_{\pi^0}^2 \le s \le 4 M_\pi^2$ ($q^2 \le 0$), ${\cal F}_0^{\pi^0} (s)$ is replaced by
a constant, equal to ${\cal F}_0^{\pi^0} (4 M_\pi^2)$.

\noindent
A more theoretically based parameterization,
adapted from the simple discussion of the cusp in $K^\pm \to \pi^0\pi^0 \pi^\pm$ given 
in Ref. \cite{Cabibbo:2004gq}, is considered in Sec. 9.4 of 
Ref. \cite{Batley:2014xwa}, though not used for the data analysis. 
As compared to Eq.~(\ref{eq:cusp_general}), its validity also
rests on additional assumptions, which, once transposed to the present situation, read:

1')~The phase $\delta (s)$ can be chosen such as to make the two functions 
${\cal F}_0^{\pi^0} (s)$ and ${\cal F}_1^{\pi^0} (s)$ simultaneously real,
so that Eq. \rf{cusp_general} takes the simpler form 
\begin{eqnarray}
\vert {F}_S^{\pi^0}(s) \vert ^2 \ =\  
\left\{
\begin{array}{l}
\vert {\cal F}_0^{\pi^0} (s) \vert^2 + \vert {\cal F}_1^{\pi^0} (s) \vert^2 \quad [ s \ge 4 M_\pi^2 ]\\
\\
{ }[ {\cal F}_0^{\pi^0} (s)  + {\cal F}_1^{\pi^0} (s) ]^2 \qquad [4 M_{\pi^0}^2 \le s \le 4 M_\pi^2]
\end{array}
\right.
.
\lbl{cusp_simple}
\end{eqnarray}

2')~${\cal F}_1^{\pi^0} (s)$ is related to the scalar form factor 
${F}_S^{\pi}(s)$ of the {\it charged} pion, multiplied by a combination of the 
two $S$-wave $\pi\pi$ scattering lengths $a_0^0$ and $a_0^2$ in the isospin limit,
\begin{equation}
{\cal F}_1^{\pi^0} (s) = - \frac{2}{3} \left( a_0^0 - a_0^2 \right) {\cal F}_S^\pi (s) {\hat\sigma} (s)
.
\lbl{F_1_Ansatz}
\end{equation}
In view of the discussion in Ref. \cite{Batley:2014xwa}, ${\cal F}_S^\pi (s)$ should be identified with
the phase-removed scalar form factor of the charged pion. The latter is given by
 $e^{-i\delta_0^\pi (s)} F_S^\pi (s)$, where the phase $\delta_0^\pi (s)$ is defined
 as $F_S^\pi (s + i0) = e^{2i\delta_0^\pi (s)} F_S^\pi (s - i0)$.

Our purpose in this Section is twofold. First, we will rewrite the two-loop representation
of the form factor ${F}_S^{\pi^0}(s)$ obtained in Ref. \cite{DescotesGenon:2012gv} in the form
\rf{cusp_general}, that makes the cusp structure explicit. Second, we will assess to which extent the 
additional features mentioned above and assumed in Ref. \cite{Batley:2014xwa} are actually
reproduced by the structure of the form factors at two loops in the low-energy expansion. In
particular, we will establish the precise relation between ${\cal F}_S^{\pi} (s)$ and ${F}_S^{\pi} (s)$
in Eq. \rf{F_1_Ansatz} at this order. In what follows, and unless otherwise stated, it will always 
be understood that $s \ge 4 M_{\pi^0}^2$. Furthermore, in practice $s \ge 4 M_\pi^2$ will actually
mean $4 M_\pi^2 \le s \le M_K^2$, where $M_K$ is the mass of the charged kaon,
so that we need not worry about thresholds other than those
produced by two-pion intermediate states.

\subsection{The cusp in the one-loop form factor}

We start with the study of the cusp using the one-loop expression
of the form factor $F_S^{\pi^0}(s)$,
\begin{equation}
F_S^{\pi^0}(s) = F_S^{\pi^0}\! (0)\! \Bigg[
1 + a_S^{\pi^0} \! s + 
16\pi \frac{\varphi_0^{00}(s)}{2}  {\bar J}_0 (s) \! \Bigg]
 \,-\,16\pi {F_S^{\pi}(0)}\, \varphi_0^{x}(s)\,{\bar J} (s)
.
\end{equation}
In this expression, $a_S^{\pi^0}$ denotes a subtraction constant, 
that we need not specify further for the time being. The loop
functions ${\bar J}_0 (s)$ and ${\bar J} (s)$ are given by
\begin{eqnarray}
{\bar J}_0 (s) &=& \frac{s}{16\pi^2}\,\int_{4M_{\pi^0}^2}^{\infty}\,\frac{dx}{x}\,\frac{1}{x-s-i0}\,\sigma_0 (x)
\nonumber\\
 {\bar J} (s) &=& \frac{s}{16\pi^2}\,\int_{4M_{\pi}^2}^{\infty}\,\frac{dx}{x}\,\frac{1}{x-s-i0}\,\sigma (x)
,
\lbl{JbarDisp}
\end{eqnarray}
with
\begin{equation}
\sigma_{0}(s)\,=\,\sqrt{1 - \frac{4M_{\pi^0}^2}{s}}\,,\ \sigma(s)\,=\,\sqrt{1 - \frac{4M_{\pi}^2}{s}} .
\lbl{def_sigma}
\end{equation}
The functions $\varphi_0^{00}(s)$ and $\varphi_0^{x}(s)$ denote
the lowest-order real parts of the $S$-wave projections of the amplitudes of
the processes $\pi^0 \pi^0 \to \pi^0 \pi^0$ and $\pi^0 \pi^0 \to \pi^+ \pi^-$,
respectively. Their expressions read
\begin{equation}
\varphi_0^{00} (s) = a_{00},
\quad
\varphi_0^{x}(s) = a_x + b_x \frac{s - 4 M_\pi^2}{F_\pi^2}
\lbl{phi_0^x}
,
\end{equation}
with \cite{Knecht:1997jw,DescotesGenon:2012gv,Bernard:2013faa} 
\begin{equation}
a_{00} = \frac{2}{3} \left( a_0^0 + 2 a_0^2 \right) \left(1 - \frac{\Delta_\pi}{M_\pi^2} \right)
,\ a_x = - \frac{2}{3} \left( a_0^0 - a_0^2 \right) + a_0^2 \frac{\Delta_\pi}{M_\pi^2}
\,, \ b_x = - \frac{1}{12} \left( 2 a_0^0 - 5 a_0^2 \right) \frac{F_\pi^2}{M_\pi^2}
\lbl{ax_bx}
,
\end{equation}
and $\Delta_\pi\equiv M_\pi^2 - M_{\pi^0}^2$.

In the range of $s$ under consideration, the function ${\bar J}_0 (s)$
is complex, but both its real and imaginary parts are smooth, 
\begin{equation}
{\bar J}_0 (s) = \frac{1}{16 \pi^2} \left[ 2 + \sigma_0 (s) L_0 (s)
+ i \pi \sigma_0 (s) \right]
,\ L_0 (s) \equiv \ln \left( \frac{1 - \sigma_0 (s)}{1 + \sigma_0 (s)} \right)
\qquad [ s \ge 4 M_{\pi^0}^2 ]
,
\lbl{J0_L0_def}    
\end{equation}
whereas ${\bar J} (s)$ may be rewritten as
\begin{eqnarray}
{\bar J} (s) \ =\  {\bar J}^{[0]} (s) + {\bar J}^{[1]} (s) \times
\left\{
\begin{array}{l}
\! -i {\hat\sigma} (s)  \quad [s \ge 4 M_{\pi}^2]\\
\\  
\, +{\hat\sigma} (s)  \quad [4 M_{\pi^0}^2 \le s \le 4 M_\pi^2]
\end{array}
\right.
.
\lbl{Jbar_decomp_1}
\end{eqnarray} 
The two functions ${\bar J}^{[0]} (s)$ and ${\bar J}^{[1]} (s)$ are smooth, and
read
\begin{equation}
{\bar J}^{[0]} (s) = \frac{1}{16 \pi^2} \left[ 2 + \sigma (s) {\hat L} (s) \right]
,\  {\bar J}^{[1]} (s) = - \frac{1}{16 \pi},
\lbl{Jbar_decomp_2}
\end{equation} 
where
\begin{equation}
 {\hat L}(s) = \ln\left(
{\displaystyle \frac{1 - \sigma(s) }{1 + \sigma(s) } }\right) \quad [s \ge 4 M_{\pi^0}^2 ]
.
\lbl{L_hat}
\end{equation}
In these expressions, the definition of $\sigma (s)$ has been extended below $s = 4 M_\pi^2$ by\footnote{This extension
follows from the usual analytical continuation resulting from the replacement $s \to s + i0$.}
\begin{eqnarray}
\sigma (s) &=& 
\left\{
\begin{array}{l} 
\sqrt{1 - {\displaystyle\frac{4 M_{\pi}^2}{s}}} = {\hat\sigma} (s) \quad\ \ [ s \ge 4 M_{\pi}^2 ]\\
\\
i \sqrt{\displaystyle{\frac{4 M_{\pi}^2}{s}} - 1} = i {\hat\sigma} (s) \quad [ 4 M_{\pi^0}^2 \le s\le 4 M_{\pi}^2] 
\end{array}
\right.
.
\lbl{sigma_def_ext}
\end{eqnarray} 
According to Eqs. \rf{F_decomp} and \rf{cusp_general}, ${\bar J} (s)$ exhibits a cusp structure at $s = 4 M_\pi^2$.
One thus obtains, at this order, the decomposition of the form \rf{F_decomp} for $F_S^{\pi^0}(s)$, with
\begin{eqnarray}
{\cal F}_0^{\pi^0} (s) &=& F_S^{\pi^0}\! (0) \Bigg\{
1 + a_S^{\pi^0} \! s +  
\frac{\varphi_0^{00}(s)}{2 \pi}  \left[ 2 + \sigma_0(s) L_0(s) \right] \! \Bigg\}
-  {F_S^{\pi}(0)} \frac{\varphi_0^{x}(s)}{\pi} \left[ 2 + \sigma (s) {\hat L}(s) \right]
\,+\,{\cal O}(E^6)
,
\nonumber\\
{\cal F}_1^{\pi^0} (s) &=&   {F_S^{\pi}(0)} \varphi_0^{x}(s) {\hat\sigma} (s)
\,+\,{\cal O}(E^6)
,
\lbl{M_0_and_M_1_1-loop}
\end{eqnarray}
provided one factorises the global phase
\begin{equation}
\delta (s) = \frac{1}{2}\,\sigma_{0}(s)\varphi_0^{00}(s) \,+\,{\cal O}(E^4)
.
\end{equation}
Therefore, up to so far unspecified higher order corrections, the
one-loop expression of the form factor can be brought into the form \rf{F_decomp}.
Both functions ${\cal F}_0^{\pi^0} (s)$ and ${\cal F}_1^{\pi^0} (s) / {\hat\sigma} (s)$
are real and smooth for $s \ge 4 M_{\pi^0}^2$ at this stage.
However, the expression for ${\cal F}_1^{\pi^0}(s)$ in \rf{M_0_and_M_1_1-loop}
does not quite comply with Eq.~\rf{F_1_Ansatz}. Whereas at this stage 
${\cal F}_S^{\pi}(s)$ is equal to the constant ${F_S^{\pi}(0)}$, 
which, at this order, can be identified with the phase-removed form factor,
the combination of scattering lengths that occurs in Eq. \rf{F_1_Ansatz} corresponds to 
$\stackrel{{ }_{\mbox{\scriptsize{$o$}}}}{\varphi} \stackrel{x}{_{0}} \! \! (4 M_\pi^2)$,
where $\stackrel{{ }_{\mbox{\scriptsize{$o$}}}}{\varphi} \stackrel{x}{_{0}} \! \! (s)$
is the expression of $\varphi_0^{x}(s)$ in the isospin limit. Thus, at this order,
the expression \rf{F_1_Ansatz} misses both the dependence with respect to $s$ in 
$\varphi^x_0 (s)$, and the isospin-breaking corrections in the scattering lengths.

For later reference, we briefly extend the discussion to
the scalar form factor of the charged pion. At one loop, it is given by 
\begin{equation}
F_S^{\pi}(s) = F_S^{\pi}(0) \!\Bigg[
1 +  a_S^{\pi}\,s + 
16\pi \varphi_0^{\mbox{\tiny{$ +-$}}}(s)   {\bar J} (s)
\Bigg]
 \,-\,16\pi {F_S^{\pi^0}(0)} \,\frac{1}{2}\,\varphi_0^{x}(s)\,
{\bar J}_0 (s)
.
\end{equation}
Besides the subtraction constant $a_S^{\pi}$, that differs from $a_S^{\pi^0}$
(they become identical in the isospin limit), this expression involves the
lowest-order real part of the $S$-wave projection of the amplitude for the
scattering process $\pi^+ \pi^- \to \pi^+ \pi^-$,
\begin{equation}
\varphi_0^{\mbox{\tiny{$ +-$}}} (s) = a_{\mbox{\tiny{$ +-$}}} + 
b_{\mbox{\tiny{$ +-$}}} \frac{s - 4 M_\pi^2}{F_\pi^2}
\lbl{phi_0^+-}
,
\end{equation}
where \cite{Knecht:1997jw,DescotesGenon:2012gv,Bernard:2013faa} 
\begin{equation}
a_{\mbox{\tiny{$ +-$}}} = \frac{1}{3} \left( 2 a_0^0 + a_0^2 \right) - 2 a_0^2 \frac{\Delta_\pi}{M_\pi^2} 
,\ b_{\mbox{\tiny{$ +-$}}} =  \frac{1}{24} \left( 2 a_0^0 - 5 a_0^2 \right) \frac{F_\pi^2}{M_\pi^2}
.
\end{equation}
After having factorised the global phase
\begin{equation}
{\tilde\delta} (s) = - \frac{1}{2}\,\sigma_{0}(s)\varphi_0^{x}(s)\,\frac{F_S^{\pi^0}(0)}{F_S^{\pi}(0)} \,+\, {\cal O}(E^4)
,
\lbl{delta_tilde}
\end{equation}
one can decompose $F_S^{\pi}(s)$ according to Eq. \rf{F_decomp}, with 
\begin{eqnarray}
{\cal F}_0^{\pi} (s) &=& F_S^{\pi}\! (0) \bigg\{
1 + a_S^{\pi} \! s +  
\frac{\varphi_0^{\mbox{\tiny{$ +-$}}}(s)}{\pi}  \left[ 2 + \sigma(s) {\hat L}(s) \right] \! \bigg\}
-  {F_S^{\pi^0}(0)} \frac{\varphi_0^{x}(s)}{2 \pi} \left[ 2 + \sigma_0 (s) L_0 (s) \right]
\,+\,{\cal O}(E^6)
,
\nonumber\\
{\cal F}_1^{\pi} (s) &=&   - {F_S^{\pi}(0)} \varphi_0^{\mbox{\tiny{$ +-$}}}(s) {\hat\sigma} (s)
\,+\,{\cal O}(E^6)
.
\lbl{charged_M_0_and_M_1_1-loop}
\end{eqnarray}
Both functions ${\cal F}_0^{\pi} (s)$ and ${\cal F}_1^{\pi} (s) / {\hat\sigma} (s)$
are real and smooth for $s \ge 4 M_{\pi^0}^2$ at this stage.

\subsection{The cusp in the two-loop form factor $F_S^{\pi^0} (s)$ of the neutral pion}

Let us now go through the same analysis, but with the two-loop expression
of the form factor. The expressions of the pion scalar form factors at two loops 
and in presence of isospin breaking have been worked out in Ref.
\cite{DescotesGenon:2012gv}
using a recursive construction based on general properties like
relativistic invariance, unitarity, analyticity, and chiral
counting. The scalar form factor of the neutral pion can be 
written as\footnote{We neglect here a tiny contribution of second order
in isospin breaking.}
\begin{eqnarray}
F_S^{\pi^0}(s) &=& F_S^{\pi^0}\! (0)\! \left(
1 + a_S^{\pi^0} \! s + b_S^{\pi^0} \! s^2 
\right)
\nonumber\\
&&
+\, 8\pi F_S^{\pi^0}\! (0) \varphi_0^{00}(s) 
\left[
1 + a_S^{\pi^0} \! s + \frac{1}{\pi} \varphi_0^{00}(s) \right] \! {\bar J}_0 (s) 
 \nonumber\\
&&
-\, 16\pi {F_S^{\pi}(0)}\, \varphi_0^{x}(s)
\left[
1 +  a_S^{\pi}\,s + \frac{2}{\pi} \varphi_0^{\mbox{\tiny{$ +-$}}}(s) \right] \! {\bar J} (s)
\nonumber\\
&&
+\, \frac{M_{\pi}^4}{F_{\pi}^4}\,F_S^{\pi^0}\! (0)\! \left [
\xi^{(0)}_{00} (s) {\bar J}_0 (s) +
\xi^{(1;0)}_{00} (s) {\bar K}_1^0 (s) +
2 \xi^{(2;0)}_{00} (s) {\bar K}_2^0 (s) +
\xi^{(3;0)}_{00} (s) {\bar K}_3^0 (s)
\right.
\nonumber\\
&&
\left.
\qquad\qquad\qquad
+\,
\xi^{(1;{\mbox{\tiny$\nabla$}})}_{00} (s) {\bar K}_1^{{\mbox{\tiny$\nabla$}}} (s)  +
\xi^{(3;{\mbox{\tiny$\nabla$}})}_{00} (s) {\bar K}_3^{{\mbox{\tiny$\nabla$}}} (s) +
2 \xi^{(2;{\mbox{\tiny{$\!\pm $}}})}_{00} (s)
 \left[ 16 \pi^2 {\bar J} (s) - 2  \right] {\bar J}_0 (s)
\right]
\nonumber\\
&&
-\, 2 \,\frac{M_{\pi}^4}{F_{\pi}^4}\,F_S^{\pi}(0) \left[
\xi^{(0)}_{x} (s) {\bar J} (s) + 2 \xi^{(2;{\mbox{\tiny{$\!\pm $}}})}_{x} (s) {\bar K}_2(s) +
\xi^{(1)}_{x} (s) K_1^x (s) +
\xi^{(3)}_{x} (s) K_3^x (s) 
\right.
\nonumber\\
&&
\left.
\qquad\qquad\qquad
+\, 2 \xi^{(2;0)}_{x} (s) \left[ 16 \pi^2 {\bar J}_0 (s) - 2  \right] {\bar J} (s) +
\Delta_1 \xi_x (s) {\bar{\cal K}}^x (s)
\right]
\,+\, {\cal O}(E^8) .
\lbl{2loop_FF}
\end{eqnarray}
In this formula, the functions 
$\xi_{00}^{(0)} (s) , \ldots , \xi_x^{(0)} (s) , \ldots$
are polynomials of at most second order in the variable $s$.
Their expressions can be found in Ref. \cite{DescotesGenon:2012gv},
except for $\Delta_1 \xi_x (s)$, that reads
\begin{equation}
\Delta_1 \xi_x (s) = 8\, \frac{\Delta_\pi}{M_\pi^2} \, b_{+0} \frac{s}{M_\pi^2}
\left[
\frac{s}{9 M_\pi^2} b_{+0} - a_{+0}  \frac{F_\pi^2}{M_\pi^2}
+ 2 b_{+0} \left( 1 + \frac{M_{\pi^0}^2}{M_\pi^2} \right) \right].
\end{equation} 
It is also useful to be aware of the relations
\begin{equation}
\xi^{(2;0)}_{00} (s) = 2 \frac{F_\pi^4}{M_\pi^4} [ \varphi_0^{00} (s) ]^2
\! ,
\ \xi^{(2;0)}_{x} (s) = 2 \frac{F_\pi^4}{M_\pi^4} \varphi_0^{00} (s) \varphi_0^x (s)
,
\ \xi^{(2;{\mbox{\tiny{$\!\pm $}}})}_{00} (s) = 4 \frac{F_\pi^4}{M_\pi^4} [ \varphi_0^x (s) ]^2
\! ,
\ \xi^{(2;{\mbox{\tiny{$\!\pm $}}})}_{x} (s) = 4 \frac{F_\pi^4}{M_\pi^4} \varphi_0^x (s) \varphi^{\mbox{\tiny{$ +-$}}} (s)
.
\lbl{xi_phi_rel}
\end{equation}

In order to achieve the decomposition \rf{F_decomp},
we need to extend the decomposition
of the function ${\bar J} (s)$ in Eqs. \rf{Jbar_decomp_1}
and \rf{Jbar_decomp_2} to the other functions, denoted generically
by ${\bar K}_n^\alpha (s)$, that appear
in the expression \rf{2loop_FF}. This may be done as follows.
First, we may observe that, like ${\bar J}_0 (s)$ or ${\bar J} (s)$, 
these functions can also be defined by a dispersive representation
of the form [for ${\bar J}_0 (s) \equiv {\bar K}_0^0 (s)$, one has 
$k_0^0 (s) = \sigma_0 (s) / 16\pi$,
whereas for ${\bar J} (s) \equiv {\bar K}_0 (s)$, $k_0 (s) = \sigma (s) / 16\pi$,
see Eq. \rf{JbarDisp}]
\begin{equation}
{\bar K}_n^\alpha (s) = \frac{s}{\pi}\,\int_{s_{\rm thr}}^{\infty}\,\frac{dx}{x}\,\frac{1}{x-s-i0}\,k_n^\alpha (x)
.
\lbl{disp_rep_K}    
\end{equation}
Explicit expressions for the functions $k_n^\alpha (s)$ are given in Appendix \ref{app:Kbar_functions}.
For the set of functions $K_n^0 (s)$ and ${\bar K}_n^{{\mbox{\tiny$\nabla $}}} (s)$, one has
$s_{\rm thr} = 4 M_{\pi^0}^2$. These functions will therefore each develop an imaginary part 
for $s \ge s_{\rm thr} = 4 M_{\pi^0}$, 
${\rm Im} {\bar K}_n^\alpha (s) = k_n^\alpha (s) \theta (s - 4 M_{\pi^0}^2)$,
while the real part displays a cusp at $s = 4 M_{\pi^0}$, but is smooth
for $s \ge 4 M_{\pi^0}$. The situation is different for the remaining functions,
${\bar K}_n (s)$, ${\bar K}_n^x (s)$, and ${\bar{\cal K}}^x (s)$, for which 
$s_{\rm thr} = 4 M_{\pi}^2$, so that, in a generic way, they have the
following structure
\begin{eqnarray}
{\bar K}_n^\alpha (s) &=&  {\rm Re}\,{\bar K}_n^\alpha (s) +
\left\{
\begin{array}{l}
i k_n^\alpha (s) \quad [s\ge 4 M_{\pi}^2] \\
\\
\ 0 \qquad\ \ \, [4 M_{\pi^0}^2 \le s \le 4 M_\pi^2]
\end{array}
\right.
\nonumber\\
&=&  {\rm Re}\,{\bar K}_n^\alpha (s) + \frac{k_n^\alpha (s)}{\sigma (s)} \times
\left\{
\begin{array}{l}
i {\hat\sigma} (s) \quad [s\ge 4 M_{\pi}^2] \\
\\
\ 0 \qquad\ \, [4 M_{\pi^0}^2 \le s \le 4 M_\pi^2]
\end{array}
\right.
.
\lbl{Kbar_Re_Im}
\end{eqnarray} 
In general, the function ${k_n^\alpha (s)}/{\sigma (s)}$, although real, is not smooth
for the whole range $s \ge 4 M_{\pi^0}^2 $, but only for
$s \ge 4 M_\pi^2$. Suppose one can find a function ${\hat k}_n^\alpha (s)$
such that it coincides with $k_n^\alpha (s)$ for $s \ge 4 M_\pi^2$,
and such that ${\hat k}_n^\alpha (s)/\sigma (s)$ is real and smooth
for all $s \ge 4 M_{\pi^0}^2$. Then one can
perform the decomposition  
\begin{eqnarray}
{\bar K} (s) \ =\  {\bar K}^{[0]} (s) + {\bar K}^{[1]} (s) \times
\left\{
\begin{array}{l}
\! - i {\hat\sigma} (s)  \quad [s\ge 4 M_{\pi}^2]  \\
\\
\, + {\hat\sigma} (s)  \quad [4 M_{\pi^0}^2 \le s \le 4 M_\pi^2]
\end{array}
\right.
,
\lbl{Kbar_decomp_1}
\end{eqnarray}  
in terms of two real and smooth functions ${\bar K}^{[0]} (s)$ and ${\bar K}^{[1]} (s)$,
given by
\begin{eqnarray}
{\bar K}_n^{\alpha [0]} (s) \ =\  
\left\{
\begin{array}{l}
{\rm Re}\,{\bar K}_n^\alpha (s)  \qquad\ \ \,\qquad [s\ge 4 M_{\pi}^2] \\
\\
{\rm Re}\,{\bar K}_n^\alpha (s) - i {\hat k}_n^\alpha (s)\quad [4 M_{\pi^0}^2 \le s \le 4 M_\pi^2]
\end{array}
\right.
,\qquad
{\bar K}_n^{\alpha [1]} (s) \ =\ - \frac{{\hat k}_n^\alpha (s)}{\sigma (s)}
.
\end{eqnarray}
Such a decomposition can indeed be achieved for the various functions considered
here, as discussed in detail in App. \ref{app:Kbar_functions}.
The decomposition \rf{F_decomp} of the form factor now follows immediately,
with
\begin{eqnarray}
e^{i \delta (s)} {\cal F}_0^{\pi^0} (s) &=& F_S^{\pi^0}\! (0)\! \left(
1 + a_S^{\pi^0} \! s + b_S^{\pi^0} \! s^2 
\right)
\nonumber\\
&&
+\, 8\pi F_S^{\pi^0}\! (0) \varphi_0^{00}(s) 
\left[
1 + a_S^{\pi^0} \! s + \frac{1}{\pi} \varphi_0^{00}(s) \right] \! {\bar J}_0 (s) 
 \nonumber\\
&&
-\, 16\pi {F_S^{\pi}(0)}\, \varphi_0^{x}(s)
\left[
1 +  a_S^{\pi}\,s + \frac{2}{\pi} \varphi_0^{\mbox{\tiny{$ +-$}}}(s)
\right] \! {\bar J}^{[0]} (s)
\nonumber\\
&&
+\, \frac{M_{\pi}^4}{F_{\pi}^4}\,F_S^{\pi^0}\! (0)\! \left [
\xi^{(0)}_{00}(s) {\bar J}_0 (s) +
\xi^{(1;0)}_{00} (s) {\bar K}_1^0(s) + 
\xi^{(1;{\mbox{\tiny$\nabla$}})}_{00}(s) {\bar K}_1^{\mbox{\tiny$\nabla$}}(s) + 
2\, \xi^{(2;0)}_{00} (s) {\bar K}_2^0(s)
\right.\nonumber\\
&&
\left.
\quad\qquad\qquad
+\, \xi^{(3;0)}_{00} (s) {\bar K}_3^0(s)  + \xi^{(3;{\mbox{\tiny$\nabla$}})}_{00}(s) {\bar K}_3^{\mbox{\tiny$\nabla$}}(s) +
2 \xi^{(2;{\mbox{\tiny{$\pm $}}})}_{00} (s)
 \left[ 16 \pi^2 {\bar J}^{[0]} (s) - 2  \right] {\bar J}_0 (s)
\right]
\nonumber\\
&&
-\, 2 \,\frac{M_{\pi}^4}{F_{\pi}^4}\,F_S^{\pi}(0) \left[
\xi^{(0)}_{x} (s) {\bar J}^{[0]} (s)  +
\xi^{(1)}_{x} (s) {\bar K}_1^{x [0]} (s) + 
2 \xi^{(2;{\mbox{\tiny{$\pm $}}})}_{x} (s) {\bar K}_2^{[0]}(s) +
\xi^{(3)}_{x} (s) {\bar K}_3^{x [0]} (s) 
\right.\nonumber\\
&&
\left.
\quad\qquad\qquad
+\, 2 \xi^{(2;0)}_{x} (s) \left[ 16 \pi^2 {\bar J}_0 (s) - 2  \right] {\bar J}^{[0]} (s) +
\Delta_1 \xi_x (s) {\bar{\cal K}}^{x [0]} (s)
\right]
\,+\, {\cal O}(E^8) 
,
\lbl{calM0_2loop}
\end{eqnarray}
and
\begin{eqnarray}
e^{i \delta (s)} {\cal F}_1^{\pi^0} (s) &=& - {\hat\sigma} (s) \bigg\{
 16\pi {F_S^{\pi}(0)}\, \varphi_0^{x}(s)
\left[
1 +  a_S^{\pi}\,s + \frac{2}{\pi} \varphi_0^{\mbox{\tiny{$ +-$}}}(s)
\right] \! {\bar J}^{[1]} (s)
\nonumber\\
&&
-\, 2\,\frac{M_{\pi}^4}{F_{\pi}^4}\,F_S^{\pi^0}\! (0)
\xi^{(2;{\mbox{\tiny{$\!\pm $}}})}_{00} (s)
\times 16 \pi^2 {\bar J}_0 (s) {\bar J}^{[1]} (s) 
\nonumber\\
&&
+\, 2 \,\frac{M_{\pi}^4}{F_{\pi}^4}\,F_S^{\pi}(0) \left[
\xi^{(0)}_{x} (s) {\bar J}^{[1]} (s)  +
\xi^{(1)}_{x} (s) {\bar K}_1^{x [1]} (s) + 
2 \xi^{(2;{\mbox{\tiny{$\!\pm $}}})}_{x} (s) {\bar K}_2^{[1]}(s) +
\xi^{(3)}_{x} (s) {\bar K}_3^{x [1]} (s) 
\right.\nonumber\\
&&
\left.
\quad\qquad\qquad
+\, 2 \xi^{(2;0)}_{x} (s) \left[ 16 \pi^2 {\bar J}_0 (s) - 2  \right] {\bar J}^{[1]} (s) +
\Delta_1 \xi_x (s) {\bar{\cal K}}^{x [1]} (s)
\right]
\bigg\}
\,+\, {\cal O}(E^8) .
\lbl{calF1_2loop}
\end{eqnarray}
Both functions $e^{i \delta (s)} {\cal F}_0^{\pi^0} (s)$ and 
$e^{i \delta (s)} {\cal F}_1^{\pi^0} (s) / {\hat\sigma} (s)$ are smooth for $s \ge 4 M_{\pi^0}^2$, but complex.
It remains to discuss the phase $\delta (s)$. If we want to make
the function ${\cal F}_0^{\pi^0} (s)$ real, while keeping it smooth,
then its choice is unique,
\begin{equation}
\delta (s) \equiv \frac{1}{2}\,\sigma_{0}(s) \left[\varphi_0^{00}(s)  + {\hat\psi}_0^{00}(s) \right] 
,
\lbl{phase_delta}
\end{equation}
with ${\hat\psi}_0^{00}(s)$ given by
\begin{eqnarray}
\frac{1}{2} \, \sigma_0 (s) {\hat\psi}_0^{00}(s)  &=&
\frac{M_{\pi}^4}{F_{\pi}^4} \bigg[
\xi^{(0)}_{00}(s) k_0^0 (s) +
\xi^{(1;0)}_{00} (s) k_1^0(s) + \xi^{(1;{\mbox{\tiny$\nabla$}})}_{00}(s) k_1^{\mbox{\tiny$\nabla$}}(s) +
\xi^{(2;0)}_{00} (s) k_2^0(s)
\nonumber\\
&&
\qquad
+ \xi^{(3;0)}_{00} (s) k_3^0(s)  + \xi^{(3;{\mbox{\tiny$\nabla$}})}_{00}(s) k_3^{\mbox{\tiny$\nabla$}}(s) +
\xi^{(2;\pm)}_{00}(s)\, \frac{1}{8 \pi} \sigma_0 (s) \sigma (s) {\hat L}(s) 
\bigg]
.
\lbl{psi0_hat}
\end{eqnarray}
Note that ${\hat\psi}_0^{00}(s)$ differs from the quantity ${\psi}_0^{00}(s)$
defined in Eq. (4.6) of Ref. \cite{DescotesGenon:2012gv}  by the presence of the function ${\hat L}(s)$ 
instead of $L (s)$ in the last term between square brackets, see Eq. \rf{def_L}. 
This makes $\sigma_0 (s) {\hat\psi}_0^{00}(s)$ a smooth function
for $s \ge 4 M_{\pi^0}^2$, whereas $\sigma (s) L(s)$, and hence ${\psi}_0^{00}(s)$, 
displays a cusp at $s = 4 M_\pi^2$. Making use of Eq. \rf{Kbar_Re_Im}, the removal
of the phase $\delta (s)$ indeed leads to a real and smooth expression for the function 
${\cal F}_0^{\pi^0} (s)$:
\begin{eqnarray}
{\cal F}_0^{\pi^0} (s) &=&  F_S^{\pi^0}\! (0)\! \left(
1 + a_S^{\pi^0} \! s + b_S^{\pi^0} \! s^2 
\right)
\nonumber\\
&&
+\, 8\pi F_S^{\pi^0}\! (0) \varphi_0^{00}(s) 
\left[
1 + a_S^{\pi^0} \! s  \right] {\rm Re} \, {\bar J}_0 (s) 
 \nonumber\\
&&
-\, 16\pi {F_S^{\pi}(0)}\, \varphi_0^{x}(s)
\left[
1 +  a_S^{\pi}\,s + \frac{2}{\pi} \varphi_0^{\mbox{\tiny{$ +-$}}}(s)
\right] \! {\bar J}^{[0]} (s)
\nonumber\\
&&
+\, \frac{M_{\pi}^4}{F_{\pi}^4}\,F_S^{\pi^0}\! (0)\! \left [
\xi^{(0)}_{00}(s) {\rm Re} \, {\bar J}_0 (s) +
\xi^{(1;0)}_{00} (s) {\rm Re} \, {\bar K}_1^0(s) + 
\xi^{(1;{\mbox{\tiny$\nabla$}})}_{00}(s) {\rm Re} \, {\bar K}_1^{\mbox{\tiny$\nabla$}}(s) +
\right.\nonumber\\
&&
\left.
\quad\qquad\qquad
+\,  \xi^{(2;0)}_{00} (s) {\rm Re} \, {\bar K}_2^0(s)
+ \xi^{(3;0)}_{00} (s) {\rm Re} \, {\bar K}_3^0(s)  
+ \xi^{(3;{\mbox{\tiny$\nabla$}})}_{00}(s) {\rm Re} \, {\bar K}_3^{\mbox{\tiny$\nabla$}}(s)
\right]
\nonumber\\
&&
-\, 2 \,\frac{M_{\pi}^4}{F_{\pi}^4}\,F_S^{\pi}(0) \left[
\xi^{(0)}_{x} (s) {\bar J}^{[0]} (s)  +
\xi^{(1)}_{x} (s) {\bar K}_1^{x [0]} (s) + 
2 \xi^{(2;{\mbox{\tiny{$\!\pm $}}})}_{x} (s) {\bar K}_2^{[0]}(s)  
\right.\nonumber\\
&&
\left.
\quad\qquad\qquad
+\, \xi^{(3)}_{x} (s) {\bar K}_3^{x [0]} (s)  +
\Delta_1 \xi_x (s) {\bar{\cal K}}^{x [0]} (s)
\right]
\nonumber\\
&&
+\,8 F_S^{\pi^0}\! (0) \left[ \varphi_0^x (s) \right]^2  \left( 16 \pi^2 {\bar J}^{[0]} (s)  - 2  \right) {\rm Re} \, {\bar J}_0 (s)
\nonumber\\
&&
-\,8 F_S^{\pi} (0) \varphi_0^x (s) \varphi_0^{00} (s) \left[ 16 \pi^2 {\rm Re} \, {\bar J}_0 (s)  - 2  \right] {\bar J}^{[0]} (s)
\nonumber\\
&&
+\,F_S^{\pi^0}\! (0) \left[ \varphi_0^{00} (s) \right]^2 \left[ 
2 {\rm Re} \, {\bar K}_2^0(s) + 8 {\rm Re} \, {\bar J}_0 (s) + \frac{1}{8} \left( 1 - \frac{4 M_{\pi^0}^2}{s} \right)
\right]
\,+\, {\cal O}(E^8) 
.
\lbl{cal_F_0}
\end{eqnarray}
As far as ${\cal F}_1^{\pi^0} (s)$ is concerned, we may even proceed
in a more direct way by noticing that, up to higher order corrections,
Eq. \rf{calF1_2loop} rewrites as
\begin{eqnarray}
e^{i \delta (s)} {\cal F}_1^{\pi^0} (s) &=& e^{\frac{i}{2} \sigma_0 (s) \varphi_0^{00} (s) }  {\hat\sigma} (s) 
\left[ \varphi_0^{x}(s) + {\hat\psi}_0^{x}(s) \right]
\nonumber\\
&&
\times
{F_S^{\pi}(0)}
\bigg\{
1 + a_S^{\pi}  s +  
16\pi \varphi_0^{{\mbox{\tiny{$+-$}}}} (s) {\bar J}^{[0]} (s)
- 8 \pi \frac{F_S^{\pi^0}(0)}{F_S^{\pi}(0)} \, \varphi_0^x (s)  {\bar J}_0 (s) \! \bigg\}
\,+\, {\cal O}(E^8) 
,
\lbl{cal_F_1}
\end{eqnarray}
with [for the notation, see Appendix \ref{app:Kbar_functions}]
\begin{eqnarray}
{\hat\psi}_0^{x}(s) &=&  2\,\frac{M_{\pi}^4}{F_\pi^4}\,
\frac{1}{\sigma(s)}\,\Bigg\{
\xi^{(0)}_{x} (s) k_0 (s) + \xi^{(2;{\mbox{\tiny{$\!\pm $}}})}_{x} (s) {\hat k}_2(s) +
\xi^{(1)}_{x} (s) k_1^x (s) +
\xi^{(3)}_{x} (s) k_3^x (s) 
\nonumber\\
&&
\qquad\qquad\qquad
+\, \xi^{(2;0)}_{x} (s) k_2^x (s) +
\Delta_1 \xi_x (s) k^x (s)
\Bigg\}
.
\lbl{psi_x_0_hat} 
\end{eqnarray}
Now, the phase that appears factored out on the right-hand side
of this equation can be identified with the phase $\delta (s)$
on the left-hand side, since the difference generates contributions
of order ${\cal O} (E^8)$, that are neglected anyway. Taking into
account Eqs. \rf{delta_tilde} and \rf{charged_M_0_and_M_1_1-loop},
one finally obtains
\begin{equation}
{\cal F}_1^{\pi^0} (s) = {\hat\sigma} (s) 
\left[ \varphi_0^{x}(s) + {\hat\psi}_0^{x}(s) \right]
{\cal F}_\pi (s)
\,+\, {\cal O}(E^8) 
,
\lbl{calF_1_pi0}
\end{equation}
with 
\begin{equation}
{\cal F}_\pi (s) \equiv e^{i {\tilde\delta} (s)} {\cal F}^\pi_0 (s)
\,+\, {\cal O}(E^6) 
.
\lbl{calF_pi}   
\end{equation}
It is possible to give a more precise interpretation of the combination
$\varphi_0^{x}(s) + {\hat\psi}_0^{x}(s)$ that occurs in \rf{calF_1_pi0}.
To this end, let us recall from Ref. \cite{DescotesGenon:2012gv} that
the $\ell = 0$ partial-wave projection $f_0^x (s)$ for the scattering 
amplitude of the process $\pi^0 \pi^0 \to \pi^+ \pi^-$ is given, at
order one loop and for $s \ge 4 M_{\pi^0}^2$, by
\begin{equation}
f_0^x (s) = \varphi_0^{x}(s) + {\psi}_0^{x}(s)
+ i \varphi^x_0 (s) \left[
\frac{1}{2} \sigma_0 (s) \varphi_0^{00} (s) + \sigma (s) \varphi_0^{{\mbox{\tiny{$+-$}}}} (s) \theta (s - 4 M_\pi^2 )
\right]
\,+\, {\cal O}(E^6) 
,
\end{equation}
where ${\psi}_0^{x}(s)$ is defined in Eq. (4.15) of Ref. \cite{DescotesGenon:2012gv}
[the contribution $\Delta_2 \psi_0^x (s)$, of second order in isospin breaking, is
numerically quite small, and is omitted for simplicity]. It differs from ${\hat\psi}_0^{x}(s)$
by the replacement of ${\hat k}_2(s)$ by $k_2 (s)$ in Eq. \rf{psi_x_0_hat}. Then applying
the decomposition \rf{Kbar_decomp_1} to $f_0^x (s)$, one finds
\begin{eqnarray}
e^{- i \delta (s) } {f_0^x} (s) \ =\  {f_0^x}^{[0]} (s) + {f_0^x}^{[1]} (s) \times
\left\{
\begin{array}{l}
\! - i {\hat\sigma} (s)  \quad [s\ge 4 M_{\pi}^2]  \\
\\
\, + {\hat\sigma} (s)  \quad [4 M_{\pi^0}^2 \le s \le 4 M_\pi^2]
\end{array}
\right.
,
\lbl{f_0^x_decomp_1}
\end{eqnarray}  
with ${f_0^x}^{[0]} (s) = \varphi_0^{x}(s) + {\hat\psi}_0^{x}(s) + {\cal O}(E^6)$, and
${f_0^x}^{[1]} (s) = - \varphi_0^{x}(s) \varphi_0^{{\mbox{\tiny{$+-$}}}} (s) + {\cal O}(E^6)$.

We may summarise this theoretical study of the cusp in the scalar
form factor of the neutral pion with a couple of remarks:
\begin{itemize}
 \item It is, in general, not possible to chose the phase $\delta (s)$ in
 Eq. \rf{F_decomp} such as to make both ${\cal F}_0^{\pi^0} (s)$ and ${\cal F}_1^{\pi^0} (s)$
 real simultaneously. A relative phase remains, see Eqs. \rf{calF_1_pi0} and \rf{calF_pi}.
 At lowest order, this phase is given by the $S$-wave projection of the inelastic
 rescattering of a pair of neutral pions through a pair of charged pions, cf. Eq. \rf{delta_tilde}.
 
 \item The structure of ${\cal F}_1^{\pi^0} (s)$ is more complicated than just
 the product of the scattering length corresponding to this rescattering amplitude
 times the phase removed scalar form factor of the charged pion. At the order we have been
 working, it involves the decomposition \rf{f_0^x_decomp_1} of the $S$-wave projection 
 of this amplitude times the part ${\cal F}^\pi_0 (s)$ of the decomposition \rf{F_decomp}
 of $F_S^\pi (s)$. This is different from the phase-removed form factor,
 as already seen at order one loop:
 \begin{equation}
  e^{-i\delta_0^\pi (s)} F_S^\pi (s) - {\cal F}^\pi_0 (s) = {\cal F}^\pi_1 (s)
  \times\left\{
\begin{array}{l}
 0  \quad [s\ge 4 M_{\pi}^2]  \\
\\
 1 \quad [4 M_{\pi^0}^2 \le s \le 4 M_\pi^2]
\end{array}
\right.
\,+\, {\cal O}(E^6) 
.
 \end{equation}
 Note however that this difference only concerns the region $4 M_{\pi^0}^2 \le s \le 4 M_\pi^2$,
 which contributes very little to the total decay rate as defined by Eqs. \rf{dist} and \rf{rates} below.

\end{itemize}

\subsection{Description of the two-loop form factor $F_S^{\pi} (s)$ of the charged pion}

We now briefly address the scalar form factor of the charged pion.
The issue here is not to describe the cusp, that occurs below
the physical threshold at $s= 4 M_\pi^2$, but to provide the
expressions that will be used in the sequel. Again, we will rely on
the results obtained in Ref. \cite{DescotesGenon:2012gv}, and rewrite
the form factor $F_S^\pi (s)$ at two loops in a way that is adapted to
our purposes. In particular, we will consider the phase-removed
form factor, which reads, in the relevant domain $s \ge 4 M_\pi^2$,
\begin{eqnarray}
e^{-i \delta_0^\pi (s)} F_S^{\pi}(s) &=& F_S^{\pi}\! (0)\! \left(
1 + a_S^{\pi}  s + b_S^{\pi}  s^2 
\right)
\nonumber\\
&&\!\!\!\!\!
-\, 8\pi F_S^{\pi^0}\! (0) \varphi_0^{x}(s) 
\left[
1 + a_S^{\pi^0} \! s + \frac{1}{\pi} \varphi_0^{00}(s) \right] \! {\rm Re} \, {\bar J}_0 (s) 
\nonumber\\
&&\!\!\!\!\!
+\, 16\pi {F_S^{\pi}(0)}\, \varphi_0^{{\mbox{\tiny{$+-$}}}}(s)
\left[
1 +  a_S^{\pi}\,s + \frac{2}{\pi} \varphi_0^{\mbox{\tiny{$ +-$}}}(s) \right] \! {\rm Re} \, {\bar J} (s)
\nonumber\\
&&\!\!\!\!\!
-\, \frac{M_{\pi}^4}{F_{\pi}^4}\,F_S^{\pi^0}\! (0)\! \left \{
\xi^{(0)}_{x} (s) {\rm Re} \, {\bar J}_0 (s) +
\xi^{(1)}_{x} (s) {\rm Re} \, {\bar K}_1^{x_0} (s) +
2 \xi^{(2;0)}_{x} (s) {\rm Re} \, {\bar K}_2^0 (s) +
\xi^{(3)}_{x} (s) {\rm Re} \, {\bar K}_3^{x_0} (s)
\right.
\nonumber\\
&&\!\!\!\!\!
\left.
\qquad\qquad\qquad
+\,\Delta_1 \xi_x (s) {\rm Re} \, {\bar{\cal K}}^{x_0} (s) +
2 \xi^{(2;{\mbox{\tiny{$\!\pm $}}})}_{x} (s)
 \left[ 16 \pi^2 {\rm Re} \, {\bar J} (s) - 2  \right] {\rm Re} \, {\bar J}_0 (s)
\right\}
\nonumber\\
&&\!\!\!\!\!
+\, 2 \,\frac{M_{\pi}^4}{F_{\pi}^4}\,F_S^{\pi}(0) \left\{
\xi_{{\mbox{\tiny{$+-$}}};S}^{(0)} (s) {\rm Re} \, {\bar J} (s) 
+  \xi^{(1;{\mbox{\tiny{$\pm$}}})}_{{\mbox{\tiny{$+-$}}};S} (s) {\rm Re} \, {\bar K}_1(s) 
+ 2 \xi^{(2;{\mbox{\tiny{$\pm$}}})}_{{\mbox{\tiny{$+-$}}};S} (s) \left[ {\rm Re} \, {\bar K}_2(s) + 
\frac{1}{32} \left( 1 - \frac{4 M_\pi^2}{s} \right) \right]
\right.
\nonumber\\
&&\!\!\!\!\!
\left.
\qquad\qquad\qquad
+\,
\xi^{(3;{\mbox{\tiny{$\pm$}}})}_{{\mbox{\tiny{$+-$}}};S} (s) {\rm Re} \, {\bar K}_3(s) +
\xi^{(1;{\mbox{\tiny$\Delta$}})}_{{\mbox{\tiny{$+-$}}}; S} (s) {\rm Re} \, {\bar K}_1^{\mbox{\tiny$\Delta$}} (s) +
\xi^{(3;{\mbox{\tiny$\Delta$}})}_{{\mbox{\tiny{$+-$}}}; S} (s) {\rm Re} \, {\bar K}_3^{\mbox{\tiny$\Delta$}} (s) +
\right.
\nonumber\\
&&\!\!\!\!\!
\left.
\qquad\qquad\qquad
+\, 2 \xi^{(2;0)}_{{\mbox{\tiny{$+-$}}}; S} (s) \left[ 
\left( 16 \pi^2 {\rm Re} \, {\bar J}_0 (s) - 2  \right) {\rm Re} \, {\bar J} (s) 
+ \frac{1}{64} \left( 1 - \frac{4 M_{\pi^0}^2}{s} \right)
\right]
\right\}
\,+\, {\cal O}(E^8) .
\lbl{2loop_FF_ch}
\end{eqnarray}
The functions ${\bar K}_n^\alpha (s)$ that appear in this expression  
have again a dispersive representation of the form displayed in
Eq. \rf{disp_rep_K}. The absorptive parts are in part given in
Appendix \ref{app:Kbar_functions}. For the remaining one, we have
\begin{equation}
k_n^{x_0} (s) = \frac{\sigma_0 (s)}{\sigma (s)} \, k_n^{x} (s)
,\quad
k^{x_0} (s) = \frac{\sigma_0 (s)}{\sigma (s)} \, k^{x} (s)
,
\end{equation}
for the functions corresponding to $s_{\rm thr} = 4 M_{\pi^0}^2$,
and, for those whose dispersive integrals start at $s_{\rm thr} = 4 M_\pi^2$,
\begin{equation}
k_1^{{\mbox{\tiny $\Delta$}}} (s) \,=\, \frac{1}{8\pi}\,\frac{\sigma (s)}{\sigma_0 (s - 4 \Delta_\pi)} \, L_0 (s - 4 \Delta_\pi)\,,
\qquad
k_3^{{\mbox{\tiny $\Delta$}}} (s) \,=\, \frac{3}{16\pi}\,\frac{M_{\pi^0}^2}{s \sigma (s)}\,L_0^2 (s - 4 \Delta_\pi)\,,
\lbl{k_n^0}
\end{equation}
with the definitions of the functions $\sigma (s)$, $\sigma_0 (s)$ and $L_0 (s)$ for $s \ge 4 M_{\pi^0}^2$
given in Eqs. \rf{def_sigma}, \rf{J0_L0_def}, and \rf{sigma_def_ext}.

\section{Generation and analysis of the pseudo-data} \label{sec:analyses}
\setcounter{equation}{0}

In order to study the effect that particular choices of phenomenological
parameterisations of the form factors can have on the output, we will
first generate numerical data sets for the scalar form factors of the neutral
and charged pions. The pseudo-data in question consist of the (unnormalized)
decay distribution defined by
\begin{equation}
 \frac{d^2 \Gamma^{\pi^0} (s , s_\ell)}{ds ds_\ell} \equiv 
 \frac{1}{2} \sqrt{1 - \frac{4 M_{\pi^0}^2}{s}} \vert F_{S ; {\mbox{\scriptsize{data}}}}^{\pi^0} (s) \vert^2
 \lambda^{3/2} ( M_K^2 , s , s_\ell)
 ,
 \lbl{dist}
 \end{equation}
with $\lambda (x,y,z) = x^2 + y^2 + z^2 - 2 x y - 2 x z - 2 y z$.
The total decay rate is obtained by  integrating the distributions \rf{dist},
convoluted with the $K_{\ell 4}$ phase space, over the whole physical range 
[we now consider the electron mode only, and set $m_e =0$]:
\begin{equation}
\Gamma^{\pi^0} = {\cal N} \int_{4 M_{\pi^0}^2}^{M_K^2} ds \int_0^{(M_K^2 - \sqrt{s})^2} \!\!\!\!\!\!\! d s_e
\,\frac{d^2 \Gamma^{\pi^0} (s , s_e)}{ds ds_e}
.
\lbl{rates}
\end{equation}
Since we consider the scalar form factor instead of $K_{\ell 4}$ form factors, 
the integration with respect to $s_\ell$ involves the phase space only.
The overall normalisation factor has been chosen to be the one of the $K_{e4}$ decay,
\begin{equation}
 {\cal N} = \frac{G_{\rm F}^2 \vert V_{us} \vert^2}{3 \cdot 2^{12} \pi^5 M_K^5} \frac{1}{\vert F^{\pi}_S(0)\vert^2}
 ,
\end{equation}
up to the factor $1/{\vert F^{\pi}_S(0)\vert^2}$, introduced for convenience.

\subsection{Form factors and input parameters used for the generation of pseudo-data}

The form factors involved in the preceding expressions are considered as known exactly
and are constructed as follows. For $F_{S ; {\mbox{\scriptsize{data}}}}^{\pi^0} (s)$,
 we will basically use the decomposition
of Eq. \rf{F_decomp}, with ${\cal F}_0^{\pi^0} (s)$ given by Eq. \rf{cal_F_0},   
and ${\cal F}_1^{\pi^0} (s)$ given by  Eqs. \rf{calF_1_pi0} and \rf{calF_pi}.
Unfortunately, for some of the functions involved in Eq. \rf{cal_F_0}, like ${\rm Re} \, {\bar K}_n^{\mbox{\tiny$\nabla $}}(s)$
or ${\rm Re} \, {\bar K}_n^x (s)$, explicit analytical expressions are not known.
For a numerical approach, we could use their dispersive representation, 
as given by Eq. \rf{disp_rep_K}. We have however found it more convenient to start from expressions
upon which we have full analytical control. For that purpose, one may replace
the functions ${\rm Re} \, {\bar K}_n^{\mbox{\tiny$\nabla $}}(s)$ by the corresponding functions 
${\rm Re} \, {\bar K}_n^0 (s)$, and likewise ${\rm Re} \, {\bar K}_n^x (s)$ by ${\rm Re} \, {\bar K}_n (s)$.
We also drop the contribution proportional to $\Delta_1 \xi_x (s)$.
In the range of $s$ we are interested in, $4 M_{\pi^0}^2 \le s \le M_K^2$, the difference
induced in ${\cal F}_0^{\pi^0} (s)$ by these changes is numerically very small. 
For the scalar form factor of the neutral pion, the resulting expression then reads
\begin{eqnarray}
{F}_{S ; {\mbox{\scriptsize{data}}}}^{\pi^0}(s) \ =\  
\left\{
\begin{array}{l}
{\cal F}_{0 ; {\mbox{\scriptsize{data}}}}^{\pi^0} (s) 
- i {\cal F}_{1 ; {\mbox{\scriptsize{data}}}}^{\pi^0}  (s)  \quad [s \ge 4 M_\pi^2 ]\\
\\ 
{\cal F}_{0 ; {\mbox{\scriptsize{data}}}}^{\pi^0} (s) 
+ {\cal F}_{1 ; {\mbox{\scriptsize{data}}}}^{\pi^0} (s)  \quad\,\, [4 M_{\pi^0}^2 \le s \le 4 M_\pi^2]
\end{array}
\right.
,
\lbl{F_data_decomp}
\end{eqnarray}
with
\begin{eqnarray}
{\cal F}_{0 ; {\mbox{\scriptsize{data}}}}^{\pi^0} (s)  &=&  F_S^{\pi^0}\! (0)\! \left(
1 + a_S^{\pi^0} \! s + b_S^{\pi^0} \! s^2 
\right)
\nonumber\\
&&
+\, 8\pi F_S^{\pi^0}\! (0) \varphi_0^{00}(s) 
\left[
1 + a_S^{\pi^0} \! s  \right] {\rm Re} \, {\bar J}_0 (s) 
 \nonumber\\
&&
-\, 16\pi {F_S^{\pi}(0)}\, \varphi_0^{x}(s)
\left[
1 +  a_S^{\pi}\,s + \frac{2}{\pi} \varphi_0^{\mbox{\tiny{$ +-$}}}(s)
\right] \! {\bar J}^{[0]} (s)
\nonumber\\
&&
+\, \frac{M_{\pi}^4}{F_{\pi}^4}\,F_S^{\pi^0}\! (0)\! \left\{
\xi^{(0)}_{00}(s) {\rm Re} \, {\bar J}_0 (s) +
\left[ \xi^{(1;0)}_{00} (s) + \xi^{(1;{\mbox{\tiny$\nabla$}})}_{00}(s) \right] {\rm Re} \, {\bar K}_1^0(s) +  
\right.\nonumber\\
&&
\left.
\quad\qquad\qquad
+\,  \xi^{(2;0)}_{00} (s) {\rm Re} \, {\bar K}_2^0(s)
+ \left[ \xi^{(3;0)}_{00} (s) + \xi^{(3;{\mbox{\tiny$\nabla$}})}_{00}(s) \right] {\rm Re} \, {\bar K}_3^0(s) 
\right\}
\nonumber\\
&&
-\, 2 \,\frac{M_{\pi}^4}{F_{\pi}^4}\,F_S^{\pi}(0) \left[
\xi^{(0)}_{x} (s) {\bar J}^{[0]} (s)  +
\xi^{(1)}_{x} (s) {\bar K}_1^{[0]} (s) + 
2 \xi^{(2;{\mbox{\tiny{$\!\pm $}}})}_{x} (s) {\bar K}_2^{[0]}(s)  
+ \xi^{(3)}_{x} (s) {\bar K}_3^{[0]} (s) 
\right]
\nonumber\\
&&
+\,8 F_S^{\pi^0}\! (0) \left[ \varphi_0^x (s) \right]^2  \left( 16 \pi^2 {\bar J}^{[0]} (s)  - 2  \right) {\rm Re} \, {\bar J}_0 (s)
\nonumber\\
&&
-\,8 F_S^{\pi} (0) \varphi_0^x (s) \varphi_0^{00} (s) \left[ 16 \pi^2 {\rm Re} \, {\bar J}_0 (s)  - 2  \right] {\bar J}^{[0]} (s)
\nonumber\\
&&
+\,F_S^{\pi^0}\! (0) \left[ \varphi_0^{00} (s) \right]^2 \left[ 
2 {\rm Re} \, {\bar K}_2^0(s) + 8 {\rm Re} \, {\bar J}_0 (s) + \frac{1}{8} \left( 1 - \frac{4 M_{\pi^0}^2}{s} \right)
\right]
,
\lbl{cal_F_0_data}    
\end{eqnarray}
and
\begin{eqnarray}
{\cal F}_{1 ; {\mbox{\scriptsize{data}}}}^{\pi^0} (s)  &=&
e^{- \frac{i}{2} \sigma_{0}(s) \varphi_0^{x}(s) \frac{F_S^{\pi^0}(0)}{F_S^{\pi}(0)} }
{\hat\sigma} (s) 
F_S^{\pi}\! (0) 
\nonumber\\
&&\times
\bigg\{
1 + a_S^{\pi} \! s +  
\frac{\varphi_0^{\mbox{\tiny{$ +-$}}}(s)}{\pi}  \left[ 2 + \sigma(s) {\hat L}(s) \right] \! 
-  \frac{{F_S^{\pi^0}(0)}}{F_S^{\pi}\! (0)} \frac{\varphi_0^{x}(s)}{2 \pi} \left[ 2 + \sigma_0 (s) L_0 (s) \right]
\!\bigg\}\lbl{cal_F_1_data}
\\
&&\times
\left\{ \varphi_0^{x}(s) + \,\frac{M_{\pi}^4}{F_\pi^4}\,
\frac{1}{8 \pi}\,\left[
\xi^{(0)}_{x} (s)  +
2 \xi^{(1)}_{x} (s) \, \frac{{\hat L} (s)}{\sigma (s)} + 
2 \xi^{(2;{\mbox{\tiny{$\!\pm $}}})}_{x} (s) \sigma (s) {\hat L} (s) +
3 \xi^{(3)}_{x} (s) \, \frac{M_\pi^2}{s - 4 M_\pi^2} \, {\hat L}^2 (s) 
\right] \right\}
\! .
\nonumber
\end{eqnarray}
Finally, in the case of $F_{S ; {\mbox{\scriptsize{data}}}}^{\pi}(s)$, we use
the expression \rf{2loop_FF_ch} of the phase-removed form factor, replacing the 
functions ${\bar K}_n^{x_0} (s)$ by ${\bar K}_n^{0} (s)$, and the functions 
${\bar K}_n^{\mbox{\tiny$\Delta $}} (s)$ by ${\bar K}_n (s)$, respectively.
This then gives
\begin{eqnarray}
F_{S ; {\mbox{\scriptsize{data}}}}^{\pi}(s) &=& F_S^{\pi}\! (0)\! \left(
1 + a_S^{\pi}  s + b_S^{\pi}  s^2 
\right)
\nonumber\\
&&\!\!\!\!\!
-\, 8\pi F_S^{\pi^0}\! (0) \varphi_0^{x}(s) 
\left[
1 + a_S^{\pi^0} \! s + \frac{1}{\pi} \varphi_0^{00}(s) \right] \! {\rm Re} \, {\bar J}_0 (s) 
\nonumber\\
&&\!\!\!\!\!
+\, 16\pi {F_S^{\pi}(0)}\, \varphi_0^{{\mbox{\tiny{$+-$}}}}(s)
\left[
1 +  a_S^{\pi}\,s + \frac{2}{\pi} \varphi_0^{\mbox{\tiny{$ +-$}}}(s) \right] \! {\rm Re} \, {\bar J} (s)
\nonumber\\
&&\!\!\!\!\!
-\, \frac{M_{\pi}^4}{F_{\pi}^4}\,F_S^{\pi^0}\! (0)\! \left \{
\xi^{(0)}_{x} (s) {\rm Re} \, {\bar J}_0 (s) +
\xi^{(1)}_{x} (s) {\rm Re} \, {\bar K}_1^{0} (s) +
2 \xi^{(2;0)}_{x} (s) {\rm Re} \, {\bar K}_2^0 (s) +
\xi^{(3)}_{x} (s) {\rm Re} \, {\bar K}_3^{0} (s)
\right.
\nonumber\\
&&\!\!\!\!\!
\left.
\qquad\qquad\qquad
+\, 2 \xi^{(2;{\mbox{\tiny{$\!\pm $}}})}_{x} (s)
 \left[ 16 \pi^2 {\rm Re} \, {\bar J} (s) - 2  \right] {\rm Re} \, {\bar J}_0 (s)
\right\}
\nonumber\\
&&\!\!\!\!\!
+\, 2 \,\frac{M_{\pi}^4}{F_{\pi}^4}\,F_S^{\pi}(0) \bigg\{
\xi_{{\mbox{\tiny{$+-$}}};S}^{(0)} (s) {\rm Re} \, {\bar J} (s) 
+  \left[ \xi^{(1;{\mbox{\tiny{$\pm$}}})}_{{\mbox{\tiny{$+-$}}};S} (s) 
+ \xi^{(1;{\mbox{\tiny$\Delta$}})}_{{\mbox{\tiny{$+-$}}}; S} (s) \right] {\rm Re} \, {\bar K}_1(s) 
\nonumber\\
&&\!\!\!\!\!
\qquad\qquad\qquad
+\,
2 \xi^{(2;{\mbox{\tiny{$\pm$}}})}_{{\mbox{\tiny{$+-$}}};S} (s) \left[ {\rm Re} \, {\bar K}_2(s) + 
\frac{1}{32} \left( 1 - \frac{4 M_\pi^2}{s} \right) \right] +
\left[
\xi^{(3;{\mbox{\tiny{$\pm$}}})}_{{\mbox{\tiny{$+-$}}};S} (s)  +
\xi^{(3;{\mbox{\tiny$\Delta$}})}_{{\mbox{\tiny{$+-$}}}; S} (s) \right] {\rm Re} \, {\bar K}_3(s)  
\nonumber\\
&&\!\!\!\!\!
\qquad\qquad\qquad
+\, 
2 \xi^{(2;0)}_{{\mbox{\tiny{$+-$}}}; S} (s) \left[ 
\left( 16 \pi^2 {\rm Re} \, {\bar J}_0 (s) - 2  \right) {\rm Re} \, {\bar J} (s) 
+ \frac{1}{64} \left( 1 - \frac{4 M_{\pi^0}^2}{s} \right)
\right]
\bigg\} .
\lbl{2loop_FF_ch_data}   
\end{eqnarray}
In the sequel, we will generate pseudo-data using the expressions presented
in this subsection, considered to provide exact descriptions of the form factors.
In particular, it is understood that higher-order contributions are considered as vanishing. 
As already mentioned, in order to work within a framework where we deal
with fully analytical expressions of the form factors, we have made
some approximations as compared to the two-loop expressions discussed
in the preceding Section. Numerically, these differences
are small, but most important is that the approximations we have made
preserve the general features of the form factors as described after
Eq. \rf{f_0^x_decomp_1}.

\begin{figure}[t]
\begin{center}
\includegraphics[width=7cm]{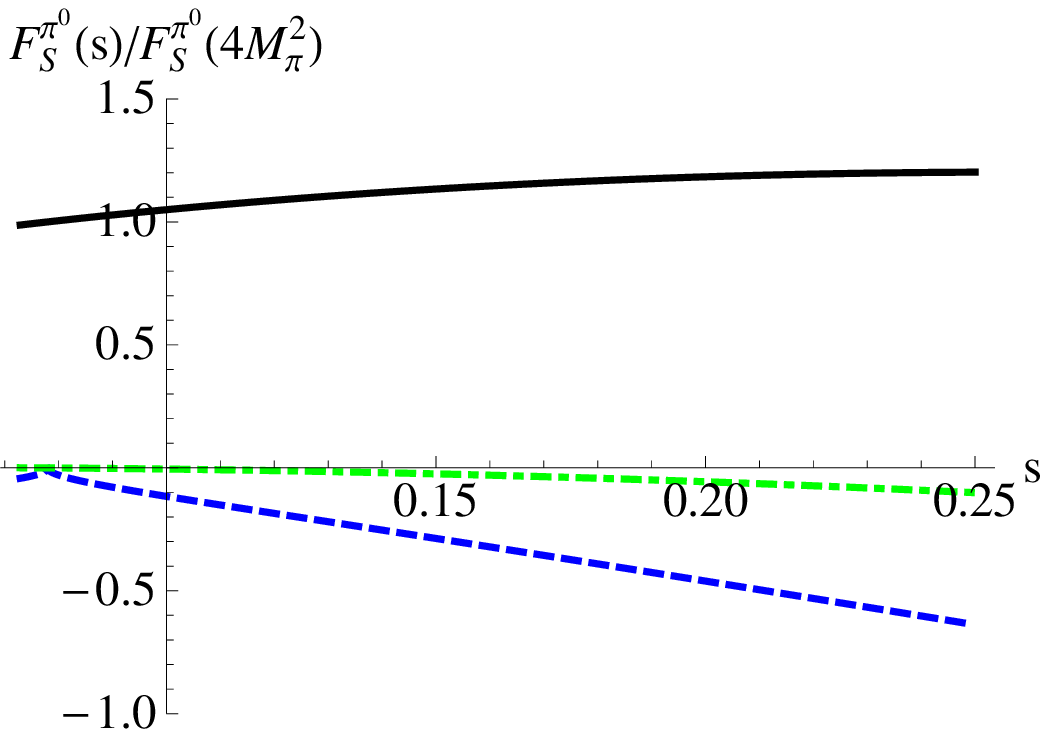}\quad \includegraphics[width=7cm]{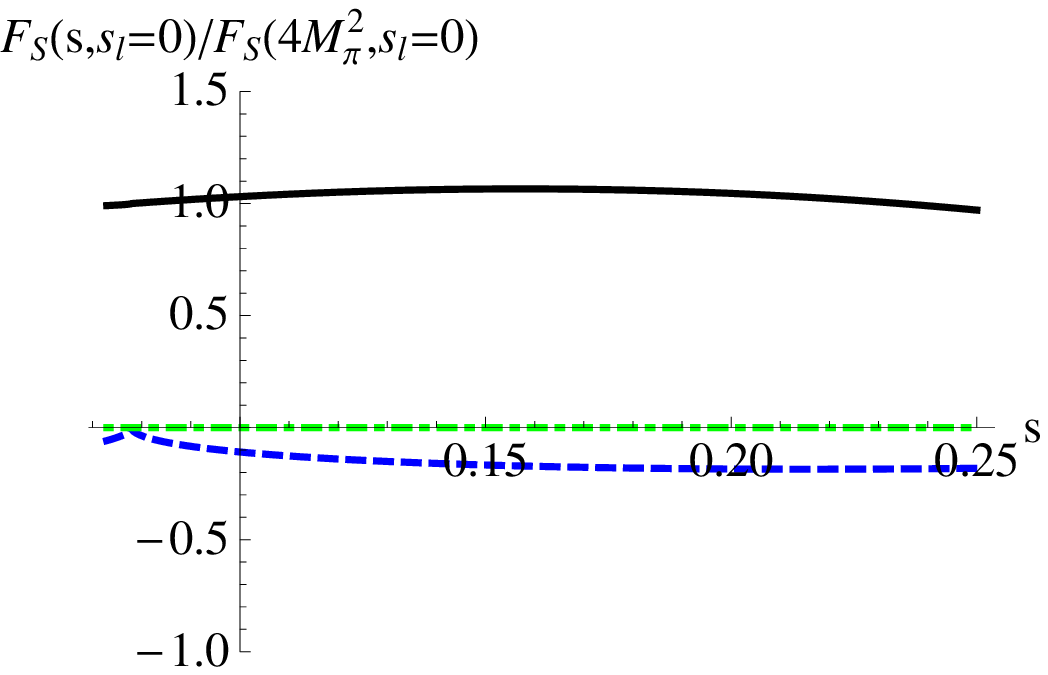}
\end{center}
\caption{Contributions, as functions of $s$ (in GeV$^2$), to $F_S^{\pi^0}(s)$ 
normalised to its value at $s = 4 M_\pi^2$: ${\cal F}_0^{\pi^0} (s)$ 
(black solid), ${\rm Re}\,{\cal F}_{1}^{\pi^0} (s)$ (blue dashed), 
and ${\rm Im}\,{\cal F}_{1 ; {\mbox{\scriptsize{data}}}}^{\pi^0} (s)$ (green dotted-dashed).
The left panel corresponds to our representation \rf{F_data_decomp}, \rf{cal_F_0_data},
and \rf{cal_F_1_data} of the exact form factors with the values of the parameters discussed  in the text. 
The right panel corresponds to the parameterisation  used by NA48/2 in Ref.~\cite{Batley:2014xwa}
for the $K_{e4}^{00}$ form factor, with the parameters chosen as described in the text after 
Eq. \rf{rates_num}.
\label{fig:Fpi0parts}}
\end{figure}

For the numerical generation of the pseudo-data, we need to fix the values of
the various parameters that occur in the expressions of the form factors.
As we want to compare different methods of analysis of the $K_{\ell 4}$ form factors,
we only aim at choosing values that are representative of the expected situation in these decays, 
with some limited arbritrariness in this choice.
In the following, we consider the case 
\begin{equation}
a_0^0=0.22  \qquad a_0^2=-0.045
. 
\end{equation}  
We fix the subtraction constants of the form factors by requesting that the charged scalar form factors 
has the typical values $r_\pi^2=0.60$ fm$^{2}$ and $c_S^\pi=10$ GeV$^{-4}$~\cite{Moussallam:1999aq}, leading to
\begin{equation}
a^\pi_S=2.63\ {\rm GeV}^{-2} \qquad b^\pi_S=2.96\ {\rm GeV}^{-4} 
\end{equation}
Using Ref.~\cite{DescotesGenon:2012gv}, one can compute the isospin-breaking shift between
$a^{\pi^0}_S$ and $a^{\pi}_S$. Assuming that $c_S^\pi=c_S^{\pi^0}$, we obtain
\begin{equation}
a^{\pi^0}_S=2.60\ {\rm GeV}^{-2} \qquad b^{\pi^0}_S=3.24\ {\rm GeV}^{-4} 
\end{equation}
leading to $r_{\pi^0}^2=0.59$ fm$^{2}$.
For the remaining parameters,
we use the same input values as in Ref.~\cite{DescotesGenon:2012gv}.
We normalise the scalar charged 
form factor to unity  at $s = 0$ and rescale the neutral one accordingly:
\begin{equation}
\frac{F_S^{\pi^0}(0)}{F_S^{\pi}(0)}=0.99
.
\lbl{FF_norm_ratio}
\end{equation}
As an illustration, we quote the values obtained for the total decay rate of Eq. \rf{rates}
from these inputs, for $V_{us} = 0.2255$:
\begin{equation}
\Gamma^{\pi^0} = 0.73 \cdot 10^{-22} \ {\rm GeV} 
.
\lbl{rates_num}
\end{equation}
We also show, on the left panel of Fig. \ref{fig:Fpi0parts},
the various contributions to the form factor $F_{S ; {\mbox{\scriptsize{data}}}}^{\pi^0}(s)$
obtained with our input values.
For comparison, the right panel shows the equivalent results in the case of the
parameterisation of the $K^{00}_{e4}$ form factor ${\cal M} (s)$
discussed in Section 9.4 of Ref. \cite{Batley:2014xwa}.
It corresponds, for ${\cal M}_1 (s)$, to the parameterization of Eqs. \rf{model_charged} 
below with $f'''_{s} = 0$ and the remaning parameters $f'_{s}/f_s$ and $f''_{s}/f_s$ fixed
at the central values given in Table 3 of Ref. \cite{Batley:2010zza}. For ${\cal M}_0 (s)$,
we take the expression of ${\cal F}_0^{\pi^0} (s)$ given in Eq. \rf{model_1}. 
The parameters, $f_s$, $f_{s0}/f_s$, $f'_{s0}/f_s$, and $f''_{s0}/f_s$ it involves 
are determined from a fit to the phase-space distribution given by Eq. (9.1) of 
Ref. \cite{Batley:2014xwa} at $s_e = 0$, and with the values of the coefficients $a$, $b$, and $d$
taken at their central values as shown in Table 1 of that same reference.
The overall normalisation $N$ is fixed such that the distribution
is equal to unity at $s = 4 M_\pi^2$, $s_e = 0$. One can observe
similar features in both plots on Fig. \ref{fig:Fpi0parts}
[the absence of an imaginary part in the right-hand plot
has already been discussed at the beginning of Sec. \ref{sec:theory}],
suggesting that our subsequent analysis, based on the scalar form factors
of the pion, has also some bearing on the $K_{e4}$ form factor.
Note that due to our choice of phase space in Eqs. \rf{dist} and \rf{rates}, 
in practice the region of interest in Fig. \ref{fig:Fpi0parts} corresponds to $s \leq 0.15$ GeV.

In summary, the form factors used in  order to generate the pseudo-data are defined by the expressions \rf{cal_F_0_data}, 
\rf{cal_F_1_data} and \rf{2loop_FF_ch_data} (with vanishing higher-order corrections), together with the values (central values
for all parameters, no error bars) of the parameters specified above.  The form factors thus defined will be 
referred to as the ``exact" form factors, considered to represent 
the ``truth'' to which we will fit different model parameterisations of the form factors, in order to obtain
a quantitative determination of the possible biases different parameterisations can have on the output of the analysis.

\subsection{Phenomenological parameterisations}\label{sect:analyses}

In order to mimic the situation in $K_{e4}$ decays, the pseudo-data generated with the 
exact scalar form factors of the pion will now be analyzed
using approximate phenomenological parameterisations.
For the analysis itself, we will consider a framework close to the experimental set up
for the $K_{e4}^{+-}$ \cite{Batley:2012rf,Batley:2007zz,Batley:2010zza} 
and the $K_{e4}^{00}$~\cite{Batley:2014xwa} decay channels.
{F}rom here on, we therefore also use $q^2 \equiv s/4M_\pi^2 - 1$
in addition to $s$, the square of the center-of-mass energy of the dipion system.
The region below the cusp corresponds to $q^2 \le0$, while positive values
of $q^2$ describe the region above the cusp.

For the charged-pion form factor, this means that we consider a parameterisation of the form
\begin{equation}
F_S^\pi (s) = f_s \left[1+\frac{f'_{s}}{f_{s}}q^2+\frac{f''_{s}}{f_{s}}q^4+ \frac{f'''_{s}}{f_{s}} q^6 \right]
,
\lbl{model_charged}
\end{equation} 
 
In the case of the neutral-pion form factor, we consider two parameterisations:
\begin{itemize}
\item Model 1: 
\begin{equation}
{\cal F}_0^{\pi^0} (s) = f_{s0} \left[1 + \frac{f'_{s0}}{f_{s0}} q^2 + \frac{f''_{s0}}{f_{s0}} q^4 \right]\qquad 
{\cal F}_1^{\pi^0} (s) = -2/3(a_0^0 - a_0^2) f_{s}\hat\sigma(s)
\lbl{model_1}
\end{equation}

\item Model 2:
\begin{equation}
\!\!
{\cal F}_0^{\pi^0} (s) = f_{s0} \left[1 + \frac{f'_{s0}}{f_{s0}} q^2 + \frac{f''_{s0}}{f_{s0}} q^4 \right]\qquad 
{\cal F}_1^{\pi^0} (s) = \hat\sigma(s)
\varphi_0^x(s)f_{s}\left[1+\frac{f'_{s}}{f_{s}}q^2+\frac{f''_{s}}{f_{s}}q^4+ \frac{f'''_{s}}{f_{s}} q^6 \right]
 e^{-\frac{i}{2} \sigma_0(s)\varphi_0^x(s)}
\lbl{model_2}
\end{equation}
\end{itemize}
In order to check the influence of possible higher-order
terms in the $q^2$ expansion, as compared to the parameterisation
considered in Ref.~\cite{Batley:2014xwa}, we have introduced a coefficient
$f'''_{s}$, which can be set to zero or kept as a free variable
in the fit.

The first parameterisation with $f'''_{s} = 0$ reproduces exactly the one that was 
considered in Sec. 9.4 in Ref.~\cite{Batley:2014xwa}. 
The second parameterisation incorporates more information gathered from
the theoretical discussion in Section \ref{sec:theory}, while remaining
sufficiently simple. Although we have chosen not to distinguish them,
the parameters $f_{s}$, $f'_{s}$, $f''_{s}$, and $f'''_{s}$
appearing in Eqs. \rf{model_1} and \rf{model_2} are not, {\it a priori}, 
identical to those occurring in the expression \rf{model_charged}.
One issue of the analysis we will present is precisely to determine
to which extent e.g. $f_s$ in Eq. \rf{model_1} should be expected
to agree with $f_s$ in Eq. \rf{model_charged}. Our first task is therefore
to provide reference values for the various parameters. This is done
by performing, in Eqs. \rf{cal_F_0_data} and \rf{2loop_FF_ch_data} the Taylor expansion 
around $q^2=0$, thus obtaining $f_{s0},f'_{s0},f''_{s0}$ from the former, and 
$f_{s},f'_{s},f''_{s},f'''_{s}$ from the latter. In the case of $F_S^\pi (s)$, we neglect the small 
half-integer powers of $q$ arising in the expansion, which do not contribute significantly in the vicinity 
of $q^2=0$. The resulting values are shown in the last column of Table \ref{tab:coeffs}.
These expansions are not supposed to provide accurate descriptions of the corresponding
form factors over the whole physical range. 

The comparison with our various fits will illustrate how different 
the parameters extracted from the fit and those describing the real Taylor expansion can be, 
and will thus give information on the possible bias introduced by the fitting procedure.
For convenience, in the following $f_{s0},f'_{s0}\ldots$ will be called ``neutral'' 
parameters, whereas $f_{s},f'_{s}\ldots$ are referred to as the ``charged'' parameters.

\subsection{Fitting procedures}\label{sect:fitting}

In order to stay close to the NA48/2 experimental set up,
we will thus assume that we have measurements of 
$|F^{\pi^0}_S(s)|^2$ at the 12 points corresponding to the barycenters of the experimental bins, 
and we assign a statistical uncertainty derived from the number of events collected in each bin 
($\sim 2900$ events in the first two bins, and $\sim 5900$ events in
all the other ones), without any correlations between the bins.
As the parameterisations given in Eqs. \rf{model_1} and \rf{model_2}
depend on the $S$-wave $\pi\pi$ scattering lengths, our $\chi^2$ will also include 
an uncertainty on these quantities in order to mock up the fact that in the real analysis these scattering lengths 
are determined from the charged form factor. Here we will use the
experimental information on these quantities, namely the latest NA48/2 combination of $K\to 3\pi$ and 
$K_{\ell 4}$ results~\cite{Batley:2010zza}
\begin{equation}\label{eq:chi2a00a02}
a_0^0=0.2210\pm 0.0047\pm 0.0040\,, \quad a_0^2=-0.0429\pm 0.0044 \pm 0.0028\,, \quad \rho_{a_0^0,a_0^2}=0.92\,,
\lbl{a00_a02_exp}
\end{equation}
where we combine statistical and systematic uncertainties in quadrature. One should notice 
that the central values are close (but not identical) to the ``true'' values used to generate our pseudo-data.

We consider the following methods to determine the coefficients of the above models.
\begin{itemize}
\item Method A:
fit of $|F^{\pi^0}_S(s)|^2$ for all points to determine all the (neutral, charged) parameters (setting $f'''_{s}=0$ 
to ensure a reasonable convergence of the fit), assuming the equality of the neutral and charged normalisation ($f_s=f_{s0}$)
\item Method B:
fit of $|F^{\pi^0}_S(s)|^2$ for all points to determine all the (neutral, charged) parameters, setting $f'''_{s}=0$
and keeping the normalisations $f_s$ and $f_{s0}$ distinct
\item Method C:
fit of $|F^{\pi^0}_S(s)|^2$ to determine the neutral  parameters, injecting information on charged 
parameters  by adding to the $\chi^2$ a contribution corresponding to a fit of the charged form factor 
$F^{\pi}_S(s)$ to the polynomial expression \rf{model_charged},
effectively identifying the charged parameters in the models for $F^{\pi^0}_S(s)$ 
with the parameters occurring in the charged scalar form factor.
\end{itemize}
In method C, we generate pseudo-data points for the charged-pion scalar form factor with energies corresponding 
to the barycenters given in ref.~\cite{Batley:2012rf}, and use the relative uncertainties  for $F_S$ (combined in quadrature) quoted for 
each bin in this reference, without correlations. In agreement with ref.~\cite{Batley:2012rf}, we add an overall 0.62\% relative 
uncertainty, completely correlated between all the charged bins. The curvature of the charged form factor $F^{\pi}_S(s)$ being 
more pronounced than that of the scalar $K_{e4}^\pm$ form factor, a
$q^6$ term must be included in the polynomial in order to obtain a good description of the form factor over the whole kinematic range.

We give the resulting $\chi^2_{\min}$ (obtained from the best-fit values of each method). Even though each model 
provides through its fit a value of $f_{s0}$, one can also determine the latter by considering the branching ratio. 
In this case, $f_{s0}$ is determined by integrating the decay distribution obtained by using as inputs
the slope parameters determined from the different methods of fitting, and fixing the normalisation
by comparison with the total decay rate $\Gamma^{\pi^0}$ defined in Eq. \rf{rates},  
and evaluated with the exact form factor ${F}_{S ; {\mbox{\scriptsize{data}}}}^{\pi^0}(s)$.
The corresponding numerical value is given in Eq. \rf{rates_num}. 
We denote by $r$ the ratio between the value of
$f_{s0}$ determined this way from the branching ratio, and the true value computed from the exact form factor,
i.e. the reference value $f_{s0} = 1.381$.

\subsection{Discussion of the results}

In order to obtain an estimate of the uncertainty attached to the coefficients of models 1-2 using methods A and B, we will perform fits 
of the models on a series of 10000 pseudo-experiments, generated by assuming that the data are random variables with a mean given by our 
theoretical model for the neutral scalar form factor and a standard deviation given by the relative uncertainty of the corresponding form 
factor measured in $K_{e4}$ decays by the NA48 experiment~\cite{Batley:2012rf,Batley:2014xwa}. We will then determine the mean and the 
variance of the resulting distribution for each coefficient of the parameterisation considered. 
The results are gathered in Table.~\ref{tab:coeffs}. The column labeled ``reference'' provides a comparison with
the coefficients obtained from the Taylor expansions of the form factors given in 
Eqs. \rf{cal_F_0_data} and \rf{2loop_FF_ch_data}, as described after Eq. \rf{model_2}.

\begin{table}
\begin{center}
\begin{tabular}{c|c|c|c}
&  A1 & A2 & Reference\\
\hline
$\chi^2/N_{dof}$  & $(9.1\pm 4.3)/9$ & $(6.9\pm 3.6)/7$  & \\
\hline
$f_{s0}/F_S^{\pi}(0)=f_{s}/F_S^{\pi}(0)$ & $1.38\pm 0.01$ &  $1.38\pm 0.01$ & (1.381,1.395)\\
$f'_{s0}/f_{s0}$  & $0.18\pm 0.03$ &  $0.23\pm 0.03$  & 0.191 \\
$f''_{s0}/f_{s0}$  & $\!\!\!\!-0.03\pm 0.03$ &  $\!\!\!\!-0.33\pm 0.44$  & $\!\!\!\!-0.059$ \\
$f'_{s}/f_{s}$  &  0 & $1.91\pm 6.31$&  0.199\\
$f''_{s}/f_{s}$  & 0 & $\!\!\!\!-1.01\pm 6.38$ & $\!\!\!\!-0.032$ \\
\hline
$r$  & $1.00\pm 0.01$ &  $0.98\pm 0.04$ & 1 \\
\\
&  B1 & B2 & Reference\\
\hline
$\chi^2/N_{dof}$  & $(8.0\pm 4.0)/8$ & $(6.0\pm 3.2)/6$ &  \\
\hline
$f_{s0}/F_S^{\pi}(0)$ & $1.38\pm 0.01$ &  $1.38\pm 0.01$ & 1.381\\
$f'_{s0}/f_{s0}$  & $0.18\pm 0.04$ &  $0.21\pm 0.12$ & 0.191 \\
$f''_{s0}/f_{s0}$  & $\!\!\!\!-0.03\pm 0.04$ &  $\!\!\!\!-0.27\pm 0.39$  & $\!\!\!\!-0.059$\\
$f_{s}/f_{s0}$  & $1.33\pm 0.40$ &  $0.97\pm 0.45$  & 1.010\\
$f'_{s}/f_{s}$  &  0 & $0.43\pm 12.8$ & 0.199\\
$f''_{s}/f_{s}$  & 0 & $0.95\pm 14.6$  & $\!\!\!\!-0.032$ \\
\hline
$r$  & $1.00\pm 0.01$ &  $0.97\pm 0.05$  & 1 \\
\\
&  C1 & C2 & Reference\\
\hline
$\chi^2/N_{dof}$  & $(3.3 \pm 0.1)\cdot 10^6/18$ & $(15.0\pm 5.6)/16$ & \\
\hline
$f_{s0}/F_S^{\pi}(0)$ & $1.38\pm 0.01$ &  $1.38 \pm 0.01$  & 1.381\\
$f'_{s0}/f_{s0}$  & $0.17\pm 0.03$ &  $0.19\pm 0.03$ & 0.191 \\
$f''_{s0}/f_{s0}$  & $\!\!\!\!-0.02\pm 0.03$ &  $\!\!\!\!-0.04\pm 0.04$ & $\!\!\!\!-0.059$\\
$f_{s}/f_{s0}$  & $1.07\pm 0.01$ &  $1.01\pm 0.01$ & 1.010\\
$f'_{s}/f_{s}$  &  0 & $0.19\pm 0.01$ & 0.199\\
$f''_{s}/f_{s}$  & 0 & $\!\!\!\!-0.03\pm 0.01$ & $\!\!\!\!-0.032$ \\
$f'''_{s}/f_{s}$  & 0 & $0.01\pm 0.01$  & 0.012\\
\hline
$r$  & $1.00\pm 0.01$ &  $1.00\pm 0.02$  & 1
\end{tabular}
\caption{Results of the different models (1-2) and methods (A-B-C), compared to the reference values,
obtained from the Taylor expansion of the exact form factors \rf{F_data_decomp}, \rf{cal_F_0_data},
and \rf{cal_F_1_data} with the values of the parameters discussed  in the text.
The value of the ratio $r$ does not result from the fit, but is obtained once the fit 
has been performed, see the last paragraph of Section \ref{sect:fitting}.
The column label A1 (A2) refers to the fit method A using Model 1 (2), and so on.}\label{tab:coeffs}
\end{center}
\end{table}

As shown by the comparison between Models 1 and 2 for methods A and B, the higher powers of $q^2$ are only 
weakly constrained. Model 1 is very rough and provides a very poor description of the charged form 
factor (modelling it as a simple constant), which explains the very bad 
$\chi^2_{\rm min}$ for method C1. Only $f_{s0}$ and 
$f'_{s0}/f_{s0}$ can be determined with a good accuracy, but there is no significant bias introduced 
by the fitting procedure with respect to the reference values.  Despite of its shortcomings, method A 
gives good results for the neutral parameters. As expected, compared to method B, method C 
provides a much better accuracy on the neutral parameters since the charged ones are constrained in this method. 
As shown by the ratio $r$, both methods yield accurate values of $f_{s0}$ (at the few percent level) obtained by
integrating over the phase space to consider the branching ratio, even methods that do not attempt at 
describing the $q^2<0$ region correctly. This can be easily understood: both methods are constrained to 
describe correctly $|F_S^{\pi^0}(s)|^2$ for small $q^2>0$ (as can be seen by their agreement 
concerning $f_{s0}$ and $f'_{s0}$), but they may differ for $q^2<0$ (which exhibit larger uncertainties). However, this region is very 
narrow ($4M_{\pi^0}^2\leq s\leq 4M_{\pi}^2$) and its contribution is further suppressed by phase space. Therefore, the impact of this 
region on the estimation of the branching ratio is very small, and the latter is completely dominated by the region $s\geq 4M_{\pi}^2$ 
where all parameterisations agree (the uncertainties reflecting mainly the uncertainties of the inputs and the lack of data at large $q^2$).

From this discussion, one thus expects that using the fit function given by Eq. (9.1) of  Ref.~\cite{Batley:2014xwa}, 
and described at the beginning of Sec.~\ref{sec:theory}, will lead to similar results for the ratio $r$,
despite the fact that this model complies with the expected structure of the cusp only if one imposes strong assumptions,
and should be considered as a mere phenomenological parametrisation to reproduce data smoothly. One indeed obtains
$r=1\pm 0.01$ and $\chi^2/N_{dof}=(8.0\pm 4.1)/9$ for method A, and $r=1\pm 0.01$ and $\chi^2/N_{dof}=(8.1\pm 4.1)/8$  
for method B, illustrating once more that a smooth parametrisation of the curve above the cusp in good agreement with the 
data is enough to obtain an accurate and unbiased value for the normalisation $f_{s0}$.

The outcome of this discussion is that the measurement of $\vert F_S^{\pi^0}\vert^2$ allows for an accurate determination of
$f_{s0}$ (at  the percent level), in the current experimental setting. As shown by the ratio $r$, 
the value of $f_{s0}$ obtained from the computation of the 
branching ratio is equal (within uncertainties) to its true value for all methods and parameterisations
considered here. Even though one has to keep in 
mind that this observation is done using the pion scalar form factors rather than the actual $K_{\ell 4}$ form factors, 
it nevertheless suggests that the fit procedure adopted in Ref.~\cite{Batley:2014xwa} does not 
bias the determination of $f_{s0}$, and thus cannot explain the surprisingly higher value of $f_{s}$ 
extracted by the NA48/2 collaboration from the $K^{00}_{e4}$ channel, as compared to 
the value for $f_s$ determined from the $K^{+-}_{e4}$ channel.

\begin{table}
\begin{center}
\begin{tabular}{c|c|c|c}
  & D2 & D2 with $\sigma[K_{\ell 4}^{00}]/ 10$ & Reference ($s0$, $s$)\\
  \hline
$\chi^2/N_{dof}$  &  $(4.8\pm 3.1)/5$ & $(5.6\pm 3.3)/7$ & \\
\hline
$f_{s0}/F_S^{\pi}(0)=f_{s}/F_S^{\pi}(0)$ &  $1.38 \pm 0.01$ &  $1.38 \pm 0.01$ &(1.381,1.395)\\
$f'_{s0}/f_{s0}$   &  $0.20\pm 0.12$ &  $0.20\pm 0.01$ & 0.191\\
$f''_{s0}/f_{s0}$  &  $0.21\pm 0.88$ &  $\!\!\!\!-0.05\pm 0.06$ & $\!\!\!\!-$0.059\\
$f'_{s}/f_{s}$   &  $\!-14\pm 234$ &  $0.62\pm 1.86$ & 0.199\\
$f''_{s}/f_{s}$  &  $~~32\pm 516$ &  $0.00\pm 3.95$ & $\!\!\!\!-$0.032\\
$a_0^0-a_0^2$ & $0.21\pm 0.16$ &  $0.25\pm 0.03$ & 0.265\\
$a_0^2$ & $\!\!\!\!-0.83\pm 2.87$ & $0.00\pm 0.43$ &  $\!\!\!\!-$0.045\\ 
\hline
$r$  &  $0.92\pm 0.15$ &  $0.99\pm 0.02$ & 1\\
\\
  & E2 & E2 with $\sigma[K_{\ell 4}^{00}]/ 10$ & Reference \\
  \hline
$\chi^2/N_{dof}$  &  $(13.2\pm 5.2)/14$ & $(12.9\pm 5.1)/14$ & \\
\hline
$f_{s0}/F_S^{\pi}(0)$ &  $1.38 \pm 0.01$ &  $1.38 \pm 0.01$ & 1.381\\
$f'_{s0}/f_{s0}$   &  $0.22\pm 0.05$ &  $0.20\pm 0.01$ & 0.191\\
$f''_{s0}/f_{s0}$  &  $\!\!\!\!-0.09\pm 0.08$ &  $\!\!\!\!-0.05\pm 0.02$ & $\!\!\!\!-$0.059\\
$f_{s}/f_{s0}$   &  $1.01\pm 0.01$ &  $1.01\pm 0.01$& 1.010\\
$f'_{s}/f_{s}$  & $0.19\pm 0.01$ & $0.19\pm 0.01$ & 0.199\\
$f''_{s}/f_{s}$   & $\!\!\!\!-0.03\pm 0.01$ & $\!\!\!\!-0.03\pm 0.01$  & $\!\!\!\!-$0.032 \\
$f'''_{s}/f_{s}$  & $0.01\pm 0.01 $ & $0.01\pm 0.01$  & 0.012\\
$a_0^0-a_0^2$ & $0.25\pm 0.10$ &  $0.26\pm 0.03$ & 0.265\\
$a_0^2$ & $0.05\pm 0.16$ & $0.07\pm 0.53$ &  $\!\!\!\!-$0.045 \\
\hline
$r$  &  $1.00\pm 0.01$ &  $1.00\pm 0.01$ & 1
\end{tabular}
\caption{Results for methods D2 and E2, where $\pi\pi$ scattering lengths are fitted in addition to data on the neutral scalar form factor 
(for method D) or for both neutral and charged scalar form factors (method E). The third column corresponds to the case where the 
uncertainties  for the neutral scalar form factor are divided by 10.}\label{tab:a02}
\end{center}
\end{table}

\subsection{Constraining the scattering lengths}\label{sect:scattering}

The presence of a cusp similar to the one observed in the three-body $K^+\to\pi^+\pi^0\pi^0$ decay 
suggests that it should, in principle, be possible to extract information on 
the scattering lengths from an accurate measurement of the $K_{e4}^{00}$ differential decay rate. 
At leading order, the cusp is related to the difference of scattering
lengths $a_0^0-a_0^2$. Going to higher orders in ${\cal F}_1^{\pi^0} $ (i.e. including $\varphi_0^{x}$) 
will also add a (weaker) dependence on $a_0^2$.
 The scattering lengths can be determined only once the relative 
normalisation of form factors involved in ${\cal F}_0^{\pi^0} $ and 
${\cal F}_1^{\pi^0} $ is fixed, which 
requires the determination of the charged parameters in some way.
We define two methods for this purpose.
Method D is exactly as method A, without including any experimental 
information on $a_0^0$ and $a_0^2$ in the $\chi^2$ [i.e. removing them from the 
$\chi^2$ as described in eq.~(\ref{eq:chi2a00a02})], and similarly
for method E with respect to method C. We proceed as before, but now also fitting the scattering lengths. 

We consider model 2, as model 1 yielded poor results in the previous section for method C. 
In order to discuss the potential impact future experimental improvements could have, 
we consider also a situation where all statistical errors are reduced by 10 (but the number of bins is 
unchanged) for the neutral channel, keeping the uncertainties unchanged for the charged channel.

The results gathered in Table~\ref{tab:a02} show that the current statistical uncertainties yield a relative 
uncertainty on $a_0^0-a_0^2$ of around 80\% for D2 and 40\% for E2. For D2, the charged parameters are only 
very poorly constrained, but this does not prevent the fit to be reasonable. Reducing the statistical 
uncertainties by 10 (for the neutral part) yields a significant reduction in the uncertainties, 
leading to a relative uncertainty on $a_0^0-a_0^2$ of 27\% for D2 and 10\% for E2. 
At this level of accuracy, there is no significant bias in the value of
$a_0^0 - a_0^2$ extracted through these various approaches.
As expected, no relevant information can be obtained on $a_0^2$, due to the very small sensitivity 
of the neutral-pion channel to this quantity. To illustrate this point, if instead we
fix the value of $a_0^2$ to its central value in Eq. \rf{a00_a02_exp}, our results
concerning the uncertainty on $a_0^0-a_0^2$ and the quality of the fit remain unchanged.

{F}rom this discussion, one can hope to get some information on $a_0^0-a_0^2$ using model 2, 
should a larger data set become available for $K_{e4}^{00}$ in the future. One has however 
to keep in mind that we have assumed the equality between the charged and neutral normalisations 
in the polynomials for the neutral scalar form factors in the case of method D, as well as the 
equality between the charged parameters in the polynomials for the charged and neutral scalar 
form factors in the case of method E. These assumptions are certainly reasonable considering 
the current uncertainties involved, but one might need to reassess them in the presence of more 
accurate data. In this context, it is also interesting to notice that
the current result from the DIRAC experiment Ref.~\cite{Adeva:2011tc} is $\vert a_0^0-a_0^2 \vert=0.253\pm 0.011$, 
i.e. a 4.3\% uncertainty, so that a substantial increase of the statistical sample of 
$K_{e4}^{00}$ decays is needed in order to reach a comparable accuracy.

\section{Radiative corrections to the $K_{e4}^{00}$ decay rate }\label{sec:rad_corr}

In this Section we now discuss radiative corrections, which were addressed
differently in the analyses of the $K^{+-}_{e4}$ and $K^{00}_{e4}$ channels so far.
In the latter case, no radiative corrections  were applied to the decay rate
measured in Ref. \cite{Batley:2014xwa}. This accounts for the unspecified factor 
$\delta_{EM}$ in Eq. \rf{exp_comp}. It is thus natural to ask
how much the observed 6.5\% discrepancy [see Eq. \rf{discrepancy}]
in the normalisation of the form factor measured in the two channels 
is due to this correction factor.
Our aim here is not to provide a complete discussion of radiative corrections
in the $K_{e4}$ decay channels at a level of sophistication
that would match the treatment of isospin breaking 
due to the difference between masses of the charged and neutral pions.
We rather want to work out these corrections in a somewhat simpler framework,
trying to reproduce a treatment of radiative
corrections in the neutral channel similar to the one that was applied in the charged channel, 
in order to make the comparison as meaningful as possible.

\subsection{Treatment of radiative corrections in $K_{e4}^{+-}$ data}\label{photos_charged}

Let us recall how radiative corrections are treated in the
charged channel \cite{Batley:2007zz,Batley:2010zza}. First, virtual photon 
exchange between all possible pairs of charged external lines are considered, 
and the corresponding Sommerfeld-Gamow-Sakharov
factors are applied. The corrections induced by emission of real photons 
are treated with PHOTOS \cite{Barberio:1993qi,Golonka:2005pn,Nanava:2006vv,XuWas10}. 
The latter also implements the wave-function
renormalisation on the external charged legs. The couplings of photons
to mesons are treated as point-like interactions, given by scalar QED.
The result is then free from infrared singularities.
Furthermore, one neglects the contributions that
vanish when the electron mass goes to zero, which is a sensible
limit to consider for the $K_{e4}$ decay channels.

Apart from the Sommerfeld-Gamow-Sakharov factors, 
some contributions that would arise within a more systematic approach,
provided by the effective low-energy theory of QCD and QED for light quarks and leptons
\cite{Gasser:1984gg,Urech:1994hd,Knecht:1999ag},
as applied in Refs. \cite{Vesna04,Cuplov:2003bj,Stoffer:2013sfa}
to the channel with two charged pions,  are not considered. 
These include, for instance, all structure-dependent
corrections, where the photon is emitted from the tree-level $K_{e4}$ vertices
or from internal charged lines.
The outcome of such a truncated calculation is affected by
an ultraviolet divergence, which is removed by renormalizing
the coupling $\vert V_{us} \vert^2 G_{\rm F}^2$
[note that in Eq. (12) of Ref. \cite{XuWas10} the factor $(\alpha/\pi)$ has 
been inadvertently omitted],
\begin{equation}
\left( \vert V_{us} \vert^2 G_{\rm F}^2 \right)^{\rm bare} 
\left( 1 - \frac{9}{4} \frac{\alpha}{\pi} \ln \frac{\Lambda^2}{M_\pi^2} \right) 
=
\vert V_{us} \vert^2 G_{\rm F}^2
.
\lbl{renorm_Gf}
\end{equation}
This same correction factor also appears in Ref. \cite{Bystritskiy:2009iv},
with the ultraviolet cut-off $\Lambda$ taken equal to $M_W$. From this
last reference, we also see that the
factor $9/4$ decomposes as $9/4 = 3 \times (1/2) - 1/4 + 1/2 + 1/2$, where
the first contribution comes from the wave-function renormalisation
of the three charged mesons, the second from the (charged) lepton wave-function
renormalisation, and the last two ones from the virtual photon loops
between the charged kaon and the charged lepton on the one hand, 
and between the two charged pions on the other hand
[the remaining divergencent contributions of this type, i.e. a photon line
connecting the external kaon to the charged-pion lines, or the
charged lepton with each of the two pions, cancel pairwise].

In the case of the $K^{00}_{e4}$
channel, we therefore expect that this factor becomes
$1 \times (1/2) - 1/4 + 1/2 = 3/4$. Since it differs from the previous
one, it cannot be absorbed by the renormalisation of the same
prefactor $\vert V_{us} \vert^2 G_{\rm F}^2$ as before. It
seems more natural instead to absorb these ultraviolet divergences
into the normalisations of the form factors
\begin{equation}
f_s^{\rm bare} \left( 1 - \frac{9}{8} \frac{\alpha}{\pi} \ln \frac{\Lambda^2}{M_\pi^2} \right) = f_s
,\qquad
f_{s0}^{\rm bare} \left( 1 - \frac{3}{8} \frac{\alpha}{\pi} \ln \frac{\Lambda^2}{M_\pi^2} \right) = f_{s0}
.
\lbl{renormphotos}
\end{equation}
This is also more in line with the effective theory approach mentioned above,
where the form factors are also corrected by (different) contributions from the low-energy constants
$K_i$ \cite{Urech:1994hd} or $X_i$ \cite{Knecht:1999ag}, which are renormalised
by the ultraviolet divergences coming from the photon loops. Using instead Eq. \rf{renorm_Gf}
in both cases would leave a remaining cut-off dependent contribution to the $K_{e4}^{00}$
amplitude. For a typical value of $\Lambda = 1$ GeV, this would modify Eq. \rf{discrepancy_corrected}
at the per mille level.

\subsection{Radiative corrections {\it \`a la} PHOTOS for the $K_{e4}^{00}$ decay rate}\label{photos_neutral}

In the following, we will try to estimate the potential impact of PHOTOS on $K_{e4}^{00}$ rather than pursuing an effective field theory approach.
If we want to reproduce the analogue of the PHOTOS treatment \cite{XuWas10} of radiative corrections
for the $K_{e4}^{00}$ decay rate, we need to consider the wave-function renormalization
of the charged lepton and of the kaon in (scalar) QED, and the vertex correction corresponding
to diagram $(a)$ in Fig. \ref{fig:virtual}. Using a Pauli-Villars regularization, 
and taking the photon propagator in the Feynman gauge, we reproduce the expressions of 
Eq. (6) in Ref. \cite{Bystritskiy:2009iv} for the former.
In order to evaluate and discuss the contribution from diagram $(a)$ in Fig. \ref{fig:virtual},
we chose to describe the tree-level $K_{\ell4}^{00}$ vertex as
\begin{equation}
A_\mu = -i \frac{1}{M_K} \left[ F^{00} (p_1 + p_2 )_\mu  + R^{00} (k - p_1 - p_2 )_\mu \right]
.
\end{equation}
with constant form factors [Bose symmetry forbids a contribution
of the form $G^{00} (p_1 - p_2 )_\mu $ with $G^{00}$ constant], so 
that the lowest-order amplitude reads
\begin{equation}
{\cal A}_0(K^{00}_{e4} ) \equiv
\frac{i}{M_K}
\frac{G_{\rm F}}{\sqrt{2}}  V_{us}
{\bar{\rm u}} (p_e) \gamma^\mu (1-\gamma_5) {\rm v}(p_\nu)
\left[ F^{00} (p_1 + p_2 )_\mu + R^{00} (k - p_1 - p_2 )_\mu \right]
.
\end{equation} 
In the limit $m_e \to 0$, we obtain
\begin{eqnarray}
{\cal A}(K^{00}_{e4} ; \ref{fig:virtual}(a)) & = &
 e^2 {\cal A}_0(K^{00}_{e4} )
 \left[
 \frac{1}{16 \pi^2} \left( \ln \frac{\Lambda^2}{M_K^2} + 1 \right)
+ 4 (k\cdot p_e ) C \left( (k-p_e)^2 ; M_K , m_e \right)
 \right. 
 \nonumber\\
 &&
 \left. 
 \qquad
 - \, 2 M_K^2 C_{11} \left( (k - p_e)^2 ;  M_K , m_e \right)
 - 4 (k \cdot p_e) C_{12} \left( (k - p_e)^2 ;  M_K , m_e \right)
 \right]
\nonumber\\
&&
\!\!\!\!
+ \, i\,\frac{e^2}{M_K}\frac{G_{\rm F}}{\sqrt{2}}  V_{us} \,
R^{00} ( p_1 + p_2 )_\nu
\times
{\bar{\rm u}} (p_e) \gamma^\nu ( 1 - \gamma_5) {\rm v}(p_{\nu_e})
\lbl{A_diag_a}\\
&&
\times
\left[
\frac{1}{32 \pi^2} \left( 3 \ln \frac{\Lambda^2}{M_K^2} + \frac{1}{2} \right)
- 4 (k \cdot p_e) C_{11} \left( (k - p_e)^2 ;  M_K , m_e \right)
+ 2 {\bar J} \left( (k-p_e)^2 ; M_K , m_e \right)
\right]
.
\nonumber
\end{eqnarray}
The various loop functions occurring in this expression are 
defined in Appendix \ref{app:loop_functions}.
Adding to it the wave-function renormalizations
on the charged external lines gives the following result,
in the framework adopted here, for the radiatively corrected amplitude
[$m_\gamma$ denotes a small photon mass, introduced as an infrared
regulator, to be sent to zero once an infrared-safe observable has
been constructed]:
\begin{eqnarray}
\sqrt{Z_e} \sqrt{Z_K} {\cal A}_0(K^{00}_{e4} ) + {\cal A}(K^{00}_{e4} ; \ref{fig:virtual}(a))
& = &
{\cal A}_0(K^{00}_{e4} ) \left[
1 + \frac{e^2}{(4 \pi )^2}
\left( \frac{3}{2} \ln \frac{\Lambda^2}{M_\pi^2}
- \frac{1}{2} \ln \frac{m_e^2}{M_\pi^2} + 2 \ln \frac{m_e^2}{m_\gamma^2} - \ln \frac{M_K^2}{M_\pi^2} - 2 \right)
\right.
\nonumber\\
&&
\left.
\qquad
\ + \, e^2 \bigg(
4 (k\cdot p_e ) C \left( (k-p_e)^2 ; M_K , m_e \right)
 \right. 
 \nonumber\\
 &&
 \left. 
 \qquad\quad
 - \, 2 M_K^2 C_{11} \left( (k - p_e)^2 ;  M_K , m_e \right)
 \right. 
 \nonumber\\
 &&
 \left. 
 \qquad\quad
 - 4 \, (k \cdot p_e) C_{12} \left( (k - p_e)^2 ;  M_K , m_e \right)
 \bigg)
\right]
\nonumber\\
&&
\!\!\!\!
+ \, i\,\frac{e^2}{M_K}\frac{G_{\rm F}}{\sqrt{2}}  V_{us} \,
R^{00} ( p_1 + p_2 )_\nu
\times
{\rm u} (p_e) \gamma^\nu ( 1 - \gamma_5) {\rm v}(p_{\nu_e})
\lbl{photos}
\nonumber\\
&&
\!\!\!
\times
\left[
\frac{1}{32 \pi^2} \left( 3 \ln \frac{\Lambda^2}{M_K^2} + \frac{1}{2} \right)
+ 2 {\bar J} \left( (k-p_e)^2 ; M_K , m_e \right)
 \right. 
 \nonumber\\
 &&
 \left. 
 \qquad\quad
- 4 (k \cdot p_e) C_{11} \left( (k - p_e)^2 ;  M_K , m_e \right)
\right]
.
\end{eqnarray}
We make a few comments about this result:

\begin{itemize}

\item
Although the result \rf {A_diag_a} holds in the Feynman gauge $\xi = 1$,
we have also computed the wave-function renormalisations
and ${\cal A}(K^{00}_{e4} ; \ref{fig:virtual}(a))$ in an arbitrary
linear and covariant $\xi$-gauge, and we have checked that the final
result \rf{photos} does actually not depend on the gauge-fixing parameter $\xi$.

\item
In order to reproduce the analogue of the PHOTOS treatment \cite{Bystritskiy:2009iv,XuWas10} 
of radiative corrections for the $K_{e4}^{00}$ decay rate, one needs to
add the emission of soft photons from the charged external lines, 
diagrams $(a)$ and $(c)$ of Fig. \ref{fig:emission} [in the $K_{e4}^{+-}$ case,
there are two more diagrams where the photon is emitted from the charged pion lines] 
so that the result is free of infrared singularities at order $\alpha$.
These corrections will be discussed later on. At this stage, we simply
note that the infrared-divergence of Eq. \rf{photos} is equal to
\begin{equation}
\frac{e^2}{(4 \pi)^2} \, {\cal A}_0(K^{00}_{e4} ) \times
\ln m_\gamma \left[ -4 - 2 (k \cdot p_e) \tau( k , p_e) \right]
,
\lbl{photos-IR}
\end{equation}
with the function $\tau (p_1 , p_2 )$ defined in Eq. \rf{tau_function}.

\item
The factor $\frac{3}{8} \frac{\alpha}{\pi} \ln \frac{\Lambda^2}{M_\pi^2}$ discussed
in Eq.~\rf{renormphotos} is indeed to be found in Eq. \rf{photos}, provided 
that one sets $R^{00}$ to zero. The only remaining contribution comes from $F^{00}$, 
which is proportional to $f_{s0}$ at this level. This indeed corresponds 
to the situation considered in Ref. \cite{Bystritskiy:2009iv}.
In the absence of radiative corrections, the form factor $R^{00}$ [or $R^{+-}$
in the charged channel] does not contribute to the decay distribution for $m_e = 0$.
In this case, one may as well take $R^{00} = 0$ from the beginning.
But once radiative corrections are switched on, taking $m_e = 0$ or $R^{00} = 0$
are no longer equivalent options. As shown by the second contribution in Eq. \rf{photos},
there is a correction to $F^{00}$ that is induced by $R^{00}$, and this
contribution is not considered in Ref. \cite{Bystritskiy:2009iv}, and is hence also
missing in Ref. \cite{XuWas10}. 

\item
At lowest order, and in the isospin limit, one has \cite{Bijnens:1989mr}
\begin{equation}
R^{00} 
= \frac{F^{00}}{2} \left[ 1 + \frac{s_\pi}{s_e - M_K^2}  \right]
= F^{00} \left[ \frac{2}{3} - \frac{1}{2} \left( \frac{1}{3} - \frac{s_\pi}{s_e - M_K^2} \right) \right]
.
\end{equation}
Actually, as shown in the second expression, the vertex in the diagram $(a)$ in Fig. \ref{fig:virtual} only
accounts for the contribution $R^{00} = (2/3) \cdot F^{00}$. 
The second factor comes from the diagram $(f)$ in Fig. \ref{fig:virtual}. 

\end{itemize}

\subsection{Additional non-factorizable radiative corrections to the $K_{e4}^{00}$ decay rate}\label{rad_neutral}

After these preliminary remarks concerning the PHOTOS-type treatment of radiative
corrections in the $K^{+-}_{e4}$ and $K^{00}_{e4}$ channels, let us now address
radiative corrections in the channel with two neutral pions in a somewhat more systematic
manner. This will allow us to estimate the size of the radiative corrections that 
are not included in the experimental analysis, as described in the previous section.
We keep on considering the limit where $m_e$ vanishes, so that in the absence
of radiative corrections the amplitude reads simply
\begin{equation}
{\cal A}_0(K^{00}_{e4} )  =
i \frac{G_{\rm F}}{\sqrt{2}}  V_{us}
{\bar{\rm u}} (p_e) \left( \not \! p_1 + \not \! p_2 \right) (1-\gamma_5) {\rm v}(p_\nu) 
\times
\frac{F^{00}}{M_K}
+ {\cal O}(m_e)
.
\end{equation}
For our purpose, it is convenient to distinguish between two types of radiative corrections, 
that we call factorisable and non-factorisable. Factorisable radiative corrections
are defined by the contributions where both ends of the virtual photon line connects
to a charged mesonic line or to the vertex with the leptonic current, or when both ends
connect to the electron line. These factorisable contributions will not modify
the structure of the matrix element, but will change the form factors $F^{00}$
and $R^{00}$. We find it convenient to express them as%
\footnote{In the $K^{+-}_{e4}$ case, there are additional structures in ${\cal A}_{\rm fact}(K^{+-}_{e4} )$,
due to the possibility, already at tree level, for a virtual photon to emit a pair of 
charged pions, see \cite{Vesna04,Stoffer:2013sfa}. Notice in this respect that the contributions
in Fig. 6 and in Fig. 8 of Ref. \cite{Stoffer:2013sfa} vanish in our case.}
\begin{equation}
{\cal A}_{\rm fact}(K^{00}_{e4} ) \equiv
i \frac{G_{\rm F}}{\sqrt{2}}  V_{us}
{\bar{\rm u}} (p_e) \left( \not \! p_1 + \not \! p_2 \right) (1-\gamma_5) {\rm v}(p_\nu) 
\times
\frac{F^{00}}{M_K}
\times
\sqrt{Z_e} \sqrt{Z_K}
+ {\cal O}(m_e) 
,
\lbl{fact_rad_corr}
\end{equation}
where we have factored out the wave-function renormalization
factors computed in QED for $Z_e$, and in scalar QED for $Z_K$: 
\begin{eqnarray}
\sqrt{Z_e} &=& 1 + e^2 \left[
{\bar\lambda} - \frac{1}{(4\pi)^2} \left(
\frac{3}{2} - \frac{1}{2} \ln \frac{m_e^2}{\mu^2} - \ln \frac{m_e^2}{m_\gamma^2}
\right)
\right]
\nonumber\\
\sqrt{Z_K} &=& 1 + e^2 \left[
- 2 {\bar\lambda} - \frac{1}{(4\pi)^2} \left( 1 + \ln \frac{M_K^2}{\mu^2} - \ln \frac{M_K^2}{m_\gamma^2}
\right)
\right]
.
\end{eqnarray}
In contrast to the preceding subsection, we use now dimensional regularization,
with the minimal subtraction of the combination
\begin{equation}
{\bar\lambda} = \frac{1}{16 \pi^2} \left[
 \frac{1}{d-4} - \frac{1}{2}
\left( \ln ( 4 \pi ) + \Gamma' (1) + 1 \right) 
\right]
.
\end{equation}
It is understood that $F^{00}$ in Eq. \rf{fact_rad_corr} now includes all
the remaining factorizable photonic corrections, together with the contributions
from the low-energy constants $L_i$ \cite{Gasser:1984gg} and $K_i$ \cite{Urech:1994hd}.
These will take care of the UV divergences due to the meson loops and to 
the photon loops, respectively, so that the product $F^{00}  \sqrt{Z_K}$
is actually UV finite. It however inherits the infrared divergence contained
in $\sqrt{Z_K}$.

\begin{figure}[t]
\begin{center}
\includegraphics[width=14cm]{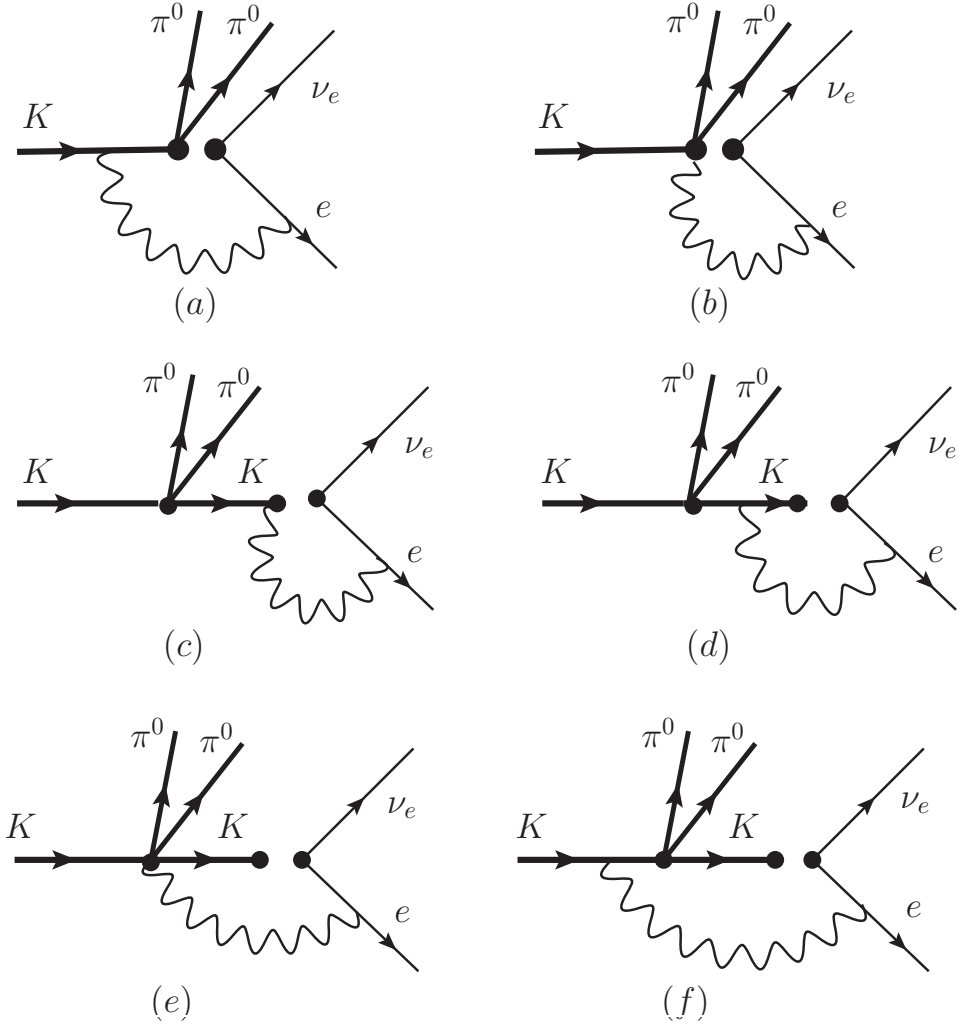}
\end{center}
\caption{The set of one-loop virtual photon exchange diagrams 
of the non-factorisable type to consider for the $K^{00}_{e4}$ decay. 
\label{fig:virtual}}
\end{figure}

\begin{figure}[ht]
\begin{center}
\includegraphics[width=15cm]{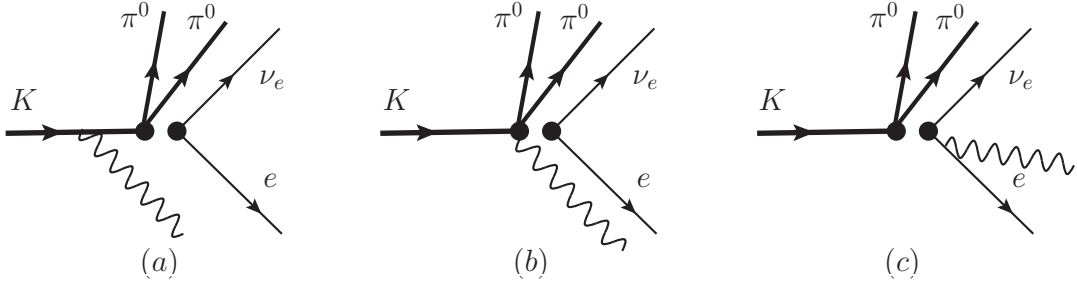}
\end{center}
\caption{The one-photon emission contributions to the $K^{00}_{e4\gamma}$ decay.
\label{fig:emission}}
\end{figure}

Let us next consider the non-factorisable contributions. As far as
the corrections to ${F}^{00}$ are concerned, they are
represented by the diagrams 
shown in Fig. \ref{fig:virtual}. On finds that the contributions coming from the
diagrams $(b)$ and $(c)$ are proportional to the lepton mass, and thus
vanish in the limit $m_e \to 0$.
There are therefore only three diagrams to compute in this approximation.
Consistently dropping terms that vanish as $m_e \to 0$, one finds that
\begin{eqnarray}
{\cal A}(K^{00}_{e4} ; \ref{fig:virtual}(a)) &=& {\cal A}_0(K^{00}_{e4} )
\times e^2
\left[
4 ( k \cdot p_e ) C\left( (k - p_e)^2 ; M_K , m_e \right)
- 2  ( k \cdot p_e )  C_{12} \left( (k - p_e)^2 ; M_K , m_e \right)
\right.
\nonumber\\
&&
\left.
-  {\bar J} \left( (k - p_e)^2 ; M_K , m_e \right) 
- 4 {\bar \lambda} 
- \frac{1}{16 \pi^2} \left(  \ln \frac{M_K^2}{\mu^2} +  \ln \frac{m_e^2}{\mu^2} - \frac{10}{3} \right)
\right]
,
\lbl{diag:a}
\end{eqnarray}
\begin{equation}
{\cal A}(K^{00}_{e4} ; \ref{fig:virtual}(e)) = {\cal A}_0(K^{00}_{e4}) \times \frac{e^2}{3}
\left[
{\bar J} \left( (k-p_1-p_2)^2, m_\gamma , M_K \right) - 2 {\bar \lambda} - \frac{1}{16 \pi^2} \left( \ln \frac{M_K^2}{\mu^2} - 1 \right) 
\right]
,
\lbl{diag:e}
\end{equation}
and
\begin{eqnarray}
{\cal A}(K^{00}_{e4} ; \ref{fig:virtual}(f)) &=&
{\cal A}_0(K^{00}_{e4}) \times \frac{e^2}{2}
\left[
2 (p_1 + p_2 )^2 C \left( (p_1 + p_2 )^2 , (k - p_1 - p_2 )^2 ; M_K , M_K \right)
\right.
\nonumber\\
&&
\left.
- (p_1 + p_2 )^2 C_{11} \left( (p_1 + p_2 )^2 , (k - p_1 - p_2 )^2 ; M_K , M_K \right)
\right.
\nonumber\\
&&
\left.
- \frac{2}{3} {\bar J} \left( (k - p_1 - p_2 )^2 ; m_\gamma, M_K \right)
+ \frac{7}{3} {\bar\lambda} + \frac{1}{16 \pi^2} \left( \frac{7}{6} \ln \frac{M_K^2}{\mu^2}
- \frac{2}{3} \right)
\right]
.
\lbl{diag:f}
\end{eqnarray}
Apart from the change of regularization, the expression for ${\cal A}(K^{00}_{e4} ; \ref{fig:virtual}(a))$
in Eq. \rf{diag:a} reproduces the one of Eq. \rf{A_diag_a} obtained previously, provided one
takes $R^{00} = (2/3) \cdot F^{00}$, as discussed at the end of Sec. \ref{photos_neutral},
and makes use of the identities
\begin{eqnarray}
( k \cdot p_e) C_{11} \left( (k - p_e)^2 ;  M_K , m_e \right) &=&
- \frac{1}{32 \pi^2} + \frac{1}{2} {\bar J} \left( (k-p_e)^2 ; M_K , m_e \right) + \ldots
,
\nonumber\\
M_K^2 C_{11} \left( (k - p_e)^2 ;  M_K , m_e \right) &=&
- \frac{1}{32 \pi^2} \left( 1 + \ln \frac{M_K^2}{m_e^2} \right)
+ \frac{1}{2} {\bar J} \left( (k-p_e)^2 ; M_K , m_e \right) 
\nonumber\\
&&
- ( k \cdot p_e) C_{12} \left( (k - p_e)^2 ;  M_K , m_e \right) + \ldots
,
\end{eqnarray}
where the ellipses denote terms that vanish in the limit $m_e \to 0$.

Adding up the contributions discussed so far, one obtains an expression for the radiative corrections
at order ${\cal O}(\alpha)$ that still contains both infrared and ultraviolet divergences. The latter
will be taken care of by the contributions from the counterterms $X_i$ introduced in \cite{Knecht:1999ag}.
Their contribution reads 
\begin{equation}
{\cal A}(K^{00}_{e4} ; {\rm cts}) =
{\cal A}_0(K^{00}_{e4}) \times e^2 \left(- \frac{8}{3} X_1 - \frac{1}{2} X_6\right)  
.
\lbl{cts}
\end{equation}
The low-energy constant $X_1$ is not renormalized, whereas
$X_6 = X_6^r (\mu) - 5 {\bar\lambda}$.
Collecting the divergent pieces from the various contributions
leads to [we recall that at this stage $F^{00} \sqrt{Z_K}$ has
already been made UV-finite through the contributions of the
low-energy constants $L_i$ and $K_i$]
\begin{equation}
{\cal A}(K^{00}_{e4} ; {\mbox{UV-div}}) =
e^2  {\cal A}_0(K^{00}_{e4}) \times  \frac{\bar\lambda}{6} \, 
\bigg[\underbrace{-\, 24\ }_{\ref{fig:virtual}(a)} \  \underbrace{-\ 4\ }_{\ref{fig:virtual}(e)}
 \  \underbrace{+\ 7\ }_{\ref{fig:virtual}(f)}
 \  \underbrace{+\, 6\ }_{\sqrt{Z_e}} \ \underbrace{+ \, 15\ }_{X_6}  \bigg]
 \,=\, 0
,
\end{equation}
which vanishes as it should. 

As to the infrared divergences, collecting the
IR-divergent pieces contained in the contributions computed so far,
one obtains
\begin{eqnarray}
{\cal A}(K^{00}_{e4} ; {\mbox{IR-div}}) &=&
\frac{e^2}{(4 \pi)^2} \, {\cal A}_0(K^{00}_{e4}) \times \ln m_\gamma
\bigg[
\!
\underbrace{(-2)}_{\sqrt{Z_e}} \,+\, \underbrace{(-2)}_{\sqrt{Z_K}} \,+\, 
\underbrace{(-2) (k \cdot p_e) \,\tau (k , p_e)}_{\ref{fig:virtual}(a)}
\!\bigg]
,
\qquad~
\end{eqnarray}
with the function $\tau ( p_1 , p_2)$ defined in Eq. \rf{tau_function}.
Besides the wave-function renormalisations, such divergences only arise from the contribution
of $C\left( (k - p_\ell)^2 ; M_K , m_e \right)$ in ${\cal A}(K^{00}_{e4} ; 1(a))$.
Notice that this infrared divergence coincides with the one of the result
\rf{photos}, given in Eq. \rf{photos-IR}.
The construction of an infrared-safe observable at order ${\cal O}(\alpha)$
requires also to consider the process with the emission of one soft photon. 
The corresponding differential decay rate is given by

\begin{eqnarray}
d \Gamma ({K^{00}_{e4\gamma}}) &=&
\frac{1}{2 M_K} \, \frac{1}{2} \sum_{\rm spins,\, pol.}
\left\vert
{\cal A}(K^{00}_{e4\gamma})
\right\vert^2
\times
(2 \pi)^4 \delta^4 ( k - p_1 - p_2 - p_{\ell} - p_\nu - q)
\nonumber\\
&&
\times
\frac{d^3 {\vec p}_1}{(2\pi)^3 2 E_1} \, \frac{d^3 {\vec p}_2}{(2\pi)^3 2 E_2} \,
\frac{d^3 {\vec p}_\ell}{(2\pi)^3 2 E_\ell} \, \frac{d^3 {\vec p}_\nu}{(2\pi)^3 2 \vert {\vec p}_\nu \vert} \,
\frac{d^3 {\vec q}}{(2\pi)^3 2 \vert {\vec q} \vert}
.
\end{eqnarray}
The amplitude ${\cal A}(K^{00}_{e4\gamma})$ for the radiative decay 
$K^\pm \to \pi^0 \pi^0 e^\pm \stackrel{_{(-)}}{\nu_e} \gamma$ can be
expanded in powers of the photon energy,
\begin{equation}
{\cal A}(K^{00}_{e4\gamma}) = {\cal A}_{-1}(K^{00}_{e4\gamma}) + {\cal A}_0(K^{00}_{e4\gamma}) + \ldots
\end{equation}
The Low approximation consists in keeping ${\cal A}_{-1}(K^{00}_{e4\gamma})$ alone. This is enough
in order to study the emission of only soft photons and
to discuss the issue of infrared divergences. Explicitly, one has
[$q_\mu$ is the momentum of the emitted (real) photon, $\varepsilon^* (q)$
the corresponding polarization vector]
\begin{equation}
{\cal A}_{-1}(K^{00}_{e4\gamma}) = e {\cal A}_0(K^{00}_{e4}) 
\left(
\frac{p_{\ell} \cdot \varepsilon^* (q)}{p_{\ell} \cdot q + \frac{m_\gamma^2}{2}}
\,-\,
\frac{k \cdot \varepsilon^* (q)}{k \cdot q - \frac{m_\gamma^2}{2}}
\right)
.
\end{equation}
Then
\begin{eqnarray}
\sum_{\rm spins,\, pol.}
\left\vert
{\cal A}_{-1}(K^{00}_{e4\gamma})
\right\vert^2 &=&
- e^2 \! \sum_{\rm spins}
\left\vert
{\cal A}(K^{00}_{e4})
\right\vert^2 \times
\left[
\frac{ m_\ell^2}{\left( p_{\ell} \cdot q + \frac{m_\gamma^2}{2} \right)^2}
+
\frac{M_K^2}{\left( k \cdot q - \frac{m_\gamma^2}{2} \right)^2}
\qquad { }
\right.
\nonumber\\
&&
\left.
\qquad
- 2\,
\frac{k \cdot p_{\ell}}{\left( p_{\ell} \cdot q + \frac{m_\gamma^2}{2} \right) %
\left( k \cdot q - \frac{m_\gamma^2}{2} \right)}
\right]
.
\end{eqnarray}
One may then perform the integration over the
undetected soft photon. In the soft-photon approximation, the photon
momentum in the delta-function of the phase-space integration is
neglected, and one takes
\begin{eqnarray}
d \Gamma^{\rm soft} ({K^{00}_{e4\gamma}}) &=&
(2 \pi)^4 \delta^4 ( k - p_1 - p_2 - p_{\ell} - p_\nu )
\times
\frac{d^3 {\vec p}_1}{(2\pi)^3 2 E_1} \, \frac{d^3 {\vec p}_2}{(2\pi)^3 2 E_2} \,
\frac{d^3 {\vec p}_\ell}{(2\pi)^3 2 E_\ell} \, \frac{d^3 {\vec p}_\nu}{(2\pi)^3 2 \vert {\vec p}_\nu \vert} 
\nonumber\\
&&
\times
\frac{1}{2 M_K} \, \frac{1}{2}
\int_{\vert{\vec q}\vert \le \Delta E}
\frac{d^3 {\vec q}}{(2\pi)^3 2 \vert {\vec q} \vert} \sum_{\rm spins,\, pol.}
\left\vert
{\cal A}_{-1}(K^{00}_{e4\gamma})
\right\vert^2
\nonumber\\
&=&
d \Gamma_0 ({K^{00}_{e4}})
\times
(- e^2)
\int_{\vert{\vec q}\vert \le \Delta E}
\frac{d^3 {\vec q}}{(2\pi)^3 2 \vert {\vec q} \vert} 
\left[
\frac{ m_\ell^2}{\left( p_{\ell} \cdot q + \frac{m_\gamma^2}{2} \right)^2}
+
\frac{M_K^2}{\left( k \cdot q - \frac{m_\gamma^2}{2} \right)^2}
\qquad { }
\right.
\nonumber\\
&&
\left.
\qquad
- 2\,
\frac{k \cdot p_{\ell}}{\left( p_{\ell} \cdot q + \frac{m_\gamma^2}{2} \right) %
\left( k \cdot q - \frac{m_\gamma^2}{2} \right)}
\right]
.
\lbl{Ke400gamma}
\end{eqnarray}
Expressions for the corresponding integrals can be found in \cite{Vesna04,Stoffer:2013sfa}.
The integral is limited to photon energies $\vert{\vec q}\vert$ below the experimental 
detection threshold $\Delta E$ in the kaon rest-frame.
As far as the infrared divergences are concerned, one has
\begin{equation}
d \Gamma^{\rm soft}_{\mbox{\tiny{IR-div}}} ({K^{00}_{e4\gamma}}) =
d \Gamma_0 ({K^{00}_{e4}})
\times
\frac{e^2}{8 \pi^2}
\times
\ln m_\gamma
\left[2 + 2 
+ 2 (k \cdot p_e) \tau(k , p_e )
\right]
.
\end{equation}
Therefore, the contributions proportional to $\ln m_\gamma$
cancel in the sum $d \Gamma ({K^{00}_{e4}}) + d \Gamma ({K^{00}_{e4\gamma}})$.
For later convenience, we rewrite Eq. \rf{Ke400gamma} in a way that
explicitly displays the IR-singular part:
\begin{eqnarray}
d \Gamma^{\rm soft} ({K^{00}_{e4\gamma}}) &=& d \Gamma_0 ({K^{00}_{e4}})
\times
\frac{e^2}{8 \pi^2} \left[ 2 + (k \cdot p_e) \tau(k , p_e ) \right] \times 2 \ln \frac{m_\gamma}{2 \Delta E}
+
d {\bar\Gamma}^{\rm soft} ({K^{00}_{e4\gamma}})
.
\lbl{Ke400gamma_bar}
\end{eqnarray}
We then add the contribution
\begin{equation}
{\cal A}_0(K^{00}_{e4}) \times 
\frac{e^2}{8 \pi^2} \left[ 2 + (k \cdot p_e) \tau(k , p_e ) \right] \times 2 \ln \frac{m_\gamma}{2 \Delta E}
\end{equation}
to the amplitudes involving virtual photons, such as to make them infrared finite.
To this end, we define the function
\begin{equation}
{\bar C} \left( (k-p_e)^2 ; M_K , m_e ; \Delta E\right)
=
C \left( (k-p_e)^2 ; M_K , m_e \right) 
+ \frac{1}{32 \pi^2} \, \tau(k , p_e ) \times \ln \frac{m_\gamma}{2 \Delta E}
.
\end{equation}

\subsection{Discussion on radiative corrections for $K_{e4}^{00}$}

We can now add the virtual and real contributions described up to now which should be involved
in a PHOTOS-like treatment of this decay. We include them as a correction of the form 
$(1+\delta_{\rm EM})$ to the determination of the form factor from the measurement of the branching ratio. 
To this end, we compute the total decay rate including the soft photon emission
\begin{equation}
 \Gamma^{\rm tot} = \Gamma ( K_{e4}^{00} ) + {\bar \Gamma}^{\rm soft} ({K^{00}_{e4\gamma}})
 ,
\end{equation}
where $\Gamma ( K_{e4}^{00} )$ includes corrections at first order in the fine-structure constant $\alpha$,
and write it in terms of the decay rate  $\Gamma_0 ( K_{e4}^{00} )$ without radiative corrections
in the form
\begin{equation}
 \Gamma^{\rm tot} = \Gamma_0 ( K_{e4}^{00} ) \times \left( 1  + 2 \delta_{EM} \right)
.
\end{equation}

Let us first discuss the corrections computed in Subsection B above. In order
to obtain a result that is as close as possible to the treatment
of radiative corrections in the $K_{e4}^{+-}$ channel, we absorb the
UV-divergent factor of Eq. \rf{photos} in $F^{00}$ and take $R^{00}$ equal
to zero. Then $\Gamma ( K_{e4}^{00} )$ is computed by performing the phase-space
integration of
\begin{equation}
{\cal A}_0(K^{00}_{e4}) \left[ {\cal A}_0(K^{00}_{e4}) + 2 e^2  \Delta {\cal A}(K^{00}_{e4}) \right]
,
\end{equation}
where
\begin{eqnarray}
\Delta {\cal A}(K^{00}_{e4}) & = & {\cal A}_0(K^{00}_{e4})
\left[ \frac{1}{(4 \pi)^2} \left(
- \frac{1}{2} \ln \frac{m_e^2}{M_\pi^2} + 4 \ln \frac{m_e}{2 \Delta E} - \ln \frac{M_K^2}{M_\pi^2} - 2 
\right)
+ 4 (k\cdot p_e ) {\bar C} \left( (k-p_e)^2 ; M_K , m_e ; \Delta E\right)
 \right. 
 \nonumber\\
 &&
 \left. 
 \qquad\quad
 - \, 2 M_K^2 C_{11} \left( (k - p_e)^2 ;  M_K , m_e \right)
 - 4  (k \cdot p_e) C_{12} \left( (k - p_e)^2 ;  M_K , m_e \right)
\right]
.
\lbl{Delta_A}
\end{eqnarray}
We take $\Delta E = 11.7$ MeV, the value corresponding to the NA48/2 experiment \cite{Batley:2014xwa},
for the real-photon detection threshold in the kaon rest frame. This gives then
\begin{equation}
\delta_{EM} = 0.018
.
\lbl{delta_photos}
\end{equation}
This value has the expected size. Moreover, it goes into the right direction, 
in the sense that it reduces the discrepancy in Eq. \rf{discrepancy} from $6.5\%$ to $4.6\%$.

As a test of the stability of the result \rf{delta_photos} we may also evaluate the  non-factorizable
radiative corrections corresponding to all the diagrams in Fig. \ref{fig:virtual}. This 
amounts to taking the expressions in Eqs. \rf{diag:a}, \rf{diag:e}, and \rf{diag:f}
for the evaluation of $\Gamma ( K_{e4}^{00} )$ (let us stress again
that these equations have been obtained in a regularisation scheme differing from 
the one discussed in Sec.~\ref{photos_neutral}). We absorb
the ultraviolet divergences, as well as the contribution \rf{cts}
into $F^{00}$, in order to build a quantity both UV and IR finite.
Constant terms have been discarded as they could be included in the contribution 
of the counterterms $X_1$ and $X_6$.
The resulting expression for $\Delta {\cal A}(K^{00}_{e4})$ then reads
\begin{eqnarray}
\Delta {\cal A}(K^{00}_{e4}) & = & {\cal A}_0(K^{00}_{e4})
\left[ \frac{1}{(4 \pi)^2} \left( - \frac{7}{4} \ln \frac{M_K^2}{M_\pi^2} - \frac{1}{2} \ln \frac{m_e^2}{M_\pi^2}
+ 2 \ln \frac{m_e}{2 \Delta E} + 2 \ln \frac{M_K}{2 \Delta E}
\right)
 \right. 
 \nonumber\\
 &&
 \left. 
 \qquad\quad 
+ 4 (k\cdot p_e ) {\bar C} \left( (k-p_e)^2 ; M_K , m_e ; \Delta E\right)
 \right. 
 \nonumber\\
 &&
 \left. 
 \qquad\quad 
 - 2  (k \cdot p_e) C_{12} \left( (k - p_e)^2 ;  M_K , m_e \right)
 - \frac{2}{3} {\bar J} \left( (k - p_e)^2 ;  M_K , m_e \right)
 \right. 
 \nonumber\\
 &&
 \left. 
 \qquad\quad 
 + (p_1 + p_2 )^2 C \left( (p_1 + p_2 )^2 , (k - p_1 - p_2 )^2 ; M_K , M_K \right)
 \right. 
 \nonumber\\
 &&
 \left. 
 \qquad\quad 
- \frac{1}{2} (p_1 + p_2 )^2 C_{11} \left( (p_1 + p_2 )^2 , (k - p_1 - p_2 )^2 ; M_K , M_K \right)
 \right. 
 \nonumber\\
 &&
 \left. 
 \qquad\quad 
- \frac{1}{3} {\bar J} \left( (k - p_1 - p_2 )^2 ; m_\gamma, M_K \right)
\right]
,
\end{eqnarray}
instead of the expression in Eq. \rf{Delta_A}. For $\Delta E = 11.7$ MeV, we obtain now
\begin{equation}
\delta_{EM} = 0.017
.
\lbl{delta_full}
\end{equation}
This value is quite close to the one obtained in Eq. \rf{delta_photos},
so that in the present case the treatment of radiative corrections
{\it \`a la} PHOTOS seems to yield stable results even after the inclusion 
of non-factorizable contributions.

\section{Summary and conclusion}

The present study is devoted to isospin-breaking effects in the semileptonic
decay of the charged kaon into two neutral pions, $K^\pm \to \pi^0 \pi^0 e^\pm \stackrel{_{(-)}}{\nu_e}$.
Because of the smallness of the electron mass and of the limited
experimental precision, this decay can be described in terms of a single form factor.
This form factor also occurs in the description of the decay into two charged
pions, $K^\pm \to \pi^+ \pi^- e^\pm \stackrel{_{(-)}}{\nu_e}$, and up
to isospin-breaking contributions, the two determinations should agree. 
The present study focuses mainly on two aspects related to this issue: 
i) to ascertain quantitatively to which extend the phenomenological
parameterizations used in order to analyse the data could impinge on the resulting
value of the normalization or on the shape of the form factor measured in
the decay $K^\pm \to \pi^0 \pi^0 e^\pm \stackrel{_{(-)}}{\nu_e}$, and ii) to obtain
a quantitative estimate of the radiative corrections to the total decay rate, 
which again might affect the normalization of the form factor.

Concerning the first issue, we have considered the form factors of the
pion as a case study. As a first step, we have discussed the structure
of the form factors, and their properties linked to the presence of a cusp,
using exact expressions of the form factors valid up to two loops in
the low-energy expansion. We have clearly established that the phenomenological
parameterizations used in order to analyse the data did not agree with the
general properties that can be infered form these exact expressions.
In a second step, we have generated pseudo-data from these
form factors, that we have then analysed
using several phenomenological parameterizations. The outcome 
of this study is that the determination of the normalization
of the form factor is actually not sensitive to the parameterizations
used. As a side product, we see that the higher orders in the Taylor expansion
of form factors are not accurately determined by a direct fit to simplified (polynomial) 
formulae. Although our study was carried out for the scalar 
form factor of the neutral pion, we expect that the conclusion also holds for
the $K_{e4}^{00}$ form factor.
We have also considered the possibility
to constrain the $\pi\pi$ $S$-wave scattering lengths from the measurement
of the decay distribution. We have found that, unfortunately, with the sample of events presently 
available, the statistical uncertainties remain large. A statistical sample
comparable to the one available in the $K_{e4}^{+-}$ channel would be required
in order to reach a precision close to that obtained by the Dirac experiment.

The second issue addressed in this paper consists in radiative corrections. We have determined
the correction factor $\delta_{EM}$ to the total decay rate in Eq. \rf{discrepancy}. 
In order to make a meaningful comparison with the value of the normalization of the 
form factor extracted from the $K_{e4}^{+-}$ channel,
we have used a simplified framework, including only those corrections that
were also included in the latter case (one-loop photonic corrections
on the wave functions and tree-level vertex). Our result $\delta_{EM} = 0.018$
leads to the replacement of Eq. \rf{discrepancy} by
\begin{equation}
\frac{f_s [K^{00}_{e4}]}{ f_s [K^{+-}_{e4}]}
= 1.046(10) 
.
\lbl{discrepancy_corrected}
\end{equation}
Note that the error bar in this equation is purely from experimental origin, and does not 
include the systematic uncertainties from the methods used for the evaluation of radiative corrections
in both channels. Such additional uncertainties can stem, for instance, from the
regularisation dependence of the PHOTOS(-like) treatment of radiative corrections, and
from neglecting the  dependence in the cut off $\Lambda$ discussed in Sec.~\ref{photos_charged}.

We have also considered additional photonic corrections  estimated within a different regularisation
scheme, and we have found that they do not modify the previous estimate in a significant way.
A few comments are in order:
\begin{itemize}

\item
The analysis of radiative corrections we have performed
provides an adequate estimate of the global factor $\delta_{EM}$
that modifies the total decay rate. It need not be suitable for
an analysis of radiative corrections to the phase-space distribution
itself.

\item
Other isospin-breaking corrections, among them factorizable exchanges of virtual photons,
but also effects due to $m_u - m_d$ or to the mass differences between charged
and neutral pions and/or kaons, are not covered by our analysis.
They could affect the normalization of the
form factors measured in the two channels in different ways. A mode elaborate
study is needed in order to reach a quantitatively meaningful interpretation 
of the result in Eq. \rf{discrepancy_corrected}.

\item
At lowest order in the chiral expansion, these additional
isospin-breaking corrections are given by Eq. \rf{R-corr}. For $R = 35.8(1.9)(1.8)$ \cite{Aoki:2013ldr}, 
and adding errors in quadrature, we obtain 
\begin{equation}
\left. \frac{f_s [K^{00}_{e4}] }{ f_s [K^{+-}_{e4}]} \right\vert_{\rm LO} = 1.042(3)
.
\end{equation}
In view of the value given in Eq. \rf{discrepancy_corrected}, the corrections
from higher orders to this ratio should therefore be small.

\item
Conversely, using the relation \rf{R-corr} in regard to the result \rf{discrepancy_corrected},
and discarding yet to be computed corrections to the former, we obtain $R = 32^{+9}_{-6}$. 

\end{itemize}

The discussion of the radiative corrections presented here is clearly only a first
step. In view of the statistical accuracy of the data, a full model-independent calculation of these 
corrections in the neutral as well as in the charged channels is certainly mandatory before a 
definite conclusion can be reached concerning the observed difference in the normalisation of the 
form factors between the neutral and the charged channels.
This task is clearly beyond the scope of the present note, and is left for future work.

\section*{Acknowledgments}

\noindent
We thank B. Bloch-Devaux from the NA48/2 Collaboration for informative discussions,
and for insightful remarks on the manuscript. 
This work is supported in part by the EU Integrated Infrastructure Initiative HadronPhysics3.

\appendix

\renewcommand{\theequation}{\Alph{section}.\arabic{equation}}

\section{Properties of the functions ${\bar K}_n^\alpha (s)$}\label{app:Kbar_functions}

In this Appendix, we wish to summarise the properties of the functions 
${\bar K}_n^\alpha (s)$ that are needed in the discussion of the cusp
in Section \ref{sec:theory}. Let us start with the functions ${\bar K}_n^0 (s)$ and ${\bar K}_n^{{\mbox{\tiny$\nabla $}}} (s)$,
defined by dispersive integral as in Eq. \rf{disp_rep_K}, with $s_{\rm thr} = 4 M_{\pi^0}^2$
and \cite{KMSF95}
\begin{eqnarray}
&& k_0^0 (s) \,=\, \frac{1}{16\pi}\,\sigma_0 (s),
\qquad
k_1^0 (s) \,=\, \frac{1}{8\pi}\,L_0(s)\,,
\nonumber\\
&& k_2^0 (s) \,=\,  \frac{1}{8\pi}\,\left(1\,-\,\frac{4 M_{\pi^0}^2}{s}\right)\,L_0(s)\,,
\qquad
k_3^0 (s) \,=\, \frac{3}{16\pi}\,\frac{M_{\pi^0}^2}{s \sigma_0 (s)}\,L_0^2(s)\,,
\nonumber\\
&& 
k_1^{{\mbox{\tiny$\nabla$}}} (s) \,=\, \frac{1}{8\pi}\,\frac{\sigma_0 (s)}{\sigma (s + 4 \Delta_\pi)} \, L (s + 4 \Delta_\pi)\,,
\qquad
k_3^{{\mbox{\tiny$\nabla$}}} (s) \,=\, \frac{3}{16\pi}\,\frac{M_{\pi}^2}{s \sigma_0 (s)}\,L^2 (s + 4 \Delta_\pi)\,,
\qquad
.
\lbl{k_n}
\end{eqnarray}
The definitions of the functions $\sigma_0 (s)$ and $L_0 (s)$ for $s \ge 4 M_{\pi^0}^2$
can be found in Eqs. \rf{def_sigma} and \rf{J0_L0_def}, respectively. The function 
$\sigma (s)$ is also to be found in Eq. \rf{def_sigma}, whereas $L(s)$ is defined as
\begin{eqnarray}
L (s) &=& 
\left\{
\begin{array}{l}
\ln \left( \frac{1 - \sigma (s)}{1 + \sigma (s)} \right) \equiv {\hat L} (s) \ \quad\ \, [s \ge 4 M_{\pi}^2 ]\\
\\
\ln \left( \frac{\sigma (s) - 1}{\sigma (s) + 1} \right) \equiv {\hat L} + i \pi \quad [ 4 M_{\pi^0}^2 \le s\le 4 M_{\pi}^2] \\
\end{array}
\right.
,
\lbl{def_L}
\end{eqnarray} 
according to the definitions \rf{L_hat} and \rf{sigma_def_ext}. 
For $s \ge 4 M_{\pi^0}^2$,
the functions $k_n^0 (s)$ and $k_n^{{\mbox{\tiny$\nabla $}}} (s)$ are real and smooth. 
In the same range of $s$, the functions ${\bar K}_n^0 (s)$ and ${\bar K}_n^{{\mbox{\tiny$\nabla $}}} (s)$
have smooth real and imaginary parts, with ${\rm Im} \, {\bar K}_n^0 (s) = k_n^0 (s)$,
${\rm Im} \, {\bar K}_n^{{\mbox{\tiny$\nabla $}}} (s) = k_n^{{\mbox{\tiny$\nabla $}}} (s)$.
Finally, the functions ${\bar K}_n^0 (s)$ for $ n \ge 1$ can be expressed in terms
of ${\bar J}_0 (s) \equiv {\bar K}_0^0 (s)$. The explicit expressions and their
derivation were given in \cite{KMSF95}.

There is not much to add as far as the functions ${\bar K}_n (s)$ are
concerned: it is sufficient to replace everywhere in Eq. \rf{k_n} $M_{\pi^0}$
by the charged pion mass $M_\pi$, and hence $\sigma_0 (s)$ by $\sigma (s)$,
and $L_0 (s)$ by $L (s)$. In the dispersive representation  \rf{disp_rep_K},
the integration starts at $s_{\rm thr} = 4 M_\pi^2$. In the case of 
${\bar K}_0 (s) \equiv {\bar J} (s)$, the decomposition \rf{Jbar_decomp_1}
then follows from \rf{Kbar_decomp_1} by noticing that $k_0 (s)/\sigma (s)$
is a constant, and that \cite{KMSF95,DescotesGenon:2012gv}
\begin{equation}
{\rm Re} \, {\bar J} (s) = \frac{1}{16 \pi^2} \left[ 2 + \sigma (s) L(s) \right]
\quad\ [ s \ge 0]
.
\end{equation}
For the remaining functions ${\bar K}_n (s)$, it is most convenient to use their
expressions in terms of ${\bar J} (s)$. One then finds
\begin{eqnarray}
16\pi^2\bar{K}_1^{[0]}=\hat{L}(s)^2-\pi^2 &\qquad & 16\pi^2\bar{K}_1^{[1]} = -2\pi \frac{\hat{L}(s)}{\sigma(s)}
\nonumber\\
16\pi^2\bar{K}_2^{[0]}=\sigma^2(\hat{L}^2(s)-\pi^2)-4 &\qquad & 
    16\pi^2\bar{K}_2^{[1]} = -2\pi \hat{L}(s)\sigma(s)
\nonumber\\
16\pi^2\bar{K}_3^{[0]}=\hat{L}(s)(\hat{L}^2(s)-2\pi^2)\frac{M_\pi^2}{s\sigma(s)}-\frac{\pi^2}{2} &\qquad & 
    16\pi^2\bar{K}_3^{[1]} = -3\pi \frac{M_\pi^2}{s} \frac{\hat{L}^2(s)}{\sigma^2(s)}
.
\end{eqnarray}
One may check that all these functions are real and smooth for $s \ge 4 M_{\pi^0}^2$
[actually, for $ s \ge 0$].

The functions ${\bar K}_n^x (s)$, $n = 1,3$, are defined by
\begin{equation}
{\bar K}_n^x (s) = \frac{s}{\pi} \int_{4 M_\pi^2}^\infty \frac{d x}{x} \frac{k_n^x (x)}{x - s - i0}
,
\lbl{K1x_disp}
\end{equation}
with [the quantities appearing in these formulae are defined in Ref. \cite{DescotesGenon:2012gv},
see Eqs. (2.13), (4.13), (4.16), and (4.17) therein] 
\begin{eqnarray}
k_1^x (s) &=& \frac{1}{8 \pi} \, \frac{1}{s \sigma_0 (s)}
\left[
\lambda^{1/2}(t_{\mbox{\tiny $-$}}(s)) {\cal L}_{\mbox{\tiny $-$}} (s) - 
\lambda^{1/2}(t_{\mbox{\tiny $+$}}(s)) {\cal L}_{\mbox{\tiny $+$}} (s)
\right]
,
\nonumber\\
k_2^x (s) &=& \frac{1}{8\pi} \sigma (s) \sigma_0 (s) L_0 (s)
,
\nonumber\\
k_3^x (s) &=& \frac{3}{16 \pi} \, \frac{M_\pi^2}{s \sigma_0 (s)}
\left[
{\cal L}_{\mbox{\tiny $-$}}^2 (s) - {\cal L}_{\mbox{\tiny $+$}}^2 (s)
\right]
.
\lbl{k1x_def}
\end{eqnarray}
The three functions ${k}_{n}^x (s)$, $n=1,2,3$, are real and smooth
for $s \ge M_{\pi}^2$, and they become purely imaginary for 
$4 M_{\pi^0}^2 \le s \le 4 M_\pi^2$. The functions ${\hat k}_{n}^x (s) \equiv {k}_{n}^x (s)/\sigma (s)$
are then smooth in the range $4 M_{\pi^0}^2 \le s\le M_K^2$, so that one obtains
\begin{equation}
{\bar K}_{n}^{x[1]} (s) = - \frac{k_{n}^x (s)}{\sigma (s)}
,
\end{equation}
and
\begin{eqnarray}
{\bar K}_{1,3}^{x[0]} (s) \ =\  
\left\{
\begin{array}{l}
{\rm Re}\,{\bar K}_{1,3}^x (s)  \qquad\qquad\qquad\quad  [s\ge 4 M_{\pi}^2] \\
\\
{\rm Re}\,{\bar K}_{1,3}^x (s) + {\displaystyle \frac{{\hat \sigma} (s)}{\sigma (s)} } 
k_{1,3}^x (s) \quad  [4 M_{\pi^0}^2 \le s \le 4 M_\pi^2]
\end{array}
\right.
.
\end{eqnarray}
Since analytical expressions for ${\rm Re}\,{\bar K}_{1,3}^x (s)$
are not available, one has to use the integral representation
given in Eq. \rf{K1x_disp} for numerical applications.

Finally, there remains to discuss the function ${\cal K}^x (s)$
whose discontinuity along the real $s$ axis for $s \ge 4 M_\pi^2$
reads
\begin{equation}
k^x (s) =
\frac{1}{16 \pi} \frac{M_\pi^2}{\Delta_\pi}\frac{\sigma(s)}{\sigma_0(s)} \frac{1}{s}
\left[
\left(\sigma(s) - \sigma_0 (s) \right)
\lambda^{1/2}(t_{\mbox{\tiny $-$}}(s)) {\cal L}_{\mbox{\tiny $-$}} (s) -\,
\left( \sigma(s) + \sigma_0 (s)\right)
\lambda^{1/2}(t_{\mbox{\tiny $+$}}(s)) {\cal L}_{\mbox{\tiny $+$}} (s)
\right]
.
\lbl{calkx_def}
\end{equation}
The function $k^x(s)/\sigma (s)$ is real and smooth for $ s \ge 4 M_{\pi^0}^2$.
Thus one infers
\begin{equation}
{\bar{\cal K}}^{x[1]} (s) = - \frac{k^x (s)}{\sigma (s)}
,
\end{equation}
and
\begin{eqnarray}
{\bar{\cal K}}^{x[0]} (s) \ =\  
\left\{
\begin{array}{l}
{\rm Re}\,{\bar{\cal K}}^x (s)  \qquad\qquad\qquad\ \ \, [s\ge 4 M_{\pi}^2] \\
\\
{\rm Re}\,{\bar{\cal K}}^x (s) + {\displaystyle \frac{{\hat \sigma} (s)}{\sigma (s)} } k^x (s) 
\quad\  [4 M_{\pi^0}^2 \le s \le 4 M_\pi^2]
\end{array}
\right.
.
\end{eqnarray}

\section{Loop functions}\label{app:loop_functions}

The computation, in Section \ref{sec:rad_corr} of the diagrams describing
the virtual photon corrections involves
a certain number of loop functions, that are briefly discussed here,
in order to make the calculation in section \ref{sec:rad_corr} self-contained. 

At the level of the two-point one-loop diagrams, one has
\begin{equation}
J \left( p^2 ; m_1 , m_2 \right) =
\frac{1}{i} \int \frac{d^4 \ell}{(2\pi)^4}
\,
\frac{1}{(\ell^2 - m_1^2) [ (\ell - p)^2 - m_2^2] }
,
\end{equation}
and
\begin{eqnarray}
J_\mu \left( p ~; m_1 , m_2 \right) &=&
\frac{1}{i} \int \frac{d^4 \ell}{(2\pi)^4}
\,
\frac{\ell_\mu}{(\ell^2 - m_1^2) [ (\ell - p)^2 - m_2^2] }
\nonumber\\
&=&
\frac{p_\mu}{2 p^2} \left[
\left( p^2 + m_1^2 - m_2^2 \right) J \left( p^2 ; m_1 , m_2 \right)
+ i A(m_1^2) - i A(m_2^2) 
\right]
,
\end{eqnarray}
where $A(m^2)$ is related to the tadpole graph,
\begin{equation}
A(m^2) = \int \frac{d^4 \ell}{(2\pi)^4}
\,
\frac{1}{\ell^2 - m^2}
=
-i m^2 \left[ 2 {\bar \lambda} + \frac{1}{16 \pi^2} \ln \frac{m^2}{\mu^2} + {\cal O}(d-4) \right]
.
\end{equation}
Other useful relations are
\begin{equation}
J(p^2 ; m_1,m_2) = {\bar J} (p^2 ; m_1,m_2)
- 2 {\bar\lambda} - \frac{1}{16 \pi^2} 
\frac{m_1^2 \ln \frac{m_1^2}{\mu^2} - m_2^2 \ln \frac{m_2^2}{\mu^2} }{ m_1^2 - m_2^2} 
,
\end{equation}
and
\begin{equation}
J(m^2 ; 0,m) = - 2 {\bar\lambda} + \frac{1}{16 \pi^2} \left[ 1 - \ln \frac{m^2}{\mu^2} \right] 
.
\end{equation}
Explicit expressions of the function ${\bar J} (p^2 ; m_1,m_2)$ can be found in Ref. \cite{Gasser:1984gg}.
Moreover, the link with the functions ${\bar J} (s)$ and ${\bar J}_0 (s)$ encountered
in Section \ref{sec:theory} is given by ${\bar J} (s) \equiv {\bar J} (s ; M_\pi, M_\pi)$
and ${\bar J}_0 (s) \equiv {\bar J} (s ; M_{\pi^0}, M_{\pi^0})$. 

As far as the three-point one-loop functions are concerned, one has
\begin{equation}
C \left( (p_1 - p_2)^2 , p^2_2 ; m_1 , m_2 \right) =
\frac{1}{i} \int \frac{d^4 \ell}{(2\pi)^4}
\,
\frac{1}{(\ell^2 - 2 \ell \cdot p_1) [(\ell - p_2)^2 - m_2^2 ] (\ell^2 - m_\gamma^2)}
,
\end{equation}
and
\begin{eqnarray}
C_\mu \left( p_1 , p_2 ; m_1 , m_2 \right) &=&
\frac{1}{i} \int \frac{d^4 \ell}{(2\pi)^4}
\,
\frac{\ell_\mu}{(\ell^2 - 2 \ell \cdot p_1) [(\ell - p_2)^2 - m_2^2 ] (\ell^2 - m_\gamma^2)}
\nonumber\\
&=&
p_{1\mu} C_{11} \left( (p_1 - p_2)^2 , p_2^2 ; m_1 , m_2 \right) + p_{2\mu} C_{12} \left( (p_1 - p_2)^2 , p_2^2 ; m_1 , m_2 \right)
,
\qquad~
\end{eqnarray}
where $p_1^2 = m_1^2$. Explicitly, one has [$\lambda(x,y,z) = x^2 + y^2 + z^2 - 2 xy - 2 xz - 2 yz$]
\begin{eqnarray}
C_{11} \left( (p_1 - p_2)^2 , p_2^2 ; m_1 , m_2 \right) &=&
\frac{2}{\lambda ( (p_1 - p_2)^2 , m_1^2 , p_2^2 )}
\left\{
p_2^2 \left[ J(p_2^2 ; m_\gamma , m_2 ) - J \left( (p_1 - p_2)^2 ; m_1 , m_2 \right) \right]
\right.
\nonumber\\
&&
\left.
- (p_1 \cdot p_2) 
\left[ J(m_1^2 ; m_\gamma , m_1 ) - J \left( (p_1 - p_2)^2 ; m_1 , m_2 \right) \right]
\right.
\nonumber\\
&&
\left.
+ (p_1 \cdot p_2) (p_2^2 - m_2^2) C \left( (p_1 - p_2)^2 , p^2_2 ; m_1 , m_2 \right)
\right\}
\nonumber\\
&=&
\frac{2}{m_1^2 - m_2^2 } \, \frac{(p_1 \cdot p_2) m_2^2 - p_2^2 m_1^2}{\lambda ( (p_1 - p_2)^2 , m_1^2 , p_2^2 )}
\frac{1}{16 \pi^2} \ln \frac{m_2^2}{m_1^2}
\nonumber\\
&&
\!\!\!\!\!\!
+\,
{2} \, \frac{(p_1 \cdot p_2) - p_2^2}{\lambda ( (p_1 - p_2)^2 , m_1^2 , p_2^2 )}
\left[ {\bar J} \left( (p_1 - p_2)^2 ; m_1 , m_2 \right)
- \frac{1}{16 \pi^2} 
\right]
\nonumber\\
&&
\!\!\!\!\!\!
+\,
{2} \, \frac{p_2^2}{\lambda ( (p_1 - p_2)^2 , m_1^2 , p_2^2 )}
\left[ {\bar J} ( p_2^2 ; m_\gamma , m_2 )
- \frac{1}{16 \pi^2} 
\right]
\nonumber\\
&&
\!\!\!\!\!\!
+
2 \frac{(p_1 \cdot p_2) (p_2^2 - m_2^2) }{\lambda ( (p_1 - p_2)^2 , m_1^2 , p_2^2 )} \,
C \left( (p_1 - p_2)^2 , p^2_2 ; m_1 , m_2 \right)
,
\end{eqnarray}
\begin{eqnarray}
C_{12} \left( (p_1 - p_2)^2 , p_2^2 ; m_1 , m_2 \right) &=&
\frac{2}{\lambda ( (p_1 - p_2)^2 , m_1^2 , p_2^2 )}
\!
\left\{
m_1^2 \left[ J(m_1^2 ; m_\gamma , m_1 ) - J \left( (p_1 - p_2)^2 ; m_1 , m_2 \right) \right]
\right.
\nonumber\\
&&
\left.
- (p_1 \cdot p_2) 
\left[ J(p_2^2 ; m_\gamma , m_2 ) - J \left( (p_1 - p_2)^2 ; m_1 , m_2 \right) \right]
\right.
\nonumber\\
&&
\left.
- m_1^2 (p_2^2 - m_2^2) C \left( (p_1 - p_2)^2 , p^2_2 ; m_1 , m_2 \right)
\right\}
\nonumber\\
&=&
\frac{2 m_1^2}{m_1^2 - m_2^2} \, \frac{(p_1 \cdot p_2) - m_2^2}{\lambda ( (p_1 - p_2)^2 , m_1^2 , p_2^2 )}
\frac{1}{16 \pi^2} \ln \frac{m_2^2}{m_1^2}
\nonumber\\
&&
\!\!\!\!\!\!
+\,
2 \frac{(p_1 \cdot p_2) - m_1^2}{\lambda ( (p_1 - p_2)^2 , m_1^2 , p_2^2 )}
\left[ {\bar J} \left( (p_1 - p_2)^2 ; m_1 , m_2 \right)
- \frac{1}{16 \pi^2} 
\right]
\nonumber\\
&&
\!\!\!\!\!\!
-\,
{2} \, \frac{(p_1 \cdot p_2 )}{\lambda ( (p_1 - p_2)^2 , m_1^2 , p_2^2 )}
\left[ {\bar J} ( p_2^2 ; m_\gamma , m_2 )
- \frac{1}{16 \pi^2} 
\right]
\nonumber\\
&&
\!\!\!\!\!\!
-
2 \frac{m_1^2 (p_2^2 - m_2^2) }{\lambda ( (p_1 - p_2)^2 , m_1^2 , p_2^2 )} \,
C \left( (p_1 - p_2)^2 , p^2_2 ; m_1 , m_2 \right)
.
\end{eqnarray}
In the case where also $p_2^2 = m_2^2$, these expressions simplify further, and one obtains
\begin{eqnarray}
C_{11} \left( t ; m_1 , m_2 \right)  
&=&
\frac{1}{16 \pi^2} \, \frac{m_2^2 - m_1^2 - t}{\lambda (t , m_1^2 , m_2^2)}
 \frac{m_2^2}{m_1^2 - m_2^2}\ln \frac{m_2^2}{m_1^2}
\nonumber\\
&&
\!\!\!\!\!\!
+\,
\frac{1}{16 \pi^2} \, \frac{m_1^2 - m_2^2 - t}{\lambda (t , m_1^2 , m_2^2)}
\left[ 16 \pi^2 {\bar J} \left( t ; m_1 , m_2 \right)
- 1  
\right]
,
\end{eqnarray}
\begin{eqnarray}
C_{12} \left( t ; m_1 , m_2 \right) 
&=&
\frac{1}{16 \pi^2} \, \frac{m_1^2 - m_2^2 - t}{\lambda (t , m_1^2 , m_2^2)}
 \frac{m_1^2}{m_1^2 - m_2^2}\ln \frac{m_2^2}{m_1^2}
\nonumber\\
&&
\!\!\!\!\!\!
+\,
\frac{1}{16 \pi^2} \, \frac{m_2^2 - m_1^2 - t}{\lambda (t , m_1^2 , m_2^2)}
\left[ 16 \pi^2 {\bar J} \left( t ; m_1 , m_2 \right)
- 1  
\right]
.
\end{eqnarray}

For $p_2^2 = m^2_2$, the function $C \left( t ; m_1 , m_2 \right) $ itself
reads \cite{Vesna04,NuovoCim}
\begin{equation}
C \left( t  ; m_1 , m_2 \right) 
=
\frac{(-1)}{32 \pi^2} 
\int_0^1 dy
\,
\frac{1}{P_y^2} \, \ln \left( \frac{P_y^2 - i \epsilon}{m_\gamma^2} \right)
,
\end{equation}
with $P_y^\mu = y p_1^\mu + (1 - y) p_2^\mu$. One may write
\begin{equation}
P_y^2 = y^2 t - 2 a t y + p_2^2 \equiv t ( y - y_{\mbox{\tiny $+$}} ) ( y - y_{\mbox{\tiny $-$}} ) 
.
\end{equation}
The two roots of $P_y^2$ are then given by
\begin{equation}
y_{\mbox{\tiny $\pm$}} = a \pm b 
.
\end{equation}
with 
\begin{equation}
a \equiv \frac{1}{2} + \frac{p_2^2 - p_1^2}{2 t} 
,\quad b^2 \equiv a^2 - \frac{p_2^2}{t} = \frac{1}{4 t^2} \lambda(t, p_1^2 , p_2^2 )
.
\end{equation}
In the case under consideration, we have $p_1^\mu = k^\mu$, $p_2^\mu = p_e^\mu$, $m_1 = M_K$, $m_2 = m_e$,
with $k^2 = M_K^2$, $p_e^2 = m_e^2$, $t = (k - p_e)^2 > 0$. 
Then, $\lambda((k-p_e)^2, M_K^2 , m_e^2 ) \ge 0$, and
\begin{equation}
a = \frac{1}{2} \left[ 1 - \frac{ (M_K - m_e) (M_K + m_e) }{t} \right]
,\quad
b = \frac{1}{2 t} \lambda^{1/2} (t, M_K^2 , m_e^2)
,
\end{equation}
with $a < 0$ and $\vert a \vert > \vert b \vert$. Therefore,
\begin{eqnarray}
 C \left( t  ; m_1 , m_2 \right)  &=&
 \frac{1}{64 \pi^2} \frac{1}{bt}
 \ln \left( \frac{m_\gamma^2}{t} \right) \times \ln \left( 
\frac{y_{\mbox{\tiny $+$}} - 1}{y_{\mbox{\tiny $-$}} - 1} \cdot
\frac{y_{\mbox{\tiny $-$}} }{y_{\mbox{\tiny $+$}} } 
\right)
\nonumber\\
&&
\!\!\!\!\!
-
 \frac{1}{128 \pi^2} \frac{1}{bt}
 \left[
 \ln^2 ( 1 - y_{\mbox{\tiny $+$}} ) - \ln^2 ( - y_{\mbox{\tiny $+$}} )
- 
 \ln^2 ( 1 - y_{\mbox{\tiny $-$}} ) + \ln^2 ( - y_{\mbox{\tiny $-$}} )
 \right]
\nonumber\\
&&
\!\!\!\!\!
+
 \frac{1}{64 \pi^2} \frac{1}{bt}
 \left[
\ln ( 1 - y_{\mbox{\tiny $+$}} ) \ln ( 1 - y_{\mbox{\tiny $-$}} ) 
-
\ln ( - y_{\mbox{\tiny $+$}} ) \ln ( - y_{\mbox{\tiny $-$}} )
- 
2 \ln ( 2 b) \ln \left( 
\frac{y_{\mbox{\tiny $+$}} - 1}{y_{\mbox{\tiny $+$}} } 
\right)
 \right]
\nonumber\\
&&
\!\!\!\!\!
+
 \frac{1}{32 \pi^2} \frac{1}{bt}
 \left[
 {\rm Li}_2 \left( \frac{y_{\mbox{\tiny $+$}} - 1}{2 b} \right)
 -
 {\rm Li}_2 \left( \frac{y_{\mbox{\tiny $+$}}}{2 b} \right)
 \right]
 .
\end{eqnarray}
Notice that
\begin{equation}
\frac{y_{\mbox{\tiny $+$}} - 1}{y_{\mbox{\tiny $-$}} - 1} \cdot
\frac{y_{\mbox{\tiny $-$}} }{y_{\mbox{\tiny $+$}} } 
=
\frac{a^2 - b^2 - a + b}{a^2 - b^2 - a - b}
=
\frac{p_1^2 + p_2^2 - t + 2 t b}{p_1^2 + p_2^2 - t - 2 t b}
,
\end{equation}
so that the infrared divergent piece of $ C \left( t  ; m_1 , m_2 \right)$
is given by
\begin{equation}
C_{\mbox{\tiny{IR-div}}} \left( t ; m_1 , m_2 \right) =
- \frac{1}{32 \pi^2} \, \tau( p_1 , p_2 ) \times \ln m_\gamma
,
\end{equation}
with
\begin{eqnarray}
\tau( p_1 , p_2 ) \equiv 
\frac{1}{b t} \,
\ln \left[ \frac{p_1^2 + p_2^2 - t - 2 t b}{p_1^2 + p_2^2 - t + 2 t b} \right]
= \frac{1}{\sqrt{(p_1 \cdot p_2)^2 - p_1^2 p_2^2}}
\, \ln \frac{(p_1 \cdot p_2) - \sqrt{(p_1 \cdot p_2)^2 - p_1^2 p_2^2}}{(p_1 \cdot p_2) + \sqrt{(p_1 \cdot p_2)^2 - p_1^2 p_2^2}}
.
\lbl{tau_function}
\end{eqnarray}

Let us now consider the case where $p_2^2 \neq m_2^2$, but
with $p_1^2 = m_1^2$ as before. 
Going through the same steps as in the previous case,
one obtains
\begin{equation}
C \left( t , p^2_2 ; m_1 , m_2 \right) =
\frac{(-1)}{16 \pi^2} 
\int_0^1 dy
\,
\frac{1}{P_y^2} \, \ln \left[ \frac{P_y^2 + (1-y) ( m_2^2 - p_2^2) - i \epsilon }{(1-y) ( m_2^2 - p_2^2) - i \epsilon } \right]
,
\end{equation}
with $P_y^\mu = y p_1^\mu + (1 - y) p_2^\mu$. As before, one has
\begin{equation}
P_y^2 = y^2 t - 2 a t y + p_2^2 \equiv t ( y - y_{\mbox{\tiny $+$}} ) ( y - y_{\mbox{\tiny $-$}} ) 
,
\end{equation}
with
\begin{equation}
y_{\mbox{\tiny $\pm$}} = a \pm b 
,\quad
a \equiv \frac{1}{2} + \frac{p_2^2 - p_1^2}{2 t} 
,\quad b^2 \equiv a^2 - \frac{p_2^2}{t} = \frac{1}{4 t^2} \lambda(t, p_1^2 , p_2^2 )
.
\end{equation}
The case under consideration here corresponds to $p_1^\mu = k$, $p_1^2 = k^2 = M_K^2$,
whereas $p_2^\mu = (k - p_1 - p_2)^\mu$, $p_2^2 = s_e$, $t \equiv (p_1 - p_2)^2 = s_\pi$,
with $m_e^2 \le s_e \le (M_K - 2 M_\pi)^2$, $4 M_\pi^2 \le s_\pi \le (M_K - \sqrt{s_e})^2$.
Then $\lambda(t, p_1^2 , p_2^2 ) = \lambda(s_\pi, s_e , M_K^2 ) \ge 0$, so that $b$ is real,
with $\vert a \vert > \vert b \vert$. 
On the other hand, one has
\begin{equation}
P_y^2 + (1-y) ( m_2^2 - p_2^2) = y^2 t - 2 {\tilde a} t y + m_2^2  
\equiv t ( y - {\tilde y}_{\mbox{\tiny $+$}} ) ( y - {\tilde y}_{\mbox{\tiny $-$}} ) 
,
\end{equation}
with ${\tilde y}_{\mbox{\tiny $\pm$}} = {\tilde a} \pm  {\tilde b}$, and
\begin{equation}
{\tilde a} \equiv \frac{1}{2} + \frac{m_2^2 - p_1^2}{2 t} 
,\quad {\tilde b}^2 \equiv {\tilde a}^2 - \frac{m_2^2}{t} = \frac{1}{4 t^2} \lambda(t, p_1^2 , m_2^2 )
.
\end{equation}
In the case at hand, this gives ${\tilde a} = 1/2$, and ${\tilde b}^2 = (s_\pi - 4 M_K^2)/(4 s_\pi) <0$,
so that
\begin{equation}
{\tilde y}_{\mbox{\tiny $\pm$}} = \frac{1}{2} \left[ 1 \pm i  \sqrt{\frac{4 M_K^2}{s_\pi} - 1} ~\right]
.
\end{equation}
Then one obtains
\begin{eqnarray}
C \left( t , p^2_2 ; m_1 , m_2 \right)
&=&
\frac{1}{32 \pi^2} \frac{1}{ b t}
\ln \left( \frac{m_2^2 - p_2^2}{t} \right) \times \ln \left[ 
\frac{ ( y_{\mbox{\tiny $+$}} - 1 ) y_{\mbox{\tiny $-$}}}{ ( y_{\mbox{\tiny $-$}} - 1 ) y_{\mbox{\tiny $+$}}}
\right]
\nonumber\\
&&
\!\!\!\!\!
-
\frac{1}{32 \pi^2} \frac{1}{ b t}
\ln \left( \frac{y_{\mbox{\tiny $+$}} - 1}{y_{\mbox{\tiny $+$}}} \right)
\left[
\ln (y_{\mbox{\tiny $+$}} - {\tilde y}_{\mbox{\tiny $+$}} ) 
+
\ln (y_{\mbox{\tiny $+$}} - {\tilde y}_{\mbox{\tiny $-$}} )
\right]
\nonumber\\
&&
\!\!\!\!\!
+
\frac{1}{32 \pi^2} \frac{1}{ b t}
\ln \left( \frac{y_{\mbox{\tiny $-$}} - 1}{y_{\mbox{\tiny $-$}}} \right)
\left[
\ln (y_{\mbox{\tiny $-$}} - {\tilde y}_{\mbox{\tiny $+$}} ) 
+
\ln (y_{\mbox{\tiny $-$}} - {\tilde y}_{\mbox{\tiny $-$}} )
\right]
\nonumber\\
&&
\!\!\!\!\!
+
\frac{1}{32 \pi^2} \frac{1}{ b t}
\left[
{\rm Li}_2 \left(\frac{1 - y_{\mbox{\tiny $+$}}}{{\tilde y}_{\mbox{\tiny $+$}} - y_{\mbox{\tiny $+$}}} \right) 
-
{\rm Li}_2 \left(\frac{- y_{\mbox{\tiny $+$}}}{{\tilde y}_{\mbox{\tiny $+$}} - y_{\mbox{\tiny $+$}}} \right) 
+{\rm Li}_2 \left(\frac{1 - y_{\mbox{\tiny $+$}}}{{\tilde y}_{\mbox{\tiny $-$}} - y_{\mbox{\tiny $+$}}} \right) 
-
{\rm Li}_2 \left(\frac{- y_{\mbox{\tiny $+$}}}{{\tilde y}_{\mbox{\tiny $-$}} - y_{\mbox{\tiny $+$}}} \right) 
\right]
\nonumber\\
&&
\!\!\!\!\!
-
\frac{1}{32 \pi^2} \frac{1}{ b t}
\left[
{\rm Li}_2 \left(\frac{1 - y_{\mbox{\tiny $-$}}}{{\tilde y}_{\mbox{\tiny $+$}} - y_{\mbox{\tiny $-$}}} \right) 
-
{\rm Li}_2 \left(\frac{- y_{\mbox{\tiny $-$}}}{{\tilde y}_{\mbox{\tiny $+$}} - y_{\mbox{\tiny $-$}}} \right) 
+
{\rm Li}_2 \left(\frac{1 - y_{\mbox{\tiny $-$}}}{{\tilde y}_{\mbox{\tiny $-$}} - y_{\mbox{\tiny $-$}}} \right) 
-
{\rm Li}_2 \left(\frac{- y_{\mbox{\tiny $-$}}}{{\tilde y}_{\mbox{\tiny $-$}} - y_{\mbox{\tiny $-$}}} \right) 
\right]
\nonumber\\
&&
\!\!\!\!\!
-
\frac{1}{32 \pi^2} \frac{1}{ b t}
\left[
{\rm Li}_2 \left(\frac{1}{1 - y_{\mbox{\tiny $+$}}} \right) 
-
{\rm Li}_2 \left(\frac{1}{1 - y_{\mbox{\tiny $-$}}} \right)
\right]
\!.
\qquad~
\end{eqnarray}


\vfill
\newpage







\begin{thebibliography}{99}

\bibitem{Batley:2012rf} 
  J.~R.~Batley {\it et al.}  [NA48/2 Collaboration],
  Phys.\ Lett.\ B {\bf 715}, 105 (2012)
  [Addendum-ibid.\ B {\bf 740}, 364 (2014)]
  [arXiv:1206.7065 [hep-ex]].       
  
\bibitem{Batley:2007zz} 
  J.~R.~Batley {\it et al.}  [NA48/2 Collaboration],
  Eur.\ Phys.\ J.\ C {\bf 54}, 411 (2008).

\bibitem{Batley:2010zza}  
  J.~R.~Batley {\it et al.}  [NA48-2 Collaboration],
  Eur.\ Phys.\ J.\ C {\bf 70}, 635 (2010).

\bibitem{Weinberg:1966kf} 
  S.~Weinberg,
  Phys.\ Rev.\ Lett.\  {\bf 17}, 616 (1966).
  
\bibitem{Gasser:1983kx} 
  J.~Gasser and H.~Leutwyler,
  Phys.\ Lett.\ B {\bf 125}, 325 (1983).
  
\bibitem{Colangelo:2000jc} 
  G.~Colangelo, J.~Gasser and H.~Leutwyler,
  Phys.\ Lett.\ B {\bf 488}, 261 (2000)
  [hep-ph/0007112].

\bibitem{Colangelo:2001df} 
  G.~Colangelo, J.~Gasser and H.~Leutwyler,      
  Nucl.\ Phys.\ B {\bf 603}, 125 (2001)
  [hep-ph/0103088].
  
\bibitem{Batley:2014xwa}
J.~R.~Batley {\it et al.}  [NA48/2 Collaboration],
  JHEP {\bf 1408}, 159 (2014)
  [arXiv:1406.4749 [hep-ex]].
  

\bibitem{Vesna04}
V. Cuplov. {\it Brisure d’isospin et corrections radiatives au processus $K_{\ell 4}$}, 
PhD thesis, Universit\'e de la M\'editerran\'ee (2004).

\bibitem{Cuplov:2003bj} 
  V.~Cuplov and A.~Nehme,
 {\it Isospin breaking in K(l4) decays of the charged kaon},
  hep-ph/0311274.
 
\bibitem{Stoffer:2013sfa} 
  P.~Stoffer,
  Eur.\ Phys.\ J.\ C { 74}, 2749 (2014)
  [arXiv:1312.2066 [hep-ph]].   
  
\bibitem{NuovoCim}
B. Morel, Quoc-Hung Do, Nuovo Cim. A 46, 253 (1978).

\bibitem{Nehme:2003bz} 
  A.~Nehme,
  Nucl.\ Phys.\ B {\bf 682}, 289 (2004)
  [hep-ph/0311113].

\bibitem{Aoki:2013ldr}
  S.~Aoki, Y.~Aoki, C.~Bernard, T.~Blum, G.~Colangelo, M.~Della Morte, S.~Dürr and A.~X.~El Khadra {\it et al.},
  Eur.\ Phys.\ J.\ C {\bf 74} (2014) 9,  2890
  [arXiv:1310.8555 [hep-lat]].
  
\bibitem{BudiniFonda61}     
P. Budini, L. Fonda, Phys. Rev. Lett 6, 419 (1961).

\bibitem{Cabibbo:2004gq} 
  N.~Cabibbo,
  Phys.\ Rev.\ Lett.\  {\bf 93}, 121801 (2004)
  [hep-ph/0405001].
  
\bibitem{Cabibbo:2005ez} 
  N.~Cabibbo and G.~Isidori,
  JHEP {\bf 0503}, 021 (2005)
  [hep-ph/0502130].

\bibitem{Gamiz:2006km} 
  E.~Gamiz, J.~Prades and I.~Scimemi,
  Eur.\ Phys.\ J.\ C {\bf 50}, 405 (2007)
  [hep-ph/0602023].

\bibitem{Colangelo:2006va} 
  G.~Colangelo, J.~Gasser, B.~Kubis and A.~Rusetsky,
  Phys.\ Lett.\ B {\bf 638}, 187 (2006)
  [hep-ph/0604084].

\bibitem{Bernard:2013faa} 
  V.~Bernard, S.~Descotes-Genon and M.~Knecht,
  Eur.\ Phys.\ J.\ C {\bf 73}, 2478 (2013)
  [arXiv:1305.3843 [hep-ph]].

\bibitem{NA48-Kpi3}
J. R. Batley et al. [NA48/2 collaboration], Phys. Lett B 633, 173 (2006) [arXiv:hep-ex/0511056].

\bibitem{DescotesGenon:2012gv} 
  S.~Descotes-Genon and M.~Knecht,
  Eur.\ Phys.\ J.\ C {\bf 72}, 1962 (2012)
  [arXiv:1202.5886 [hep-ph]].

\bibitem{Colangelo:2008sm} 
  G.~Colangelo, J.~Gasser and A.~Rusetsky,
  Eur.\ Phys.\ J.\ C {\bf 59}, 777 (2009)
  [arXiv:0811.0775 [hep-ph]].
  
\bibitem{Bystritskiy:2009iv} 
  Y.~M.~Bystritskiy, S.~R.~Gevorkyan and E.~A.~Kuraev,
  Eur.\ Phys.\ J.\ C {\bf 64}, 47 (2009)
  [arXiv:0906.0516 [hep-ph]].

\bibitem{Isidori:2007zt} 
  G.~Isidori,
  Eur.\ Phys.\ J.\ C {\bf 53}, 567 (2008)
  [arXiv:0709.2439 [hep-ph]].
  
\bibitem{Knecht:1997jw}
  M.~Knecht, R.~Urech,
  Nucl.\ Phys.\  B {\bf 519}, 329 (1998)
  [arXiv:hep-ph/9709348].

\bibitem{Moussallam:1999aq}
  B.~Moussallam,
  Eur.\ Phys.\ J.\ C {\bf 14} (2000) 111
  [hep-ph/9909292].
  
\bibitem{Adeva:2011tc}
  B.~Adeva {\it et al.},
  Phys.\ Lett.\ B {\bf 704} (2011) 24
  [arXiv:1109.0569 [hep-ex]].

\bibitem{Barberio:1993qi} 
  E.~Barberio and Z.~Was,
  Comput.\ Phys.\ Commun.\  {\bf 79}, 291 (1994).
  
\bibitem{Golonka:2005pn} 
  P.~Golonka and Z.~Was,
  Eur.\ Phys.\ J.\ C {\bf 45}, 97 (2006)
  [hep-ph/0506026].
  
\bibitem{Nanava:2006vv} 
  G.~Nanava and Z.~Was,
  Eur.\ Phys.\ J.\ C {\bf 51}, 569 (2007)
  [hep-ph/0607019].

\bibitem{XuWas10}
Q. Xu, Z. Was, Chin. Phys. C {\bf 34}, 889 (2010).

\bibitem{Bijnens:1989mr} 
  J.~Bijnens,
  Nucl.\ Phys.\ B {\bf 337}, 635 (1990).
  
\bibitem{Gasser:1984gg} 
  J.~Gasser and H.~Leutwyler,
  Nucl.\ Phys.\ B {\bf 250}, 465 (1985).
  
\bibitem{Urech:1994hd} 
  R.~Urech,
  Nucl.\ Phys.\ B {\bf 433}, 234 (1995)
  [hep-ph/9405341].

\bibitem{Knecht:1999ag} 
  M.~Knecht, H.~Neufeld, H.~Rupertsberger and P.~Talavera,
  Eur.\ Phys.\ J.\ C {\bf 12}, 469 (2000)
  [hep-ph/9909284].

\bibitem{KMSF95}
   M. Knecht, B. Moussallam, J. Stern, N. H. Fuchs,
  Nucl. Phys.  B {\bf 457}, 513 (1995)  [hep-ph/9507319].
  

  
  
\end{thebibliography}
\end{document}